\documentclass[11pt,a4paper]{article}
\usepackage{jheppub,xcolor}
\usepackage[utf8]{inputenc}
\graphicspath{{./figs/}}
\usepackage[a4paper,top=5cm,bottom=0.5cm,left=5.0cm,right=-0.0cm,bindingoffset=0mm]{geometry}

\usepackage{amssymb}
\usepackage{dcolumn}	% Align table columns on decimal point
\usepackage{bm}			% bold math
\usepackage{bbm} 		% blackboard math for \mathbbm{1}
\usepackage{multirow}
\usepackage{slashed}
\usepackage{blkarray}
\usepackage{booktabs}
\usepackage{makecell}
\usepackage{longtable}
\usepackage{caption}
\usepackage{subcaption}
\usepackage{placeins}
\usepackage{graphicx}
\usepackage{float}
\usepackage{amsmath}
\usepackage{array}
\usepackage{multirow}
\usepackage{verbatim}
\usepackage{hyperref}
\usepackage{lscape} % For tables in landscape
\usepackage{tablefootnote} % for table footnotes
\usepackage[font=small,labelfont=bf]{caption}
\usepackage{tikz-feynman,amsmath}

\newcommand\sw{s_\text{w}}
\newcommand\cw{c_\text{w}}
\newcommand{\code}[1]{\mbox{\texttt{#1}}}
\newcommand\im{\mathtt{IM}}
\newcommand\re{\mathtt{RE}}

\newcommand{\Oth}{\mcO_{t \varphi}}
\newcommand{\OtZ}{\mcO_{t Z}}
\newcommand{\OtW}{\mcO_{t W}}
\newcommand{\Otg}{\mcO_{t G}}
\newcommand{\OpW}{\mcO_{\varphi W}}

\newcommand{\OpWtil}{\mcO_{\varphi \tilde{W}}}
\newcommand{\OpBtil}{\mcO_{\varphi \tilde{B}}}
\newcommand{\OpWB}{\mcO_{\varphi W B}}

\newcommand{\OpWBtil}{\mcO_{\varphi \tilde{W}B}}

\newcommand{\Opgtil}{\mcO_{\varphi \tilde{G}}}
\newcommand{\op}{\mathcal{O}}

\newcommand{\cth}{c_{t \varphi }}
\newcommand{\ctg}{c_{t G}}

\newcommand{\cpgtil}{c_{\varphi \tilde{G}}}
\newcommand{\ctW}{c_{t W}}
\newcommand{\ctB}{c_{t B}}
\newcommand{\ctZ}{c_{t Z}}

\newcommand{\cpBtil}{c_{\varphi \tilde{B}}}
\newcommand{\cpW}{c_{\varphi W}}
\newcommand{\cpWtil}{c_{\varphi \tilde{W}}}
\newcommand{\cpWBtil}{c_{\varphi \tilde{W}B}}
\newcommand{\cpWB}{c_{\varphi W B}}

\newcommand{\mcO}{\mathcal{O}}
\newcommand{\sss}{\scriptscriptstyle}
\newcommand{\pdp}{\ensuremath{\varphi^\dagger\varphi}}

\newcommand\logsmt{\text{log}\left(\frac{s}{m_t^2} \right)}

\title{CP violation in loop-induced diboson production}

\author[a]{Marion~O. A.~Thomas}
\author[a]{and Eleni~Vryonidou}
\affiliation[a]{Dept. of Physics and Astronomy, University of Manchester, Manchester M13 9PL, UK}

\emailAdd{marion.thomas@manchester.ac.uk}
\emailAdd{eleni.vryonidou@manchester.ac.uk}

\date{\today}

\abstract{
We consider the impact of CP-violating Higgs and top interactions on diboson production from gluon fusion within the Standard Model Effective Field Theory framework. We systematically study differential distributions for double Higgs, double $W$ and $Z$ production and compare their features to those obtained from CP-conserving interactions. For electroweak gauge boson production, we explore the impact of the new interactions on the angular distributions of the leptonic decay products and the associated gauge boson polarisation fractions both inclusively and differentially with the transverse momentum of the gauge boson. 
}

\begin{document}
\maketitle
\newpage

\section{Introduction}

The existence of CP violation is key to address the mystery of matter-antimatter asymmetry in the Universe. As the amount of CP violation in the Standard Model (SM) is insufficient, the search for new sources of CP violation is of high priority for the particle physics community and constitutes a key goal of the LHC programme. As such, several phenomenological  studies have focused on proposing appropriate observables to maximise sensitivity to CP violation at the LHC. 
These analyses often employ the Standard Model Effective Field Theory (SMEFT) \cite{1706.08945}, an extension of the SM  allowing us to parametrise the effect of heavy new physics in a model-independent manner.  

In the precision programme of indirect searches for new physics effects, both CP-conserving and CP-violating, it is imperative to explore different processes and benefit from the plethora of differential measurements taking place at the LHC. In this effort the class of processes involving two electroweak gauge bosons i.e. $WW$, $ZZ$ and double Higgs production $HH$, plays a special role in probing the gauge symmetries and thus the  EW couplings of the SM, as well as the Higgs potential. Electroweak gauge boson pair production is dominated by the quark-initiated contributions, but also receives a very important loop-induced contribution which provides a crucial bridge to top quark interactions. On the other hand, for double Higgs production the leading production mode is loop-induced and dominated by top quark loops.

The loop-induced $gg \rightarrow VV, HH \, (V= W,Z)$ processes are therefore particularly well suited to probe BSM signatures in the top quark, electroweak and Higgs sectors. Furthermore, it has been shown that these processes are particularly sensitive to new interactions, with some of the helicity amplitudes for these processes growing with the center-of-mass energy \cite{1608.00977,2004.02031,2306.09963} when modified by SMEFT operators.  Therefore diboson production from gluon fusion can benefit from the high energies probed at the HL-LHC and at future high-energy proton  colliders to reveal signs of new physics. 

In this work we focus on the loop-induced diboson production processes and explore systematically how these can be affected by the presence of CP-violating interactions within the Standard Model Effective Field Theory. Whilst CP violation has been explored extensively for the quark anti-quark initiated diboson processes, these studies typically focus on CP violation in the triple gauge couplings \cite{1612.01808,1804.01477,1810.11657,2009.13394,2102.01115}. Our study will enable for the first time a comprehensive study of CP-violating effects in diboson production, including also CP violation in top quark interactions. This will allow us to establish whether this class of processes can set competitive constraints on CP-violating interactions. These constraints can then be compared to those obtained through low energy measurements of electric dipole moments (EDMs), a class of observables extremely sensitive to CP-violating new physics leading to stringent constraints often found to be complementary to LHC constraints \cite{1503.01114,1510.00725,1603.03049,1903.03625,2109.15085}.

In order to explore these effects, an implementation of these processes in Monte Carlo generators is imperative. Whilst the effects of CP-conserving SMEFT operators in loop-induced processes can be explored with public Monte Carlo tools \cite{2008.11743}, such an implementation is not available for CP-violating operators. 
CP violation can be studied at tree-level in dimension-six SMEFT coefficients using the \code{SMEFTsim} \cite{2012.11343} package. A few next-to-leading order (NLO) studies exist, with so far the CP-odd purely bosonic operators having been considered for $WZ, W\gamma$ and $WW$ production \cite{1901.04821,2405.19083}. Additionally, CP-violating top-Higgs interactions at one-loop can be probed with the Higgs Characterisation framework \cite{1306.6464,1407.5089}, and CP-violating EFT effects in Higgs pair production at NLO QCD were calculated in the heavy top quark limit in \cite{1705.05314}. In this work we take another step towards more comprehensive  CP violation studies in the SMEFT by extending the \code{SMEFTatNLO} UFO \cite{2008.11743} with the CP-odd operators entering in $gg \rightarrow HH, VV$. 
The rest of this article is organised as follows. In Section \ref{sec:UFO}, after setting our conventions, we discuss the implementation of the relevant CP-odd operators in the \code{SMEFTatNLO} model. In particular we give the results for the counterterms needed for the calculation, both rational terms and UV counterterms. Section \ref{2to2} is dedicated to studying the impact of CP-violating operators entering gluon-induced double Higgs, double $Z$ and double $W$ production. We perform a phenomenological analysis of $ZZ$ and $WW$ production in terms of angular and polarisation observables in Section \ref{2to4} before concluding in Section \ref{sec:Conclusions}.

\section{CP-odd operators in \code{SMEFTatNLO}}
\label{sec:UFO}

\subsection{Conventions and Methodology}
\label{sec:Notation}

\renewcommand{\arraystretch}{1.5}
\begin{table}
    \centering
    \begin{tabular}{lll | lll} 
    \toprule
    \multicolumn{3}{c}{Hermitian operators}& \multicolumn{3}{c}{Non-hermitian operators}\\
    \midrule
    \midrule
 
    $\mathcal{O}_i$ & $c_i$ & Definition & $\mathcal{O}_i$ & $c_i$ & Definition  \\
    \midrule

    $\Opgtil$&$\cpgtil$ & $(\pdp) \widetilde{G}^{\mu\nu}_{\sss A}\,G_{\mu\nu}^{\sss A}$&

    $\op_{tG}$&$\ctg$ & $g_s\,\big(\bar{Q}\sigma^{\mu\nu}\,T_{\sss A}\,t\big)\,\tilde{\varphi}\,G^A_{\mu\nu} + \text{h.c.}$
    \\
    
    $\op_{\varphi \tilde{B}}$& $\cpBtil$ &
     $(\pdp) \widetilde{B}^{\mu\nu}\,B_{\mu\nu}$&

     $\op_{t\varphi}$ & $\cth$ & $(\pdp) \, (\bar{Q}\,t\,\tilde{\varphi}) + \text{h.c.}$    
     \\
     
    $\op_{\varphi \tilde{W}}$&$\cpWtil$ &$(\pdp)\widetilde{W}^{\mu\nu}_{\sss I}\,W_{\mu\nu}^{\sss I}$ &
     $\op_{tW}$&$\ctW$ & $\big(\bar{Q}\sigma^{\mu\nu}\,\tau_{\sss I}\,t\big)\, \tilde{\varphi}\,W^I_{\mu\nu} + \text{h.c.}$
     \\

      $\op_{\varphi \tilde{W}B}$&$\cpWBtil$&
     $(\varphi^\dagger \tau_{\sss I}\varphi)\,B^{\mu\nu}\widetilde{W}_{\mu\nu}^{\sss I}\,$&

      $\op_{tB}$&$\ctB$&
    $\big(\bar{Q}\sigma^{\mu\nu}\,t\big)
    \,\tilde{\varphi}\,B_{\mu\nu}
    + \text{h.c.}$
    \\

    &&& $\mathcal{O}_{tZ}$& $\ctZ$ & $-\sin\theta_W\, \ctB + \cos\theta_W\, \ctW$     
     \\
    \bottomrule    
    \end{tabular}
    \caption{ CP-odd dimension-$6$ operators $\mathcal{O}_i$ and their associated Wilson Coefficients $c_i$.
    }
    \label{operators}
\end{table}

In this paper we extend the \code{SMEFTatNLO} model \cite{Degrande:2020evl} to include the CP-odd operators relevant for diboson production from gluon fusion. Using the Warsaw basis of dimension-$6$ operators along with a U$(2)_q\times$U$(3)_d\times$U$(2)_u\times(\text{U}(1)_\ell\times\text{U}(1)_e)^3$ flavour assumption, we find that there are $9$ CP-odd operators relevant for the processes considered. These are presented in Table~\ref{operators}. The left column shows the CPV hermitian operators for which the Wilson coefficients $\cpgtil, c_{\varphi \tilde{B}}, c_{\varphi \tilde{W}} $  and $c_{\varphi \tilde{W}B}$ are real. The right column defines the non-hermitian operators and their associated Wilson coefficients which can be complex. CP violating effects arise when the imaginary part of $\ctg$, $\cth$, $\ctW$ or $\ctB$ is non-zero. We thus use the following notation for the Wilson coefficients of the non-hermitian operators:
\begin{equation}
    c_i = \mathtt{RE}c_i+ i \, \mathtt{IM}c_i
\end{equation}
while for the Wilson coefficients of the hermitian operators $c_i = \mathtt{RE}c_i$ and we drop the $\mathtt{RE}$ label. Furthermore the left-handed third-generation quark doublet is denoted by $Q$ while $t$ denotes a right-handed top-quark field. $T^A = \frac{1}{2} \lambda^A$ are the $SU(3)$ generators where $\lambda^A$ are the Gell-Mann matrices and $\sigma^{\mu\nu} = \frac{i}{2} [\gamma^\mu,\gamma^\nu]$. The Higgs doublet, $\varphi$,  has a vacuum expectation value $v/\sqrt{2}$. $G^A_\mu$, $W_\mu$ and $B_\mu$ are the gauge bosons fields such that $G^A_{\mu\nu}$, $W_{\mu\nu}$ and $B_{\mu\nu}$ stand for the $SU(3)_C$, $SU(2)_L$ and $U(1)_Y$ field strength tensors respectively. Covariant derivatives are defined with the following convention:
\begin{equation}
    D_\mu \varphi = \Big(\partial_\mu -i \frac{g}{2} \tau_I W_\mu^I -i \frac{g'}{2}B_\mu \Big) \varphi
\end{equation}
where $\tau_{I}$ are the Pauli sigma matrices, and the dual tensors are given by:

\begin{equation}
    \widetilde{X}_{\mu \nu} = \frac{1}{2} \varepsilon_{\mu \nu \rho \sigma} X^{\rho \sigma}.
\end{equation}
In order to extend the \code{SMFTatNLO} model to study $gg \rightarrow HH, ZZ, WW$ modified by the CP-odd operators from Table~\ref{operators} we had to implement in the UFO model the following ingredients:

\begin{itemize}
    \item The Feynman rules for the vertices modified by the CP-violating operators. These have been checked against the implementation in \code{SMEFTsim} \cite{2012.11343}.

    \item The Feynman rules for the rational terms $R_2$ \cite{0609007}. These are needed within the Ossola-Papadopoulos-Pittau (OPP) reduction method \cite{0609007}, as implemented in \code{CutTools} allowing the automation of one-loop computations in \code{MadLoop} \cite{1103.0621}. Their calculation and implementation is discussed in details in Sec.~\ref{sec:UFO_R2}.

    \item The UV counterterms for the UV-divergent contributions arising in the presence of $\im\ctg$. Their calculation and implementation is discussed in Sec.~\ref{sec:UFO_UV}. 
\end{itemize}

\subsection{Feynman rules for the rational part}
\label{sec:UFO_R2}

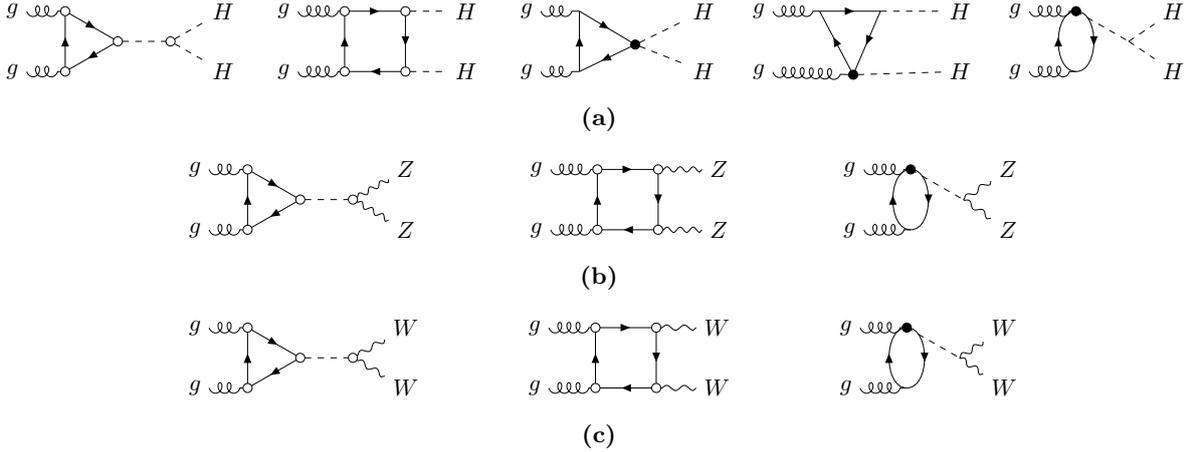
\begin{figure}
    \centering
%%%%%%%%%%%%%%%%%%%%%%%%%%% gg > HH %%%%%%%%%%%%%%%%%%%%%%%%%%%%%%%
    \begin{subfigure}[h]{\textwidth}
    \centering
%%%%% ggHH SM triangle
\begin{centering}
\scalebox{0.8}{
\begin{tikzpicture}
\begin{feynman}
\vertex (i1) {$g$};
\vertex[below=1cm of i1] (i2) {$g$};
\vertex[right=0.866cm of i1, empty dot] (l1) {};
\vertex[below=1cm of l1, empty dot] (l3) {};
\vertex[below right=0.5cm and 0.866cm of l1, empty dot] (l2) {};
\vertex[right=0.866cm of l2, empty dot] (v2) {};
\vertex[right=2.6cm of l1] (f1) {$H$};
\vertex[right=2.6cm of l3] (f2) {$H$};
\diagram* {{[edges={fermion, arrow size=1pt}]
(l1) -- (l2) -- (l3) -- (l1),},
(i1) -- [gluon] (l1),
(i2) -- [gluon] (l3),
(l2) -- [scalar] (v2),
(v2) -- [scalar] (f1),
(v2) -- [scalar] (f2) };
\end{feynman}
\end{tikzpicture}}
\hfill
%%%%% ggHH SM box
\hfill
\scalebox{0.8}{
\begin{tikzpicture}
\begin{feynman}
\vertex (i1) {$g$};
\vertex[below=1cm of i1] (i2) {$g$};
\vertex[right=1cm of i1, empty dot] (l1) {};
\vertex[right=1cm of l1, empty dot] (l2) {};
\vertex[below=1cm of l2, empty dot] (l3) {};
\vertex[below=1cm of l1, empty dot] (l4) {};
\vertex[right=1cm of l3] (f1) {$H$};
\vertex[right=1cm of l2] (f2) {$H$};
\diagram* {{[edges={fermion, arrow size=1pt}]
(l1) -- (l2) -- (l3) -- (l4) -- (l1),},
(i1) -- [gluon] (l1),
(i2) -- [gluon] (l4),
(l3) -- [scalar] (f1),
(l2) -- [scalar] (f2) };
\end{feynman}
\end{tikzpicture}}
\hfill
%%%%% ggHH ctp ttHH triangle
\hfill
\scalebox{0.8}{
\begin{tikzpicture}
\begin{feynman}
\vertex (i1) {$g$};
\vertex[below=1cm of i1] (i2) {$g$};
\vertex[right=0.866cm of i1] (l1);
\vertex[below=1cm of l1] (l3);
\vertex[below right=0.5cm and 0.866cm of l1, dot] (l2) {};
\vertex[right=1.732cm of l1] (f1) {$H$};
\vertex[right=1.732cm of l3] (f2) {$H$};
\diagram* {{[edges={fermion, arrow size=1pt}]
(l1) -- (l2) -- (l3) -- (l1),},
(i1) -- [gluon] (l1),
(i2) -- [gluon] (l3),
(l2) -- [scalar] (f1),
(l2) -- [scalar] (f2) };
\end{feynman}
\end{tikzpicture}}
%%%%% ggHH ctG ttgH triangle
\hfill
\scalebox{0.8}{
\begin{tikzpicture}
\begin{feynman}
\vertex (i1) {$g$};
\vertex[below=1cm of i1] (i2) {$g$};
\vertex[right=1cm of i1] (l1);
\vertex[right=1cm of l1] (l2);
\vertex[below right = 1cm and 0.5cm of l1, dot] (l3) {};
\vertex[right=1cm of l2] (f2) {$H$};
\vertex[below=1cm of f2] (f1) {$H$};
\diagram* {{[edges={fermion, arrow size=1pt}]
(l1) -- (l2) -- (l3) -- (l1),},
(i1) -- [gluon] (l1),
(i2) -- [gluon] (l3),
(l3) -- [scalar] (f1),
(l2) -- [scalar] (f2) };
\end{feynman}
\end{tikzpicture}}
%%%%% ggHH ctG blob
\hfill
\scalebox{0.8}{
\begin{tikzpicture}
\begin{feynman}
\vertex (i1) {$g$};
\vertex[below=1cm of i1] (i2) {$g$};
\vertex[right=1cm of i1] (l1);
\vertex[right=1cm of i1, dot] (ph1) {};
\vertex[right=1cm of i2] (l2);
\vertex[below right=0.5 cm and 0.866cm of l1] (v1);
\vertex[right=2.6cm of i1] (f1) {$H$};
\vertex[right=2.6cm of i2] (f2) {$H$};
\diagram* {{[edges={fermion, arrow size=1pt}]
(l1) -- [out=0, in=0] (l2) -- [out=180, in=180] (l1),},
(i1) -- [gluon] (l1),
(i2) -- [gluon] (l2),
(l1) -- [scalar] (v1),
(v1) -- [scalar] (f1),
(v1) -- [scalar] (f2) };
\end{feynman}
\end{tikzpicture}}
\hfill
\end{centering}
\subcaption{}
\label{diagHH}
\end{subfigure}
\par\medskip 
%%%%%%%%%%%%%%%%%%%%%%%%%%% gg > ZZ %%%%%%%%%%%%%%%%%%%%%%%%%%%%%%%
\begin{subfigure}{0.7\textwidth}
  \centering
%%%%%SM topology, gg to triangle to H and ZZH coupling
\scalebox{0.8}{
\begin{tikzpicture}
\begin{feynman}
\vertex (i1) {$g$};
\vertex[below=1cm of i1] (i2) {$g$};
\vertex[right=0.866cm of i1, empty dot] (l1){};
\vertex[below=1cm of l1, empty dot] (l3){};
\vertex[below right=0.5cm and 0.866cm of l1, empty dot] (l2) {};
\vertex[right=0.866cm of l2, empty dot] (v2) {};
\vertex[right=2.6cm of l1] (f1) {$Z$};
\vertex[right=2.6cm of l3] (f2) {$Z$};
\diagram* {{[edges={fermion, arrow size=1pt}]
(l1) -- (l2) -- (l3) -- (l1),},
(i1) -- [gluon] (l1),
(i2) -- [gluon] (l3),
(l2) -- [scalar] (v2),
(v2) -- [photon] (f1),
(v2) -- [photon] (f2) };
\end{feynman}
\end{tikzpicture}}
%%%%%SM topology, gg to box and Z emitted from vertices.
\hfill
\scalebox{0.8}{
\begin{tikzpicture}
\begin{feynman}
\vertex (i1) {$g$};
\vertex[below=1cm of i1] (i2) {$g$};
\vertex[right=1cm of i1, empty dot] (l1){};
\vertex[right=1cm of l1, empty dot] (l2) {};
\vertex[below=1cm of l2, empty dot] (l3) {};
\vertex[below=1cm of l1, empty dot] (l4){};
\vertex[right=1cm of l3] (f1) {$Z$};
\vertex[right=1cm of l2] (f2) {$Z$};
\diagram* {{[edges={fermion, arrow size=1pt}]
(l1) -- (l2) -- (l3) -- (l4) -- (l1),},
(i1) -- [gluon] (l1),
(i2) -- [gluon] (l4),
(l3) -- [photon] (f1),
(l2) -- [photon] (f2) };
\end{feynman}
\end{tikzpicture}}
%%%%% ctg topology, gg to blob to H propagator, ggtt coupling.
\hfill
\scalebox{0.8}{
\begin{tikzpicture}
\begin{feynman}
\vertex (i1) {$g$};
\vertex[below=1cm of i1] (i2) {$g$};
\vertex[right=1cm of i1] (l1);
\vertex[right=1cm of i1, dot] (ph1) {};
\vertex[right=1cm of i2] (l2);
\vertex[below right=0.5 cm and 0.866cm of l1] (v1);
\vertex[right=2.6cm of i1] (f1) {$Z$};
\vertex[right=2.6cm of i2] (f2) {$Z$};
\diagram* {{[edges={fermion, arrow size=1pt}]
(l1) -- [out=0, in=0] (l2) -- [out=180, in=180] (l1),},
(i1) -- [gluon] (l1),
(i2) -- [gluon] (l2),
(l1) -- [scalar] (v1),
(v1) -- [photon] (f1),
(v1) -- [photon] (f2) };
\end{feynman}
\end{tikzpicture}}
\subcaption{}
    \label{diagZZ}
\end{subfigure}
\par\medskip 
%%%%%%%%%%%%%%%%%%%%%%%%%%% gg > WW %%%%%%%%%%%%%%%%%%%%%%%%%%%%%%%
\begin{subfigure}{0.7\textwidth}
  \centering
%%%%%SM topology, gg to triangle to H and ZZH coupling
\scalebox{0.8}{
\begin{tikzpicture}
\begin{feynman}
\vertex (i1) {$g$};
\vertex[below=1cm of i1] (i2) {$g$};
\vertex[right=0.866cm of i1, empty dot] (l1){};
\vertex[below=1cm of l1, empty dot] (l3){};
\vertex[below right=0.5cm and 0.866cm of l1, empty dot] (l2) {};
\vertex[right=0.866cm of l2, empty dot] (v2) {};
\vertex[right=2.6cm of l1] (f1) {$W$};
\vertex[right=2.6cm of l3] (f2) {$W$};
\diagram* {{[edges={fermion, arrow size=1pt}]
(l1) -- (l2) -- (l3) -- (l1),},
(i1) -- [gluon] (l1),
(i2) -- [gluon] (l3),
(l2) -- [scalar] (v2),
(v2) -- [photon] (f1),
(v2) -- [photon] (f2) };
\end{feynman}
\end{tikzpicture}}
%%%%%SM topology, gg to box and Z emitted from vertices.
\hfill
\scalebox{0.8}{
\begin{tikzpicture}
\begin{feynman}
\vertex (i1) {$g$};
\vertex[below=1cm of i1] (i2) {$g$};
\vertex[right=1cm of i1, empty dot] (l1){};
\vertex[right=1cm of l1, empty dot] (l2) {};
\vertex[below=1cm of l2, empty dot] (l3) {};
\vertex[below=1cm of l1, empty dot] (l4){};
\vertex[right=1cm of l3] (f1) {$W$};
\vertex[right=1cm of l2] (f2) {$W$};
\diagram* {{[edges={fermion, arrow size=1pt}]
(l1) -- (l2) -- (l3) -- (l4) -- (l1),},
(i1) -- [gluon] (l1),
(i2) -- [gluon] (l4),
(l3) -- [photon] (f1),
(l2) -- [photon] (f2) };
\end{feynman}
\end{tikzpicture}}
%%%%% ctg topology, gg to blob to H propagator, ggtt coupling.
\hfill
\scalebox{0.8}{
\begin{tikzpicture}
\begin{feynman}
\vertex (i1) {$g$};
\vertex[below=1cm of i1] (i2) {$g$};
\vertex[right=1cm of i1] (l1);
\vertex[right=1cm of i1, dot] (ph1) {};
\vertex[right=1cm of i2] (l2);
\vertex[below right=0.5 cm and 0.866cm of l1] (v1);
\vertex[right=2.6cm of i1] (f1) {$W$};
\vertex[right=2.6cm of i2] (f2) {$W$};
\diagram* {{[edges={fermion, arrow size=1pt}]
(l1) -- [out=0, in=0] (l2) -- [out=180, in=180] (l1),},
(i1) -- [gluon] (l1),
(i2) -- [gluon] (l2),
(l1) -- [scalar] (v1),
(v1) -- [photon] (f1),
(v1) -- [photon] (f2) };
\end{feynman}
\end{tikzpicture}}

\subcaption{}
    \label{diagWW}
\end{subfigure}
    \caption{Diagram topologies that enter in the computation of (a) $gg\to HH$, (b) $gg\to ZZ$ and (c) $gg\to WW$  in the SMEFT at one-loop. The empty dots represent couplings that could be either SM-like or modified by CP-violating dimension-6 operators. The filled dots represent vertices generated only by CP-violating dimension-6 operators. Only one insertion of dimension-6 operators is allowed per diagram. }
    \label{fig:FeynmanDiags}
\end{figure}

The Feynman rules for the $R_2$ rational part of the one-loop CP-violating amplitudes are calculated following the method outlined in \cite{0802.1876, 0903.0356, 0910.3130}. 
The relevant Feynman rules were implemented in the \code{SMEFTatNLO} FeynArts model, and the $R_2$ terms contributing to $gg \rightarrow HH$, $WW$, $ZZ$ in the presence of the CP-violating operators from Table~\ref{operators} were computed in Mathematica. Representative diagrams for the processes considered are displayed in Fig.~\ref{fig:FeynmanDiags}.  %Interestingly ,
 We summarise in Table~\ref{R2Table2} which $R_2$ terms are non-zero for the different processes considered. We note that the SM, the CP-even, and the CP-odd coefficients can have different contributions to the $R_2$ of a given process. While this paper focuses on $HH, ZZ$ and $WW$ production, we also consider $gg \rightarrow \gamma Z, \gamma \gamma$ as these processes enter in $4$-lepton production which we discuss in Sec.~\ref{2to4}. 

\begin{table}
   \centering
   \renewcommand{\arraystretch}{1.1}
   \begin{tabular}{c|c|c|c|c|c|c}
   \toprule
   Coefficients & $ggH$ & $ggHH$ & $ggZZ$ & $ggWW$& $gg\gamma Z$ & $gg \gamma \gamma$ \\
   \midrule
   \midrule
$SM$ &$\checkmark$ & $\checkmark$ & $\checkmark$ & $\checkmark$ & $\checkmark$ &  $\checkmark$
\\
\midrule
%%%%%%%%%%%%%%%%%%%%%%%%%%%%%%%%%%%%%%%%%%%%%%%%%%%%%%%%%%%%%%%%%%%%%%%%%%%%%%%%%%%%%%%%%%%%%%%%%%%%%%%%%%%%%%%
$\re\cth$
& $\checkmark$ & $\checkmark$ & $-$ &  $-$ &  $-$ &  $-$
\\
%%%%%%%%%%%%%%%%%%%%%%%%%%%%%%%%%%%%%%%%%%%%%%%%%%%%%%%%%%%%%%%%%%%%%%%%%%%%%%%%%%%%%%%%%%%%%%%%%%%%%%%%%%%%%%%
$\im\cth$
&$0$ & $0$ & $-$ &  $-$ &  $-$ &  $-$
\\
%%%%%%%%%%%%%%%%%%%%%%%%%%%%%%%%%%%%%%%%%%%%%%%%%%%%%%%%%%%%%%%%%%%%%%%%%%%%%%%%%%%%%%%%%%%%%%%%%%%%%%%%%%%%%%%
\midrule
%%%%%%%%%%%%%%%%%%%%%%%%%%%%%%%%%%%%%%%%%%%%%%%%%%%%%%%%%%%%%%%%%%%%%%%%%%%%%%%%%%%%%%%%%%%%%%%%%%%%%%%%%%%%%%%
$\re\ctg$
& $\checkmark$ & $\checkmark$ & $0$& $0$& $0$& $0$
\\
%%%%%%%%%%%%%%%%%%%%%%%%%%%%%%%%%%%%%%%%%%%%%%%%%%%%%%%%%%%%%%%%%%%%%%%%%%%%%%%%%%%%%%%%%%%%%%%%%%%%%%%%%%%%%%%
$\im\ctg$
& $\checkmark$ & $\checkmark$ & $0$& $0$& $0$& $0$
\\
%%%%%%%%%%%%%%%%%%%%%%%%%%%%%%%%%%%%%%%%%%%%%%%%%%%%%%%%%%%%%%%%%%%%%%%%%%%%%%%%%%%%%%%%%%%%%%%%%%%%%%%%%%%%%%%
\midrule
%%%%%%%%%%%%%%%%%%%%%%%%%%%%%%%%%%%%%%%%%%%%%%%%%%%%%%%%%%%%%%%%%%%%%%%%%%%%%%%%%%%%%%%%%%%%%%%%%%%%%%%%%%%%%%%
$\re\ctZ$
&  $-$ &  $-$ & $0$& $-$ &$0$ & $0$ 
\\
%%%%%%%%%%%%%%%%%%%%%%%%%%%%%%%%%%%%%%%%%%%%%%%%%%%%%%%%%%%%%%%%%%%%%%%%%%%%%%%%%%%%%%%%%%%%%%%%%%%%%%%%%%%%%%%
$\im\ctZ$
&  $-$ &  $-$ & $0$& $-$ & $0$ & $0$ 
\\
%%%%%%%%%%%%%%%%%%%%%%%%%%%%%%%%%%%%%%%%%%%%%%%%%%%%%%%%%%%%%%%%%%%%%%%%%%%%%%%%%%%%%%%%%%%%%%%%%%%%%%%%%%%%%%%
\midrule
%%%%%%%%%%%%%%%%%%%%%%%%%%%%%%%%%%%%%%%%%%%%%%%%%%%%%%%%%%%%%%%%%%%%%%%%%%%%%%%%%%%%%%%%%%%%%%%%%%%%%%%%%%%%%%%
$\re\ctW$
&  $-$ &  $-$ & $-$& $0$ & $0$& $0$ 
\\
%%%%%%%%%%%%%%%%%%%%%%%%%%%%%%%%%%%%%%%%%%%%%%%%%%%%%%%%%%%%%%%%%%%%%%%%%%%%%%%%%%%%%%%%%%%%%%%%%%%%%%%%%%%%%%%
$\im\ctW$
&  $-$ &  $-$ & $-$& $0$ & $0$ & $0$ 
\\
%%%%%%%%%%%%%%%%%%%%%%%%%%%%%%%%%%%%%%%%%%%%%%%%%%%%%%%%%%%%%%%%%%%%%%%%%%%%%%%%%%%%%%%%%%%%%%%%%%%%%%%%%%%%%%%
   \bottomrule
   \end{tabular}
   \caption{Overview indicating which Wilson coefficients return a non-zero $R_2$ ($\checkmark$) or a vanishing one ($0$) for the effective vertices considered. A $-$ indicates that the coefficient does not enter in the effective vertex. }
   \label{R2Table2}
\end{table}

\begin{table}
    \centering
    \renewcommand{\arraystretch}{2}
    \begin{tabular}{c l}

         \begin{tikzpicture}[baseline={(current bounding box.center)}, vertical=e to f] {a [nudge=(-30:5mm)]
         \begin{feynman}
         \vertex (i1) {$p_1$, $a$, $\alpha$};
         \vertex[below=2cm of i1] (i2) {$p_2$, $b$, $\beta$};
         \vertex[below right=1cm and 1cm of i1,dot,minimum size=2.5mm] (l1){};
 
         \vertex[right=1.2cm of l1] (f1) {$H$};

         \diagram* {
         (i1) -- [gluon](l1),
         (i2) -- [gluon] (l1),
         (l1) -- [scalar] (f1)};
 
         \end{feynman} }
        \end{tikzpicture} &
         
             $=\,\,\frac{\mathtt{IM}\ctg}{\Lambda^2} \frac{\sqrt{2}\, i\, g_s^2\, m_t }{3\pi^2} \varepsilon^{\alpha \beta p_1 p_2 }  \delta_{ab} $
        \\
        \\
         
\begin{tikzpicture}[baseline={(current bounding box.center)}, vertical=e to f] {a [nudge=(-30:5mm)]
\begin{feynman}
\vertex (i1) {$p_1$, $a$, $\alpha$};
\vertex[below=2cm of i1] (i2) {$p_2$, $b$, $\beta$};
\vertex[below right=1cm and 1cm of i1,dot,minimum size=2.5mm] (l1){};
\vertex[right=1.8cm of i1] (f1) {$H$};
\vertex[right=1.8cm of i2] (f2) {$H$};
\diagram* {
(i1) -- [gluon] (l1),
(i2) -- [gluon] (l1),
(l1) -- [scalar] (f1),
(l1) -- [scalar] (f2) };
\end{feynman}}
\end{tikzpicture}&
         
         $=\,\,\frac{\mathtt{IM}\ctg}{\Lambda^2} \frac{i\, g_s^2\, m_t}{\sqrt{2} \pi^2 v}\varepsilon^{\alpha \beta p_1 p_2 }  \delta_{ab} $

    \end{tabular}
    \caption{Effective vertices contributing to the $R_2$ rational terms. All momenta are taken to be outgoing. }
    \label{R2Table}
\end{table}

The tensor operator $\Otg$ is the only operator leading to a non-zero rational term for the processes considered in this paper and special care is needed in the treatment of 
 the $\gamma_5$ matrices entering the computation. We will comment on this in the following. As it is well known $\gamma_5 = \frac{i}{4} \varepsilon_{\mu \nu \rho \sigma} \gamma^\mu \gamma^\nu \gamma^\rho \gamma^\sigma$ is well defined only in four dimensions due to the presence of the antisymmetric tensor $ \varepsilon_{\mu \nu \rho \sigma}$ \cite{1106.5483}. In dimensional regularisation algebraic consistency problems arise if  the four-dimensional  $\gamma_5$ rules and the cyclicity of Dirac $\gamma_\mu$ traces are both kept \cite{Kreimer:1989ke}. In this paper we consider two strategies for the treatment of $\gamma_5$. The first one was proposed by Kreimer, Körner and Schilcher \cite{Kreimer:1989ke,Korner:1991sx,9401354}, abbreviated as the KKS scheme. In this scheme $\gamma_5$ anti-commutes with the Dirac matrices in $d$ dimensions, denoted by $\bar{\gamma}_{\bar{\mu}}$, and the cyclicity of the traces of Dirac matrices is abandoned. The second scheme was introduced by Breitenlohner, Maison, 't-Hooft and Veltman and we abbreviate it as the BMHV scheme \cite{tHooft:1972tcz,Breitenlohner:1976te,Breitenlohner:1977hr}. Contrary to KKS, in BMHV the cyclicity of the trace is preserved but the anticommutation properties of $\gamma_5$ are modified: the Dirac matrices are explicitly divided into a $4$-dimensional part ($\gamma_\mu$) and a $(d-4)$-dimensional part ($\tilde{\gamma}_{\tilde{\mu}}$) such that:
\begin{equation}
    \bar{\gamma}_{\bar{\mu}} = \gamma_\mu +  \tilde{\gamma}_{\tilde{\mu}} \,\,\,\, \text{and} \,\,\,\, \{ \gamma_\mu, \gamma_5\} = 0, \,\,\, [\tilde{\gamma}_{\tilde{\mu}}, \gamma_5] = 0
\end{equation}

The tensor operator $\Otg$ needs to be treated carefully. Indeed it is known that the calculation of loop diagrams with tensor operator insertions involves traces with an odd number of $\gamma_5$ which are not well-defined in the KKS scheme \cite{2111.11449, 2211.01379, 2304.00985}. This is because since the trace is not cyclic in this scheme, the result depends on the \textit{reading point}, i.e. which $\gamma$ matrix is first in the non-cyclic trace. 
The difference between different reading points is of order $\mathcal{O}(\epsilon)$ and is thus only relevant in divergent diagrams 
\cite{2012.08506,2211.09144}. 
It has been argued in \cite{2012.08506,2211.09144} that these ambiguities cancel when the matching to a renormalizable UV theory is performed as long as the reading point is chosen identically in the matching and in the EFT calculations. 
The amplitudes for $gg \rightarrow H, HH$ in the presence of $\mathtt{IM}\ctg$ are indeed UV-divergent and therefore reading point ambiguities arise when using the KKS prescription. We specify here our chosen reading point for these calculations, such that the matching calculations can be performed in a consistent manner to cancel the ambiguities. We choose to use as the reading point the $\gamma_5$ vertex, that is the vertex with an insertion of $\mathtt{IM}\ctg$. As multiple gluons are present, $\mathtt{IM}\ctg$ can enter in several vertices for the same process. However, each Feynman diagram only has one insertion of $\mathtt{IM}\ctg$ and we start reading the trace at this vertex.  The results for the non-zero $R_2$s are presented in Table~\ref{R2Table}. To be consistent with the UFO conventions, all momenta are taken to be outgoing \cite{1108.2040}. Additionally we calculated the Feynman rules for the $R_2$ terms from Table~\ref{R2Table} using the BMHV $\gamma_5$ scheme and we obtained the same results. 
The full amplitude calculations of $gg \rightarrow HH, WW, ZZ$ in the presence of CP-violating operators were performed in the KKS scheme with the reading point specified above. We stress again that only the amplitudes involving $\mathtt{IM}\ctg$ in $ggH$ and $ggHH$ loops depend on the reading point selected. 

We added the non-zero $R_2$ terms in the \code{SMEFTatNLO} UFO, and we validated our implementation by calculating analytically the full amplitude of the $gg \rightarrow HH, WW, ZZ, \gamma Z, \gamma \gamma$ processes in the presence of one operator at a time using the FeynCalc~\cite{Mertig:1990an,1601.01167,2001.04407}, FeynHelpers~\cite{1611.06793}, Package-X~\cite{1503.01469}, LoopTools \cite{10.1007/BF01621031, 9807565} and FeynArts~\cite{hep-ph/0012260} packages. The one-loop calculations performed with Package-X and LoopTools do not require the addition of external rational terms and as such the analytical amplitude calculated in Mathematica is independent of our implementation of the $R_2$ terms. 
We compared the numerical values obtained from our analytical predictions with the numerical predictions given by the modified version of \code{SMEFTatNLO} in \code{Madgraph5\_aMC@NLO}. This was done for all the processes and operators considered, finding perfect agreement between the two computations.
 
\subsection{The UV counterterms}
\label{sec:UFO_UV}

The processes we consider are loop-induced in the presence of all operators except for $\Opgtil$, which introduces a contact term between the Higgs boson and the gluons and hence generates tree-level diagrams for $gg \rightarrow HH, ZZ, WW$. All the loops considered here are finite except for those arising from $\mathtt{IM}\ctg$. Similarly to the CP-even case, the UV divergence can be reabsorbed in the renormalisation of $\Opgtil$. The complete one-loop renormalisation group structure was calculated in \cite{1308.2627, 1310.4838, 1312.2014}, from which we get for ($\cpgtil$, $\im \ctg$):

\begin{equation}
    \frac{d \,\cpgtil}{d \log{\mu}} = \frac{g_s^2}{4\pi^2} \left( -\frac{7}{2} \cpgtil + \frac{\sqrt{2}\, m_t}{v}\, \im \ctg\  \right)
    \label{eq:rge}
\end{equation}
where $\mu$ is the renormalisation scale and $m_t$ is the top quark mass. In the $\overline{MS}$ scheme, $\cpgtil$ is renormalised as:
\begin{equation}
    \cpgtil^0 = \cpgtil(\mu_{EFT}) + \delta\cpgtil
\end{equation}
 where $\mu_{EFT}$ is the renormalisation scale of the EFT and $\cpgtil^0$ can be obtained from Eq.~\eqref{eq:rge}:
\begin{equation}
    \cpgtil^0 = \cpgtil \left(1-\frac{7 g_s^2}{16 \pi^2}\frac{1}{\epsilon} \,\Delta\right) + \im\ctg \frac{g_s^2\, m_t}{4 \sqrt{2} \pi^2 \, v}\frac{1}{\epsilon}\,\Delta,  \,\,\,\,\,\, \Delta \equiv \Gamma(1+\epsilon)\left(\frac{4 \pi \mu^2}{\mu_{EFT}^2}\right)^\epsilon
    \label{eq:counterterm}
\end{equation} 

Using Eq.~\eqref{eq:counterterm}, we calculated the UV counterterm vertices necessary for the processes considered in  this paper. For completeness they are presented in Table~\ref{UVTable}. We validated our results by calculating analytically the UV poles coming from an insertion of $\im \ctg$ in $gg \rightarrow HH, ZZ, WW$ and verifying that the poles cancel against the UV divergent part of the counterterm. In addition, as was explained in Sec.~\ref{sec:UFO_R2}, the finite parts of the amplitudes, including the finite contribution from the $\overline{MS}$ counterterm, were obtained analytically in Mathematica and numerically with \code{Madgraph5\_aMC@NLO} and the two calculations agree for all cases.

\begin{table}[H]
    \centering
    \renewcommand{\arraystretch}{2}
    \begin{tabular}{c l}

         \begin{tikzpicture}[baseline={(current bounding box.center)}, vertical=e to f] {a [nudge=(-30:5mm)]
         \begin{feynman}
         \vertex (i1) {$p_1$, $a$, $\alpha$};
         \vertex[below=2cm of i1] (i2) {$p_2$, $b$, $\beta$};
         \vertex[below right=1cm and 1cm of i1,dot,minimum size=2.5mm] (l1){};
 
         \vertex[right=1.2cm of l1] (f1) {$H$};

         \diagram* {
         (i1) -- [gluon](l1),
         (i2) -- [gluon] (l1),
         (l1) -- [scalar] (f1)};
 
         \end{feynman} }
        \end{tikzpicture} &
         
             $=\,\,\frac{\mathtt{IM}\ctg}{\Lambda^2} \frac{i\, g_s^2\, m_t }{\sqrt{2}\pi^2} \left(\frac{1}{\epsilon} - \log\left(\frac{\mu_{EFT}^2}{\mu_R^2} \right) \right)\varepsilon^{p_1 p_2 \alpha \beta }  \delta_{ab} $
        \\
        \\
         
\begin{tikzpicture}[baseline={(current bounding box.center)}, vertical=e to f] {a [nudge=(-30:5mm)]
\begin{feynman}
\vertex (i1) {$p_1$, $a$, $\alpha$};
\vertex[below=2cm of i1] (i2) {$p_2$, $b$, $\beta$};
\vertex[below right=1cm and 1cm of i1,dot,minimum size=2.5mm] (l1){};
\vertex[right=1.8cm of i1] (f1) {$H$};
\vertex[right=1.8cm of i2] (f2) {$H$};
\diagram* {
(i1) -- [gluon] (l1),
(i2) -- [gluon] (l1),
(l1) -- [scalar] (f1),
(l1) -- [scalar] (f2) };
\end{feynman}}
\end{tikzpicture}&
         
         $=\,\,\frac{\mathtt{IM}\ctg}{\Lambda^2} \frac{i\, g_s^2\, m_t}{\sqrt{2} \pi^2 v} \left(\frac{1}{\epsilon} - \log\left(\frac{\mu_{EFT}^2}{\mu_R^2} \right) \right)\varepsilon^{p_1 p_2 \alpha \beta}  \delta_{ab} $

    \end{tabular}
    \caption{UV counterterm vertices in the $\overline{MS}$ scheme. All momenta are taken to be outgoing. }
    \label{UVTable}
\end{table}

\section{Inclusive diboson production}
\label{2to2}

In this section we consider the effect of the CP-odd operators in double Higgs and double Z/W production. In particular we analyse the impact of CP-even and CP-odd operators on the invariant mass distributions of the final states. The independent helicity configurations of the different processes are shown in Table~\ref{HelConfigTable}. The other possible helicity configurations are related to those shown in the table through Bose symmetry and CP transformations.

The numerical results presented in this section are obtained with our extended version of the \code{SMEFTatNLO} UFO interfaced with \code{Madgraph5\_aMC@NLO} v$3.4$ for the $13$ TeV LHC. We perform our calculations with the NNPDF30\_lo\_as\_0118 PDF \cite{1410.8849} and use the following numerical inputs:
\begin{equation}
\begin{gathered}
        m_Z = 91.1876\, \text{GeV} \, , \quad m_W = 79.8244\, \text{GeV}  \, , \quad m_H= 125\, \text{GeV}\\
        m_t= 172\, \text{GeV} \, , \quad G_F= 1.166370 \cdot 10^{-5}\, \text{GeV}^{-2}
\end{gathered}
\end{equation}
The SMEFT coefficients are all taken to have the value $c/\Lambda^2 = 1 \text{TeV}^{-2}$ unless specified otherwise.  A fixed-scale choice is made for the renormalisation and factorisation scales of $\mu_R=\mu_F=\sum_i m_{i}/2$, where the sum runs over the final state gauge bosons. The renormalisation scale of the EFT, $\mu_{EFT}$, is also set to the same scale. As we have employed a fixed scale we are not considering the impact of renormalisation group running and mixing of the operators. We note however that it has been shown that these effects can be important when considering differential distributions \cite{2406.06670}. A similar study for the case of CP-odd operators is left for future work.

\begin{table}
    \centering
    \renewcommand{\arraystretch}{1.6}
    \begin{tabular}{l | l}
    \toprule
    Process & Independent helicity configurations\\
    \midrule
    \midrule
    $gg \rightarrow HH$& $(+ + 0\,0)$, $(+ - 0\,0)$ \\
    \midrule
    \multirow{2}{*}{ $gg \rightarrow ZZ$}& $(+ + + +)$, $(+ + + -)$, $(+ + + 0)$, $(+ + - -)$,$(+ + - 0)$,\\
    &  $(+ + 0\,0)$, $(+ - - +)$, $(+ - - - )$, $(+ - - 0)$,$(+ - 0\,0)$ \\
    \midrule
    \multirow{3}{*}{ $gg \rightarrow W^+ W^-$}& $(+ + + +)$, $(+ + + -)$, $(+ + - +)$, $(+ + - -)$, $(+ + + 0)$,\\
    &$(+ + 0 +)$, $(+ + - 0)$, $(+ + 0 -)$, $(+ + 0\,0)$, $(+ - - +)$,\\
    &$(+ - + -)$, $(+ - - - )$, $(+ - - 0)$, $(+ - 0 -)$, $(+ - 0\,0)$ \\
    \bottomrule
    \end{tabular}
    \caption{Independent helicity configurations for each process studied in this paper. The helicities are given for $g_1$, $g_2$, $V_1/S_1$, $V_2/S_2$ respectively.}
    \label{HelConfigTable}
\end{table}

\subsection{$gg \rightarrow HH$} \label{sec:ggHH}
\begin{figure}
    \centering
    \begin{subfigure}[h]{0.49\textwidth}
    \centering
        \includegraphics[width=\textwidth]{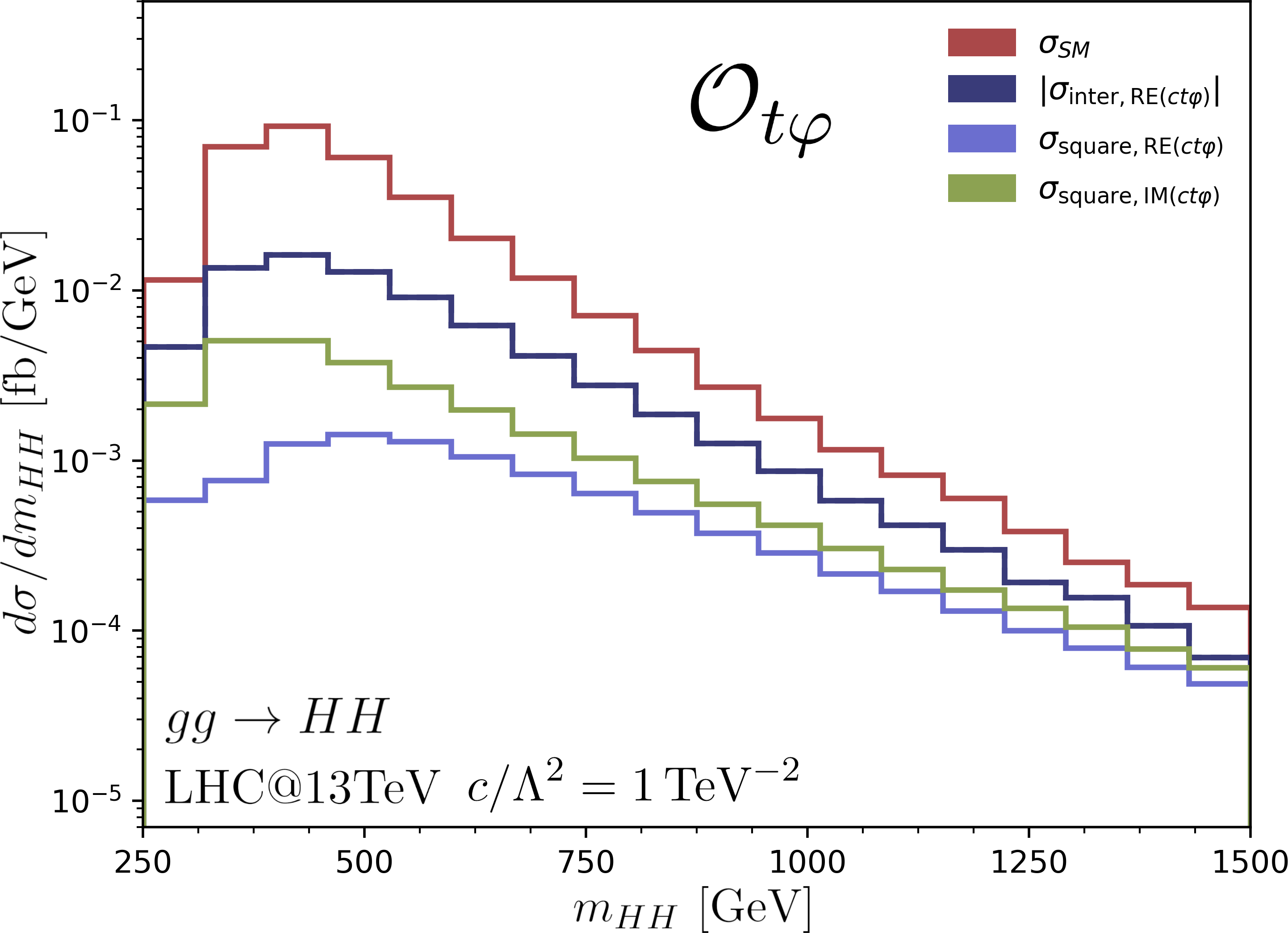}
        \subcaption{}
        \label{HHdista}
    \end{subfigure}
    \hfill
    \begin{subfigure}[h]{0.49\textwidth}
    \centering
        \includegraphics[width=\textwidth]{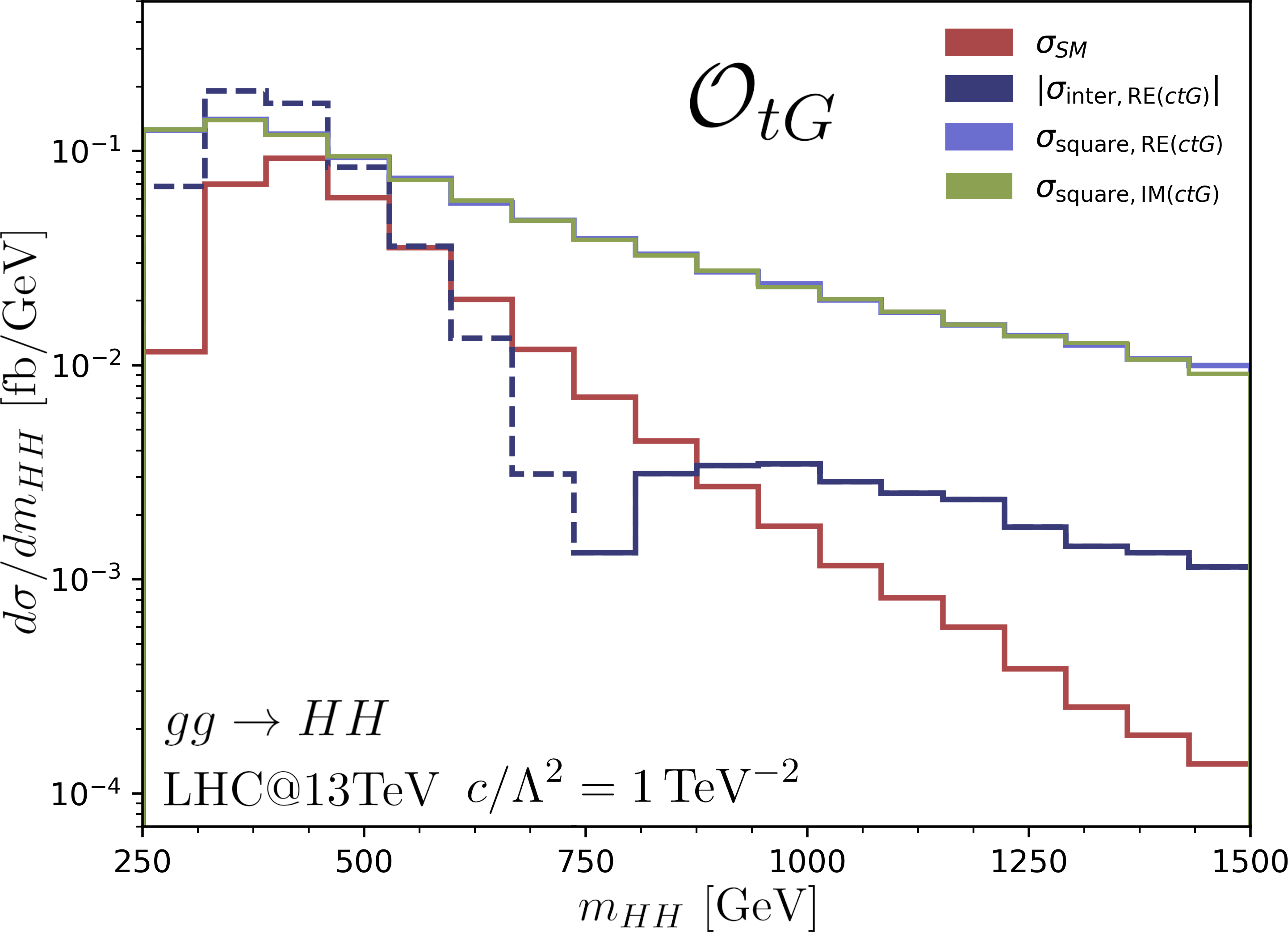}
        \subcaption{}
        \label{HHdistb}
    \end{subfigure}
\caption{Invariant mass distribution of the Higgs pair in the presence of (a) $\mathcal{O}_{t \varphi}$ and (b) $\mathcal{O}_{tG}$. The distributions were obtained for $c/\Lambda^2 = 1 \text{TeV}^{-2}$. In both cases the interference of the imaginary part of the coefficient with the SM vanishes and in (b) the squared distributions overlap, see the text for details. For the interferences, a dashed line denotes a negative contribution.}
\label{HHdist}
\end{figure}

The first process we consider is double Higgs production, which can be modified by three CP-odd operators: $\Otg$, $\Oth$ and 
$\Opgtil$. 
In this section, as in the rest of this paper, we focus on the operators which enter at loop-level and thus we do not consider $\Opgtil$. The invariant mass distributions of the Higgs pair in the presence of $\ctg$ and $\cth$ are shown in Fig.~\ref{HHdist}. The predictions are obtained with the numerical setup described above.

For both operators, the interference of the imaginary part of the coefficient with the SM is zero, and this was verified at amplitude level. More specifically, the individual helicity amplitudes have a non-zero interference with the SM  but these contributions cancel out due to the CP properties of the helicity amplitudes: for example under CP transformation $I_\text{CPV}(++0\,0) = -I_\text{CPV}(--0\,0)$, where $I_\text{CPV}$ denotes a CP-violating interference for a specific helicity configuration given in the brackets.
A similar observation is made for all the $2 \rightarrow 2$ 
processes and CP-violating operators considered in this section.

We first comment on the tensor operator $\Otg$, which modifies the $t\bar{t}g$ vertex and introduces a $t\bar{t}gh$ vertex. Interestingly, despite the different Lorentz structure of the $\bar{t}tg$ and $\bar{t}tgh$ vertices when modified by the real or imaginary part of the Wilson coefficient, the quadratic contributions of $\mathtt{IM}\ctg$ and $\mathtt{RE}\ctg$ coincide. 
This can be understood by looking at the individual helicity amplitudes. The process $gg \rightarrow HH$ has two independent helicity configurations: gluons with the same  polarisations or gluons with opposite polarisations as shown in Table~\ref{HelConfigTable}, and their high energy limit in the presence of CP-even SMEFT coefficients have been studied in \cite{2306.09963}.
In the presence of $\mathtt{RE}\ctg$ both  the $(+ + 0\, 0)$ and $(+ - 0\, 0)$ amplitudes grow quadratically with energy, but the configuration with both gluons having the same polarisation dominates in the sum over helicities. 
In the presence of $\mathtt{IM}\ctg$, the $(+ - 0\,0)$ amplitude evaluates to zero, however the $(+ + 0\,0)$ amplitude is the same up to an overall phase compared to the $\mathtt{RE}\ctg$ amplitude. This degeneracy in the $(+ + 0\,0)$ amplitude amplitude is exact at all energies. Combined with the small contribution of the $(+ - 0\,0)$ amplitude for $\mathtt{RE}\ctg$, this degeneracy leads to the two quadratic cross-sections being very similar. 

The situation is different for the top Yukawa operator $\Oth$ which rescales the $\bar{t}tH$ vertex and introduces a new $\bar{t}tHH$ vertex. As can be seen in Fig.~\ref{HHdista}, the invariant mass distribution of the Higgs pair in the presence of $\mathtt{RE}\cth$ is distinct from the one in the presence of $\mathtt{IM}\cth$. In the $(+ - 0\, 0)$ helicity configuration, the $\mathtt{IM}\cth$ amplitude is equal to zero while the $\mathtt{RE}\cth$ amplitude tends to a constant in the high energy limit \cite{2306.09963}. The leading contribution to the total amplitude comes from the $(+ + 0\, 0)$ configuration in which both the $\mathtt{IM}\cth$ and $\mathtt{RE}\cth$ amplitudes grow logarithmically with energy and the amplitude contributions not vanishing at high-energy take the form:
\begin{align}
    &\mathtt{RE}\cth : \frac{m_t\,v\,g_s^2\, \delta_{ab}}{32 \pi^2} \left[- 3\,\left(\logsmt- i \pi\right)^2 +20 \right]\\
    &\mathtt{IM}\cth : \frac{m_t\,v\,g_s^2\, \delta_{ab}}{32 \pi^2} \left[-3i\,\left(\logsmt- i \pi\right)^2 \right]
\end{align}
where $\delta_{ab}$ is the colour factor of the amplitude and $a,b$ are the colours of the incoming gluons.

The two amplitudes differ by a constant factor which is only significant at low energy, explaining the converging amplitudes at high energies.

\subsection{$gg \rightarrow ZZ$} \label{sec:ggZZ}

\begin{figure}[htp]
    \centering
    \begin{subfigure}[h]{0.49\textwidth}
    \centering
        \includegraphics[width=\textwidth]{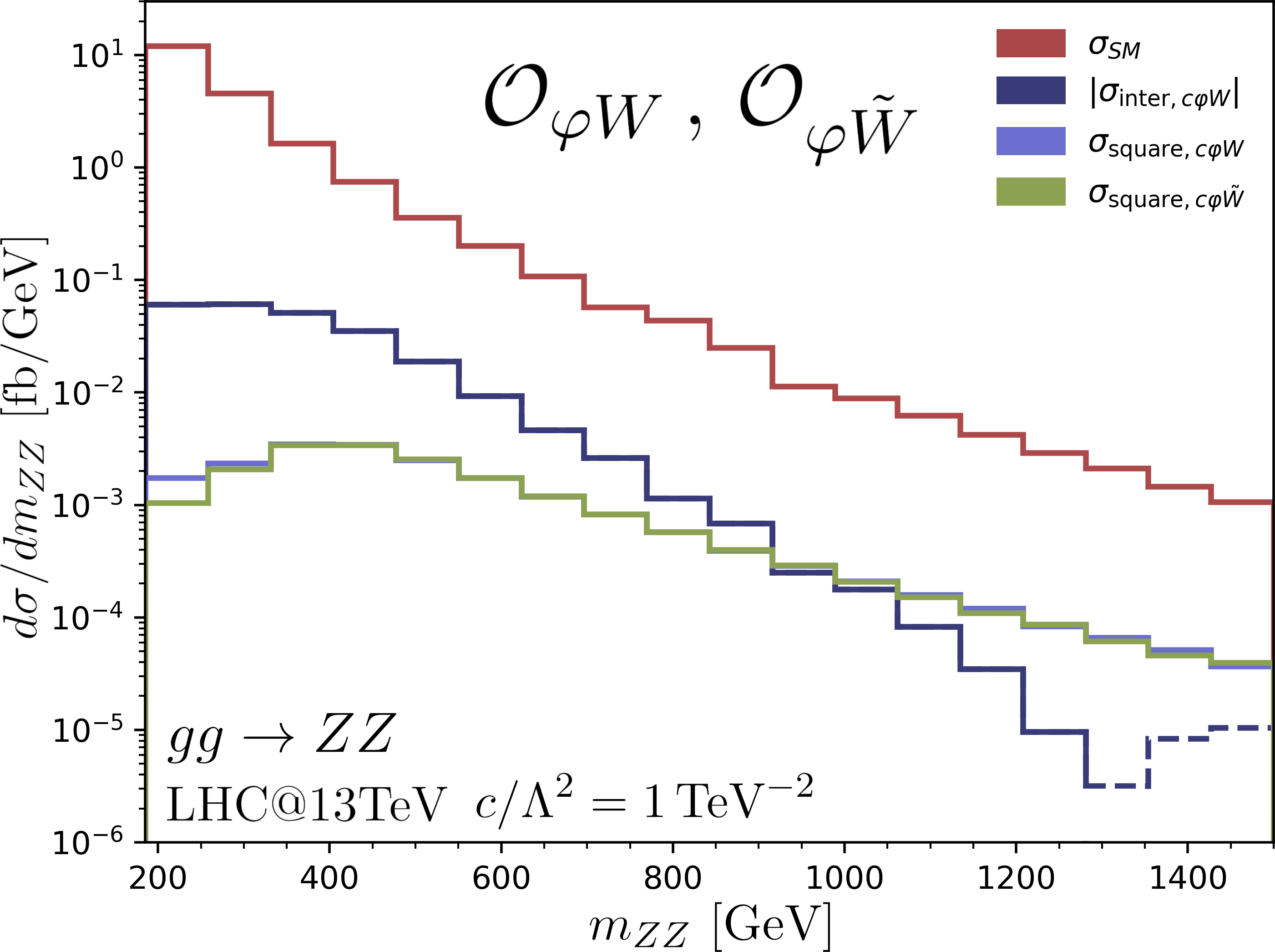}
        \subcaption{}
        \label{SubFig1}
    \end{subfigure}
    \hfill
    \begin{subfigure}[h]{0.49\textwidth}
    \centering
        \includegraphics[width=\textwidth]{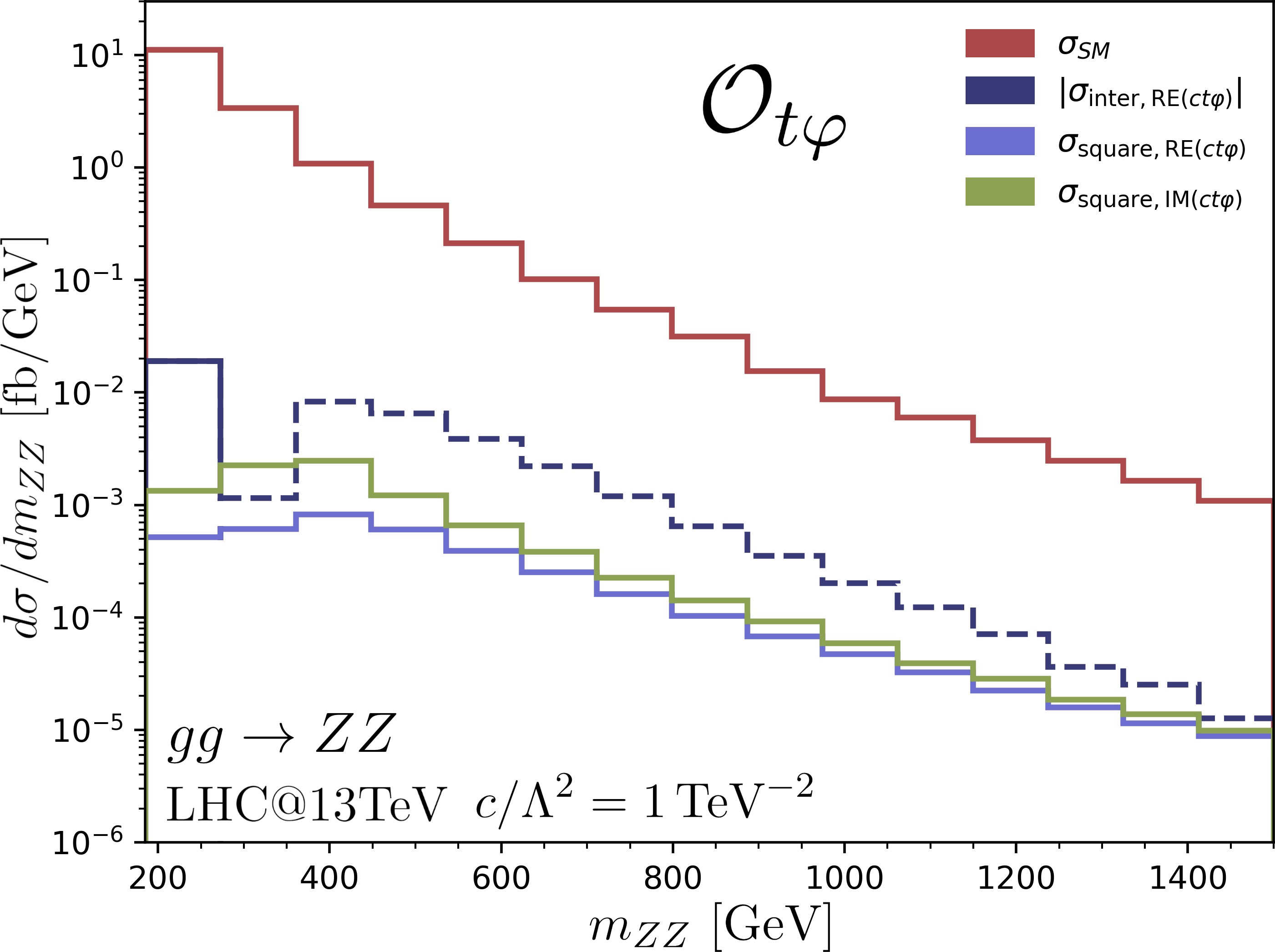}
        \subcaption{}
        \label{SubFig1}
    \end{subfigure}

    \medskip
    
    \begin{subfigure}[h]{0.49\textwidth}
    \centering
        \includegraphics[width=\textwidth]{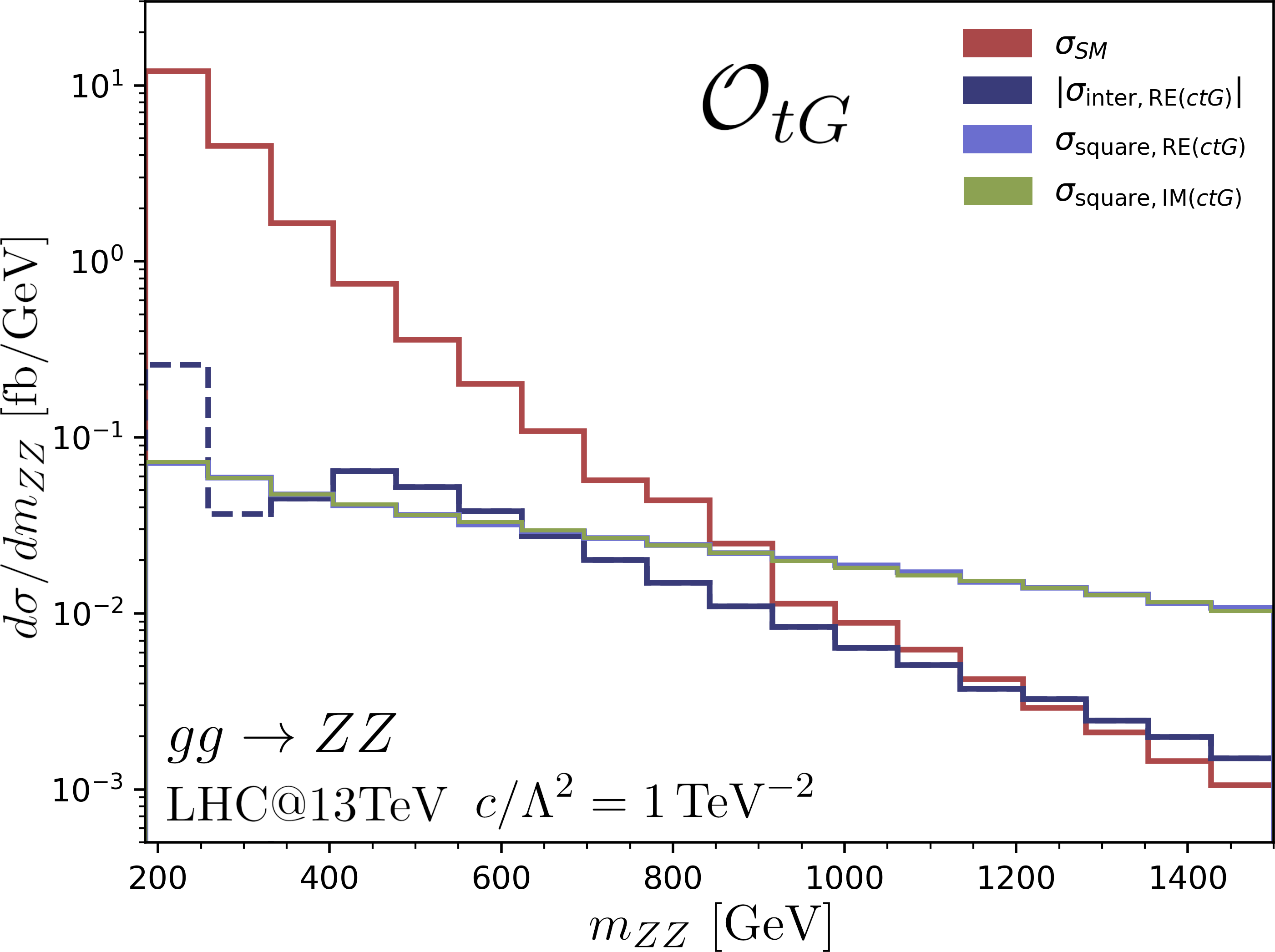}
        \subcaption{}
        \label{SubFig1}
    \end{subfigure}
    \hfill
    \begin{subfigure}[h]{0.49\textwidth}
    \centering
        \includegraphics[width=\textwidth]{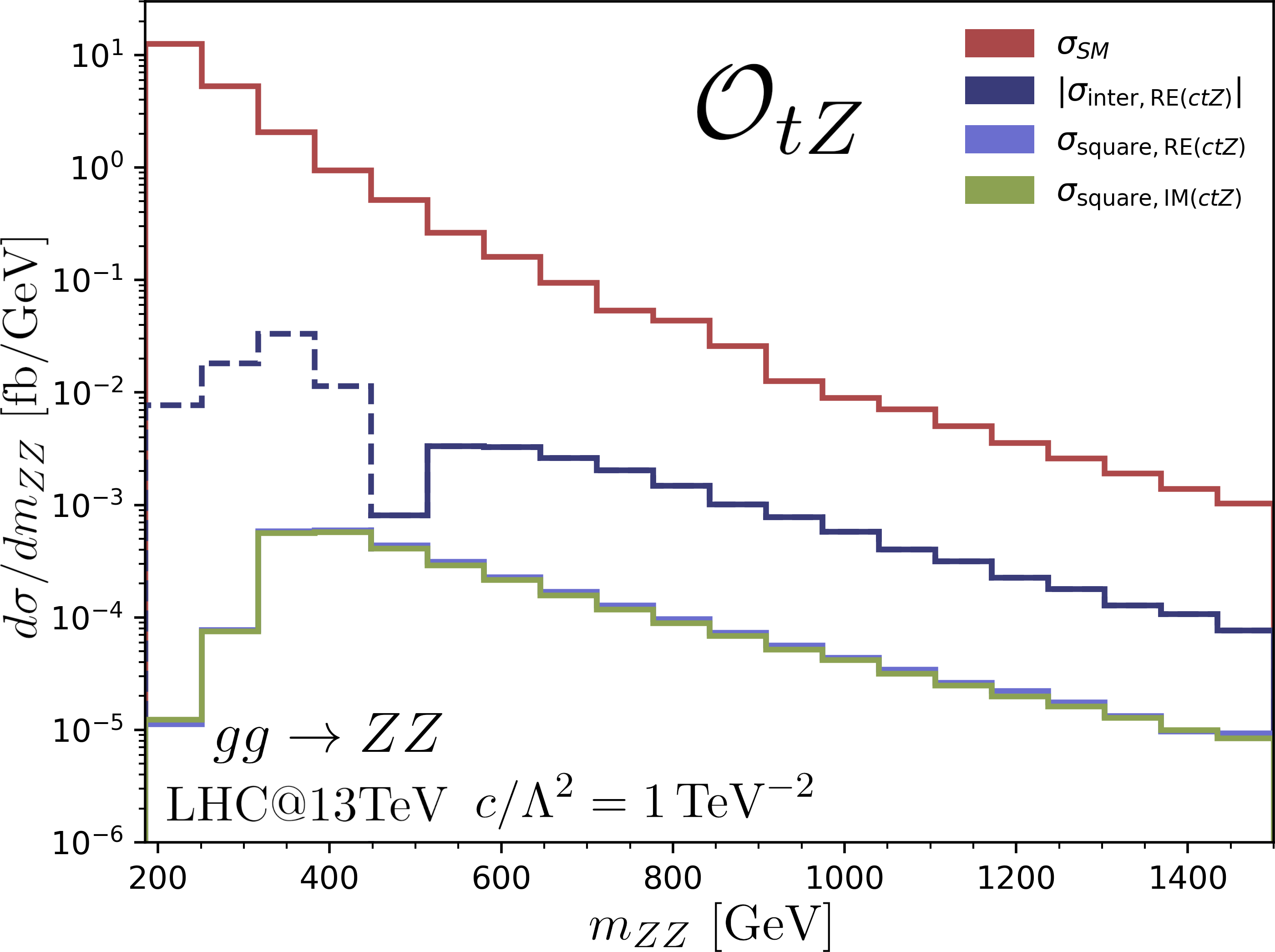}
        \subcaption{}
        \label{SubFig2}
    \end{subfigure}

\caption{Invariant mass distribution of the $Z$ pair in the presence of (a) $\OpW$ and $\OpWtil$, (b) $\mathcal{O}_{t \varphi}$, (c) $\mathcal{O}_{tG}$ and (d) $\mathcal{O}_{tZ}$. In (c) and (d) the squared distributions overlap. For the interferences, a dashed line denotes a negative contribution.}
\label{ZZdist}
\end{figure}

We now turn our attention to double $Z$ production, which can be modified by seven CP-odd operators: the top Yukawa $\Oth$, the top dipoles $\Otg$ and $\OtZ$, the gauge operators $\OpWtil$, $\OpBtil$ and $\OpWBtil$\footnote{The CP-even equivalent of $\OpWBtil$, $\OpWB$, induces a kinetic mixing term between the $SU(2)$ $W_{\mu \nu}^3$ and the hypercharge $B_{\mu \nu}$ gauge fields. Rotating away this mixing term leads to shifted parameters such as the weak angle and the $Z$ boson mass, which depend on the value of the coefficient $\cpWB$.} and the contact operator $\Opgtil$, the latter entering at tree-level in $gg \rightarrow ZZ$. As we want to study the effects of CP-odd operators compared to CP-even ones in loop processes, we focus on the four following operators:  $\Oth$, $\OtZ$, $\Otg$ and $\OpWtil$. The impact of $\OpBtil$ and $\OpWBtil$ can be inferred from the results for  $\OpWtil$ as they modify the $HZZ$ vertex in the same manner ~\cite{2409.00168}. 
The invariant mass distributions of the $Z$ pair in the presence of the SMEFT coefficients, already presented in \cite{2203.02418} for the CP-even coefficients, are shown in Fig.~\ref{ZZdist}.

We first comment on the gauge operators $\OpW$, $\OpWtil$, which modify the interactions of the $Z$ bosons with the Higgs. The quadratic distributions of the CP-even and the CP-odd coefficient differ slightly at the threshold but become indistinguishable at relatively low energies. This can be understood from the helicity amplitudes, which are non-zero in only three out of the ten independent helicity configurations: $(+ + 0\,0)$, $(+ + + +)$ and $(+ + - - )$ , since the gauge operators enter solely in the triangle diagrams with a Higgs propagator. The  $(+ + + +)$ and $(+ + - - )$ helicity amplitudes grow logarithmically with energy and dominate the total amplitudes for both coefficients. While their exact expressions differ slightly for $\OpW$ and $\OpWtil$ close to the production threshold region, the CP-even and CP-odd helicity amplitudes converge quickly and differ by only a few percent at $\sqrt{s} = 300$ GeV. 

Similarly, in the presence of $\Oth$, the quadratic contributions of $\mathtt{RE}\cth$ and $\mathtt{IM}\cth$ converge in the high energy limit. The Yukawa operator also only enters in the triangle diagrams with a Higgs propagator, leading to the same three non-vanishing helicity amplitudes as for the gauge operators. Considering the high energy behaviour of the helicity amplitudes, we find that when both $Z$ bosons are transversely polarised both the $\mathtt{RE}\cth$ and $\mathtt{IM}\cth$ amplitudes decrease following $ \left[\logsmt- i \pi\right]^2/s$ and their magnitudes differ by a term $\propto 1/s$. When both $Z$ bosons are longitudinally polarised the $\re\cth$ and $\im\cth$ amplitudes grow logarithmically with the energy and their high energy behaviour is given by:
\begin{align}
    &\mathtt{RE}\cth : \frac{m_t\,v^3\,e^2\,g_s^2 \delta_{ab}}{128 \pi^2\, m_Z^2 \,\cw^2\, \sw^2} \left[\left(\logsmt- i \pi\right)^2 - 4 \right]\\
    &\mathtt{IM}\cth : i\, \frac{m_t\,v^3\,e^2\,g_s^2 \delta_{ab}}{128 \pi^2\, m_Z^2\, \cw^2\, \sw^2} \left(\logsmt- i \pi\right)^2 \label{eq:ggZZ_IMctp}
\end{align} 
such that the two amplitudes differ by a constant which becomes negligible at high energy.

The tensor operator $\Otg$ leads to rapidly growing amplitudes, which with  $ \re\ctg = \im\ctg = 1 \text{TeV}^{-2}$ even surpass the SM invariant mass distribution for $m_{ZZ} > 900$ GeV and $m_{ZZ} > 1200$ GeV in the case of the real and imaginary quadratic contributions and real interference respectively. The energy behaviour of the helicity amplitudes modified by $\re\ctg$ was calculated in \cite{2306.09963} and it was shown that out of all the operators entering $gg \rightarrow ZZ$ at one loop, $\Otg$ leads to the most extreme energy growths, as it is the only one inducing quadratic growths for some helicity configurations, namely  $(+ + 0\,0)$ and  $(+ - 0\,0)$. Studying the energy behaviour of the $\im\ctg$ helicity amplitudes, we find that for most helicity configurations, the $\im\ctg$ and $\re\ctg$ amplitudes are either identical or converge at high energy. The total amplitudes for both $\re\ctg$ and $\im\ctg$ are completely dominated by the  $(+ + 0\,0)$ contribution, which is the only amplitude growing quadratically for $\im\ctg$ and is larger by $2-3$ orders of magnitude than the $(+ - 0\,0)$ one for $\re\ctg$ for the energies considered in Fig.~\ref{ZZdist}. The $\re\ctg$ and $\im\ctg$ $(+ + 0\,0)$  amplitudes are exactly the same explaining the overlapping quadratic distributions.

We conclude our discussion of double $Z$ production by considering the weak dipole operator $\OtZ$ which modifies the $t\bar{t}Z$ vertex. While the total amplitude is not the same for real and imaginary coefficients, it is dominated by the three helicity amplitudes which grow with energy, namely $(+ + + -)$, $(+ + - -)$ and $(+ - - -)$. The first helicity amplitude is exactly the same for $\re\ctZ$ and $\im\ctZ$ and the other two converge quickly: they match below $1\%$ around $\sqrt{s} = 500$ GeV for  $(+ + - -)$ and $\sqrt{s} = 1$ TeV for $(+ - - -)$. At low energy the $(+ + + +)$ configuration also contributes to the total amplitude. This helicity amplitude tends to the same constant for $\re\ctZ$ and $\im\ctZ$ in the high energy limit and both the real and imaginary squared contributions match within a few percent already at  $\sqrt{s} = 300$ GeV.

\subsection{$gg \rightarrow WW$} \label{sec:ggWW}

\begin{figure}[h]
    \centering
    \begin{subfigure}[h]{0.49\textwidth}
    \centering
        \includegraphics[width=\textwidth]{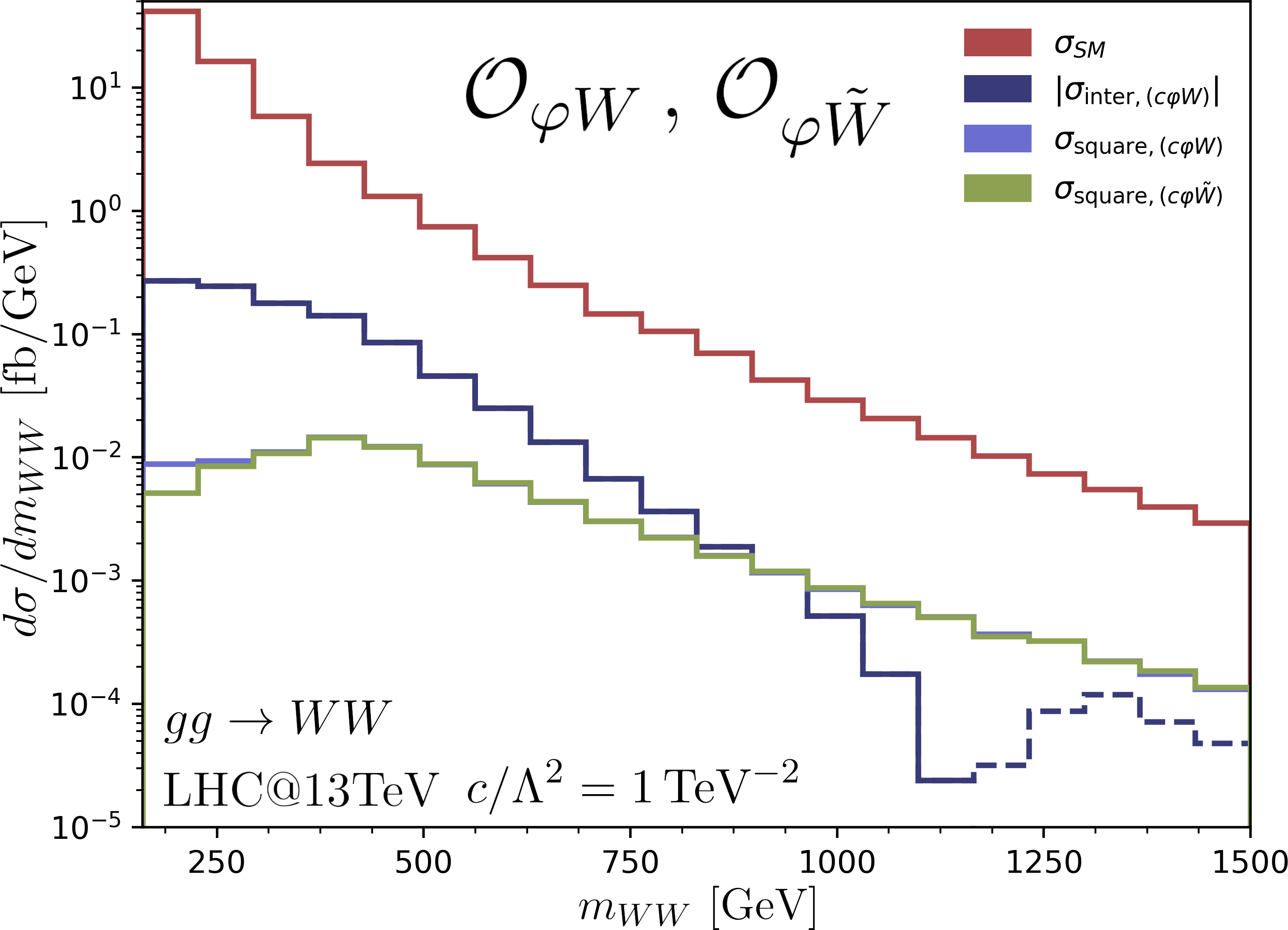}
        \subcaption{}
        \label{SubFig1}
    \end{subfigure}
    \hfill
    \begin{subfigure}[h]{0.49\textwidth}
    \centering
        \includegraphics[width=\textwidth]{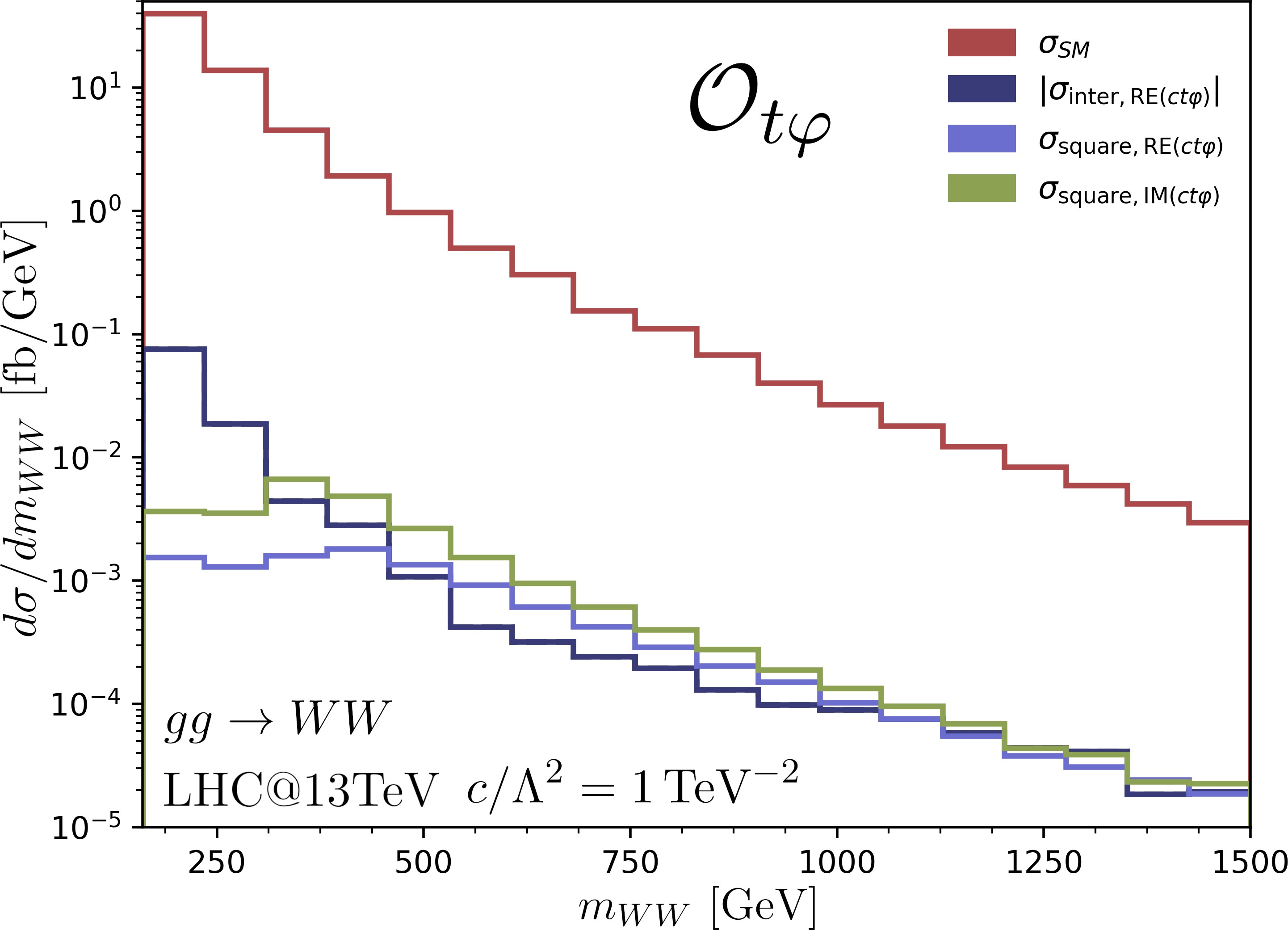}
        \subcaption{ }
        \label{SubFig2}
    \end{subfigure}

    \medskip
    
    \begin{subfigure}[h]{0.49\textwidth}
    \centering
        \includegraphics[width=\textwidth]{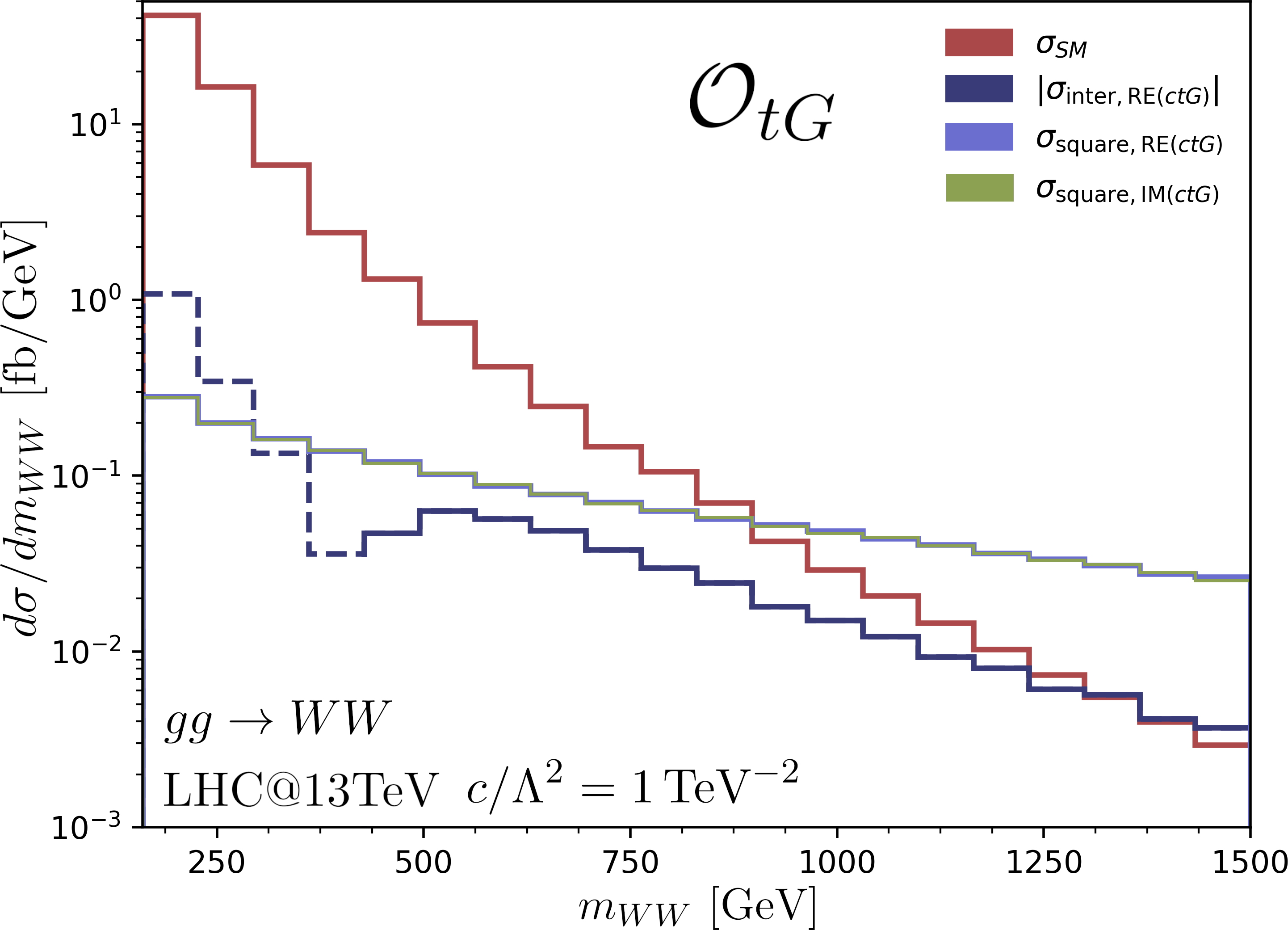}
        \subcaption{}
        \label{SubFig2}
    \end{subfigure}
    \hfill
    \begin{subfigure}[h]{0.49\textwidth}
    \centering
        \includegraphics[width=\textwidth]{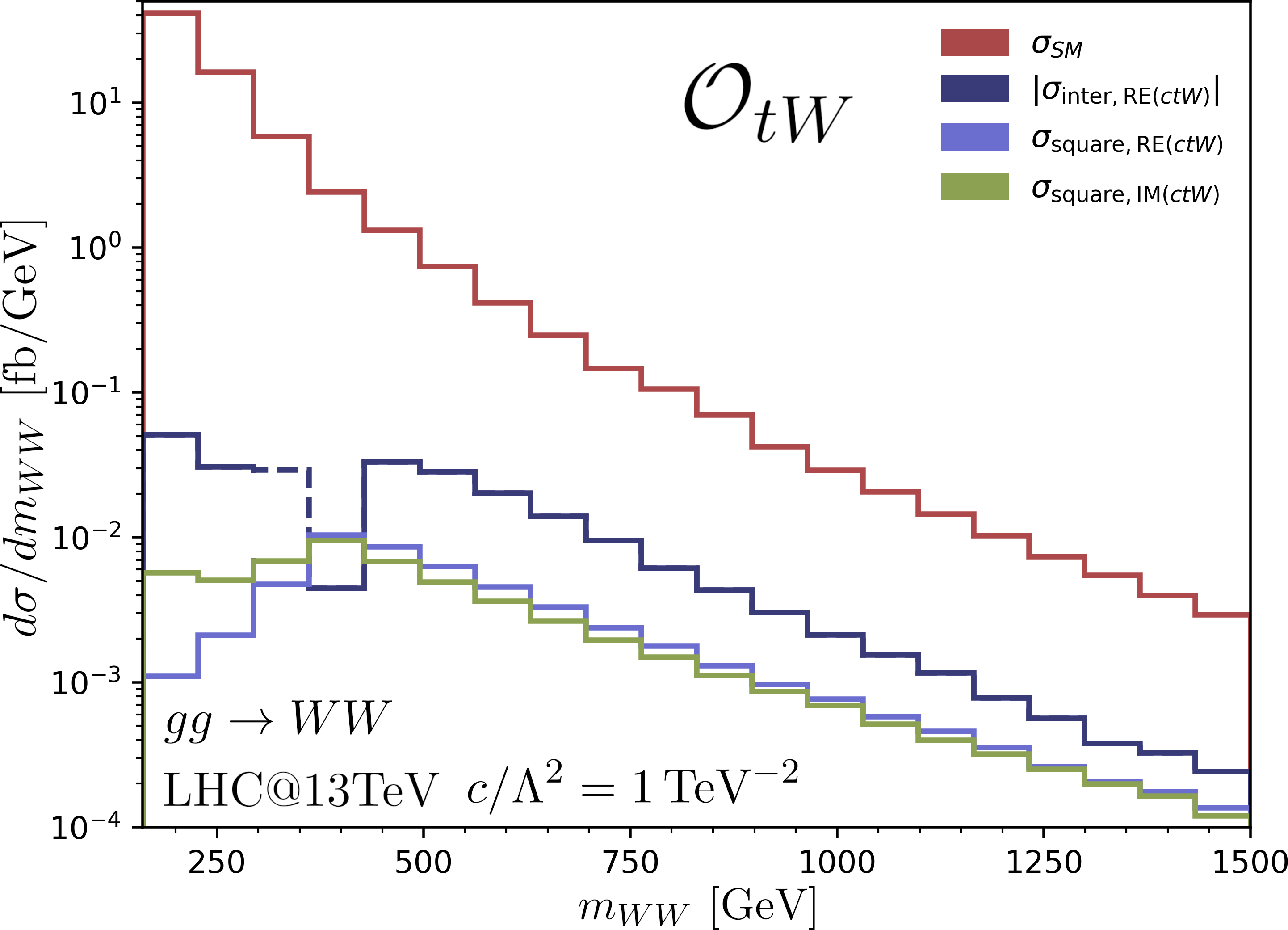}
        \subcaption{}
        \label{SubFig2}
    \end{subfigure}
\caption{Invariant mass distribution of the $W$ pair in the presence of (a) $\OpW$ and $\OpWtil$, (b) $\Oth$, (c) $\Otg$, and (d) $\OtW$. In (c) the squared distributions overlap. For the interferences, a dashed line denotes a negative contribution. }
\label{WWdist}
\end{figure}

We finally turn our attention to double $W$ production, which can be modified by the following five CP-violating operators: the top Yukawa $\Oth$, the top dipoles $\Otg$ and $\OtW$, the gauge operator $\OpWtil$ and the contact operator $\Opgtil$.~\footnote{ In principle operators modifying $Z$ boson interactions could contribute through the $gg \rightarrow Z \rightarrow W^+W^-$ diagram, however the amplitude vanishes for on-shell $Z$ bosons both in the SM and EFT.}  The invariant mass distributions of the $W$ boson pair in the presence of the $\Oth$, $\Otg$, $\OtW$ and $\OpWtil$ operators are shown in Fig.~\ref{WWdist}.

The results for $WW$ share the same features as those shown for $ZZ$ with the common operators leading to similar effects for the two processes. The only exception is for the CP-even interference of the Yukawa operator, which is positive for all the energies considered in $WW$ production, while it is negative for $m_{ZZ} \gtrapprox 300$ GeV in $ZZ$ production. For both processes the interference of $\im\cth$ with the SM triangle (box) diagrams is negative (positive). However, in $WW$ production the absolute value of the interference of $\im\cth$ with the SM box diagrams is slightly larger than the absolute value of the interference with the SM triangle diagrams while this is only the case for $m_{ZZ} \lessapprox 300$ GeV in $ZZ$ production. 
Similarly to $ZZ$ and $HH$, the interference of the CP-odd coefficients with the SM vanishes at the amplitude level while the quadratic contributions of the CP-odd coefficients and their CP-even counterpart either overlap or converge in the high energy limit. Such a behaviour does not allow the distinction between CP-even and CP-odd contributions, and motivates studying angular and polarisation observables of the decay products of the $Z$ and $W$ bosons, as we will discuss in the following section.

\section{Angular observables in four-lepton production }
\label{2to4}

The kinematic distributions of the $2 \rightarrow 2$ processes presented in the previous section do not allow the distinction between the CP-even and CP-odd contributions. In addition in a realistic experimental setup the final-state bosons are only detected through their decay products. Therefore in this section, we  explore the potential of probing CP-violating SMEFT operators in the processes $gg \rightarrow Z Z \rightarrow 4l$  and $gg \rightarrow WW \rightarrow 2l2\nu$ at the $13$ TeV LHC with angular observables. 

\subsection{Analysis strategy} \label{sec::strategy}

We consider the general case of a vector boson $V=Z,W$ that decays leptonically at the $13$ TeV LHC. In an inclusive setup the angular distribution of the vector boson decay products in the $V$ rest frame is given by \cite{1204.6427}:
\begin{equation}
\begin{split}
    \frac{1}{\sigma} \frac{d\sigma}{d\text{cos}\theta^*}  = & \frac{3}{8} \left(1+\text{cos}^2\theta^* - \frac{2(c_L^2 -c_R^2)}{(c_L^2 + c_R^2)}\text{cos}\theta^* \right) f_R+\frac{3}{8} \left(1+\text{cos}^2\theta^* + \frac{2(c_L^2 -c_R^2)}{(c_L^2 + c_R^2)}\text{cos}\theta^* \right) f_L \\
    &+ \frac{3}{4} \text{sin}^2\theta^* f_0
\end{split}
\label{eq::angular_dist_Z}
\end{equation}
where $c_L$ and $c_R$ are the left and right-handed couplings of the fermions to $V$ and $f_{L,R,0}$ are the polarisation fractions. From Eq.~\eqref{eq::angular_dist_Z}, we can write the polarization fractions as:
\begin{align}
    &f_R = -\frac{1}{2} - \frac{(c_L^2 + c_R^2)}{(c_L^2 - c_R^2)} \langle\text{cos}\theta^*\rangle + \frac{5}{2}\langle\text{cos}^2\theta^*\rangle
    \label{eq::polfrac1}\\
    &f_L = -\frac{1}{2} + \frac{(c_L^2 + c_R^2)}{(c_L^2 - c_R^2)} \langle\text{cos}\theta^*\rangle + \frac{5}{2}\langle\text{cos}^2\theta^*\rangle\\
    &f_0 = 2 - 5 \langle\text{cos}^2\theta^*\rangle
    \label{eq::polfrac2}
\end{align}

For the leptonic decays of a $Z$ boson, $gg \rightarrow ZZ \rightarrow e^+ e^- \mu^+ \mu^-$, we define $\theta^*$ as the opening angle between the three-momentum of the produced electron in the rest frame of the $Z$ boson and the $Z$ boson three-momentum in the center-of-mass frame of the $ZZ$ pair. The  SM values of $c_L$ and $c_R$ are given by:
\begin{equation}
    c_L = \frac{ie}{2 \sw \cw} (1-2 \sw^2), \qquad  \qquad c_R=-\frac{ie \sw^2}{\cw \sw}.
\end{equation}
In the case of $V=W$, we consider $gg \rightarrow W^+ W^- \rightarrow \mu^+\, \nu_\mu \,e^-\, \bar{\nu}_e$ and the expressions above simplify as $c_R = 0$. In this case, we define $\theta^*$ as the opening angle between the three-momentum of the produced electron in the rest frame of the $W^-$ and the $W^-$ three-momentum in the center-of-mass frame of the $W^+W^-$ pair \footnote{Eqs.~\eqref{eq::angular_dist_Z}-\eqref{eq::polfrac2} reproduce Ref.~\cite{1204.6427} with $f_L$ and $f_R$ interchanged as in this work $\theta^*$ is defined with respect to the momentum of the particle produced by a $Z$ boson, while in \cite{1204.6427} $\theta^*$ is defined with respect to the momentum of the antiparticle. Similarly for the $W$ boson, $f_L$ and $f_R$ are interchanged as in this work $\theta^*$ is defined with respect to the charged particle produced by a $W^-$, while in \cite{1204.6427} $\theta^*$ is defined with respect to the charged particle produced by a $W^+$.}.

We obtain predictions for $gg \rightarrow e^+ e^- \mu^+ \mu^-$ and  $gg \rightarrow \mu^+ \nu_\mu e^- \bar{\nu}_e$ in \code{Madgraph5\_aMC@NLO} with the same numerical inputs as detailed in Sec.~\ref{2to2}. To focus on the $Z$ ($W$) on-shell region, we impose a cut on the invariant mass of the same flavour leptons of $66<m_{ll}<116$ GeV  ($65<m_{l\nu}<105$ GeV).  The lower bound is chosen to prevent the cross-section from being dominated by the Higgs signal at $125$ GeV while for the upper bound we use the on-shell region defined by experimental collaborations ~\cite{2311.09715}.  In the case of $W$ production, such a cut cannot be imposed experimentally due to the presence of neutrinos in the final state and particular care is needed to deal with neutrino reconstruction, see for example Ref. \cite{2409.16731}. Here we apply such a naive cut as we interested in a first exploration of CP violation in $2l2\nu$ production.
We impose no other kinematic cuts, except on the transverse momentum of the lepton pairs $p_{Tll} > 0.5$ GeV ($p_{Tl\nu} > 0.5$ GeV) to prevent the $Z$ ($W$) boson from being produced along the beamline and avoid numerical instabilities in the loop amplitudes. In order to validate our analysis, we compared our Standard Model predictions for $gg \rightarrow 4l$ with the results from \cite{2004.02031} and found good agreement.

Finally, as mentioned in Sec.~\ref{sec:ggWW}, $2l2\nu$ production can in principle be mediated by a triangle diagram with a $Z$ propagator, which vanishes in the SM and the EFT when the $W$ bosons are exactly on-shell. Our chosen cuts allow the $W$ bosons to be slightly off-shell, leading to a non-zero contribution from the triangle diagrams with a $Z$ propagator. We verified that this contribution is negligible both for the SM and in the presence of SMEFT operators. \footnote{In practice the contribution from the triangle diagrams with a $Z$ propagator always vanish since when the $W$ bosons are off-shell, a second type of triangle diagrams with a $Z$ propagator arises with only one $W$ boson but a $2l 2\nu$ final state, and the two contributions cancel each other out \cite{1107.5569}.}

\subsection{Angular distributions}\label{sec:4l_angdist}

\begin{figure}
\centering

\begin{subfigure}{0.49\columnwidth}
\centering
\includegraphics[width=\textwidth]{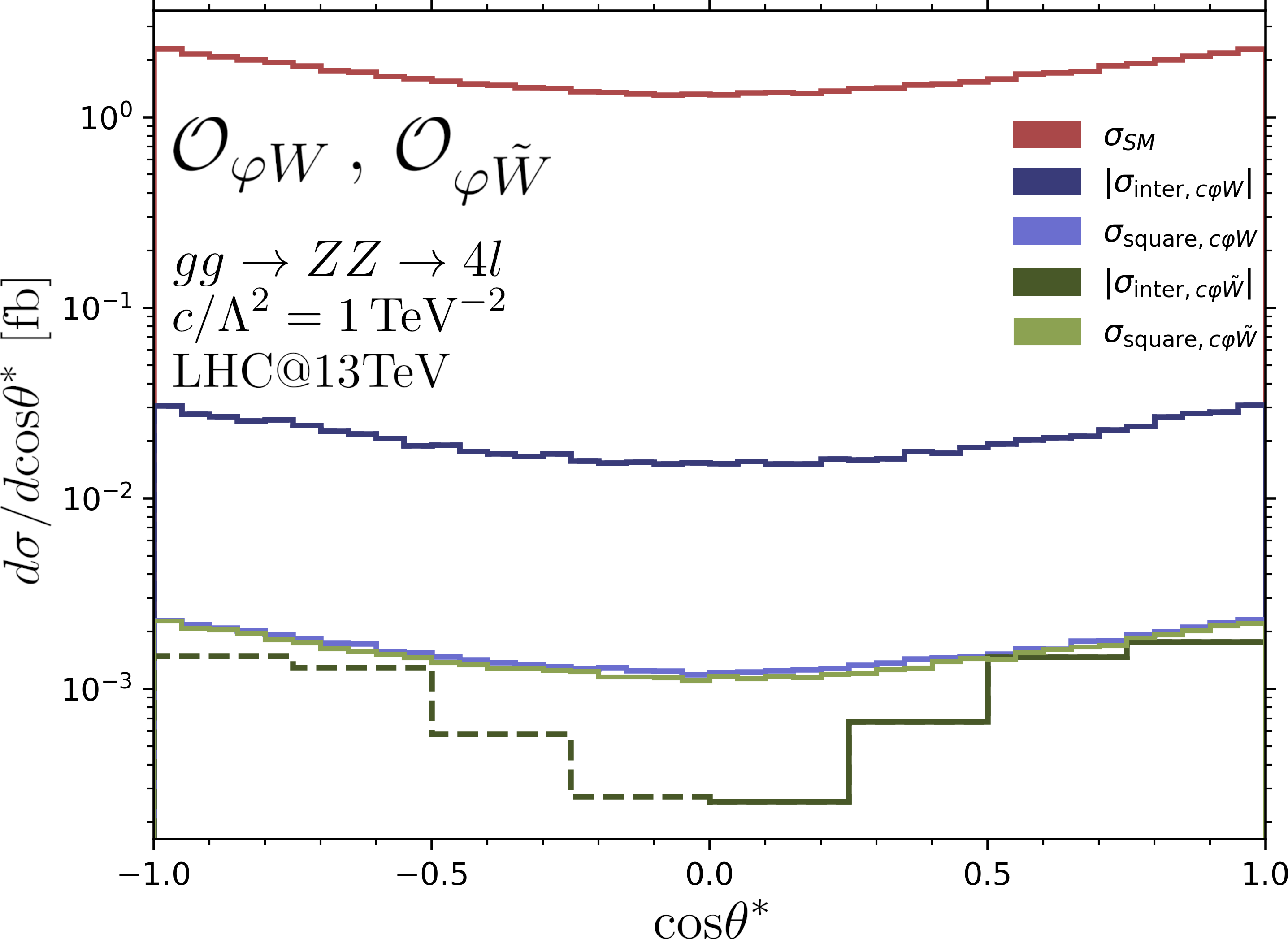}
\caption{}
\label{fig:time1}
\end{subfigure}\hfill
\begin{subfigure}{0.49\columnwidth}
\centering
\includegraphics[width=\textwidth]{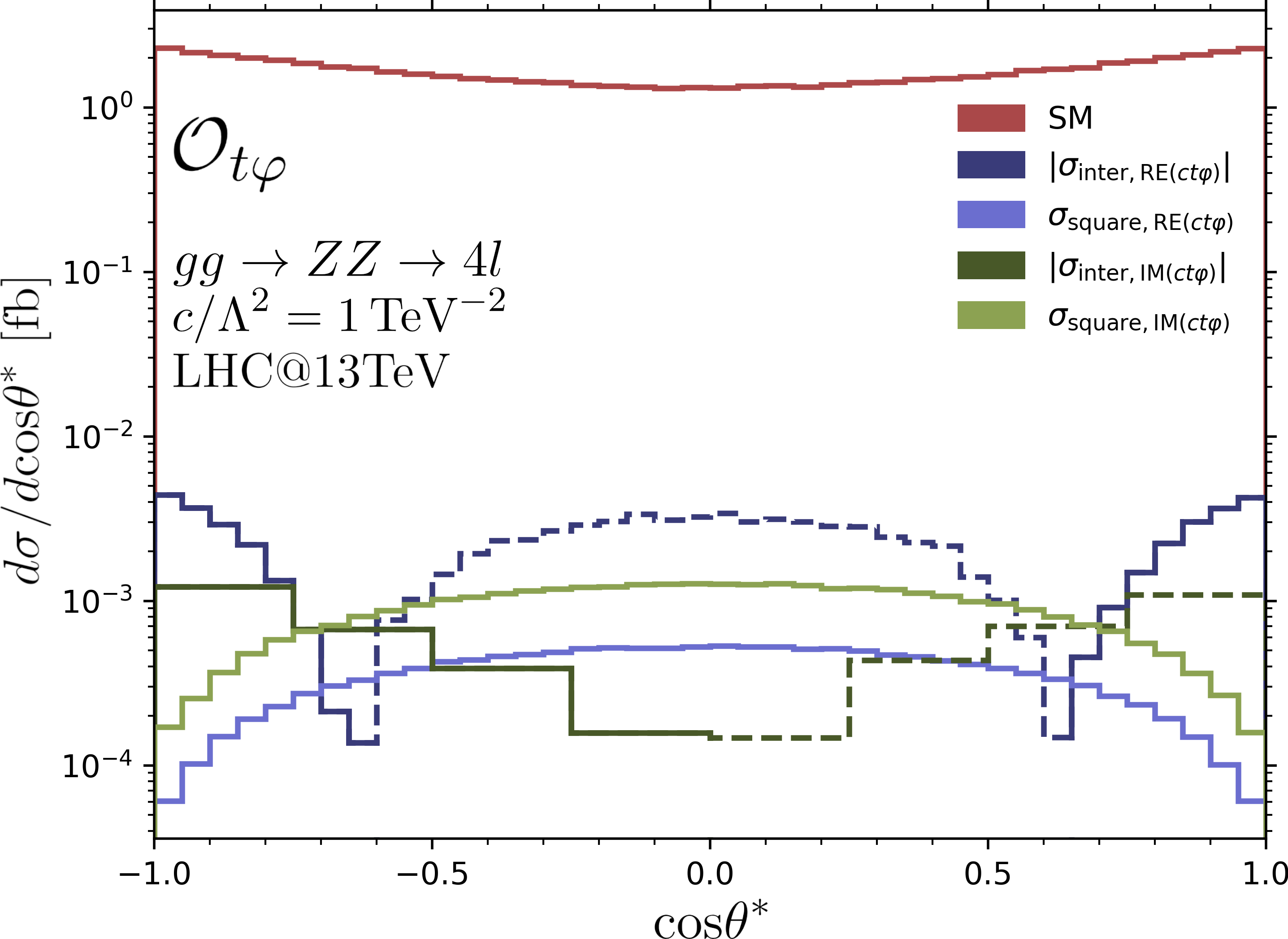}
\caption{}
\label{fig:time2}
\end{subfigure}

\medskip

\begin{subfigure}{0.49\columnwidth}
\centering
\includegraphics[width=\textwidth]{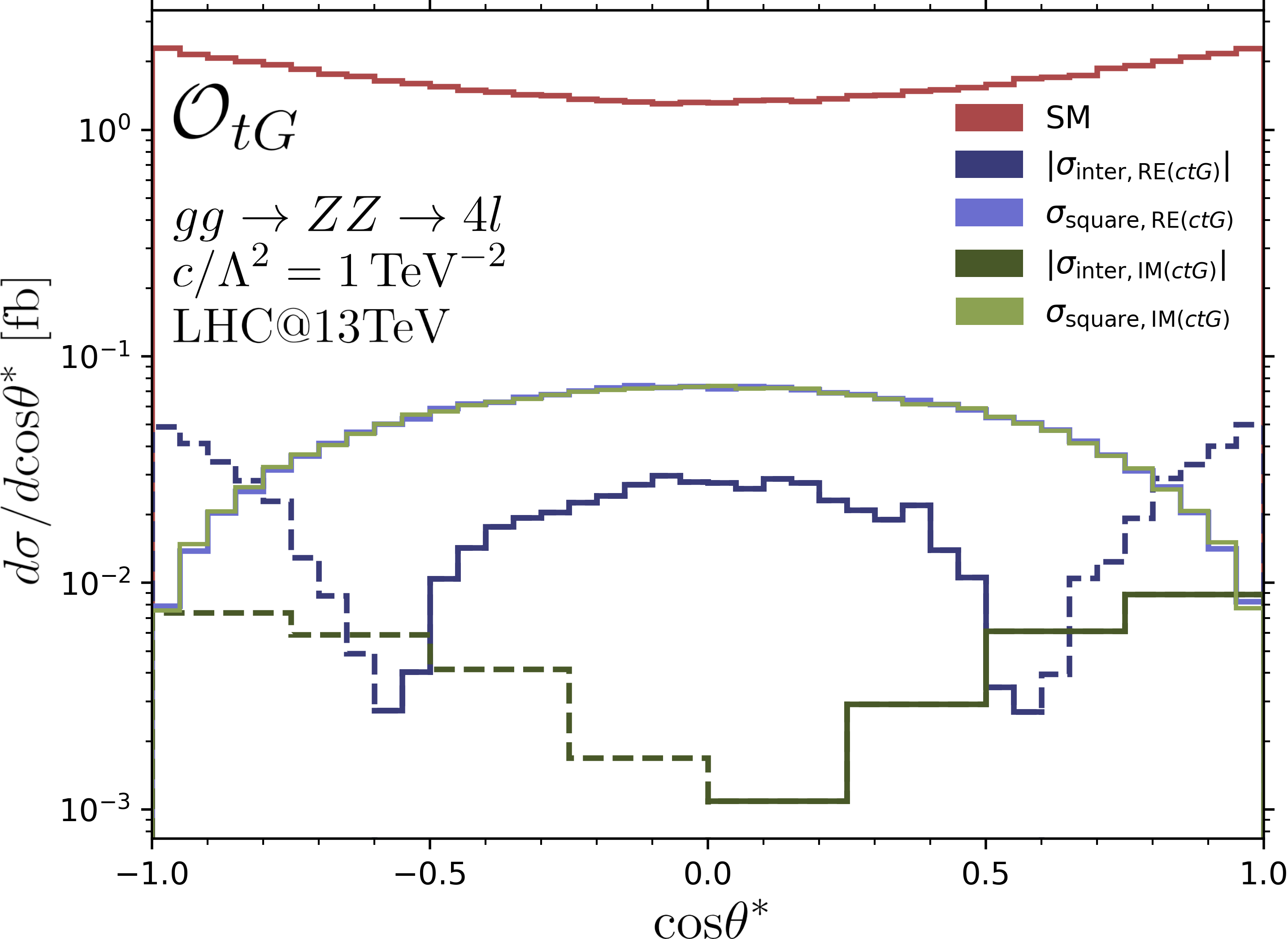}
\caption{}
\label{fig:time3}
\end{subfigure}\hfill
\begin{subfigure}{0.49\columnwidth}
\centering
\includegraphics[width=\textwidth]{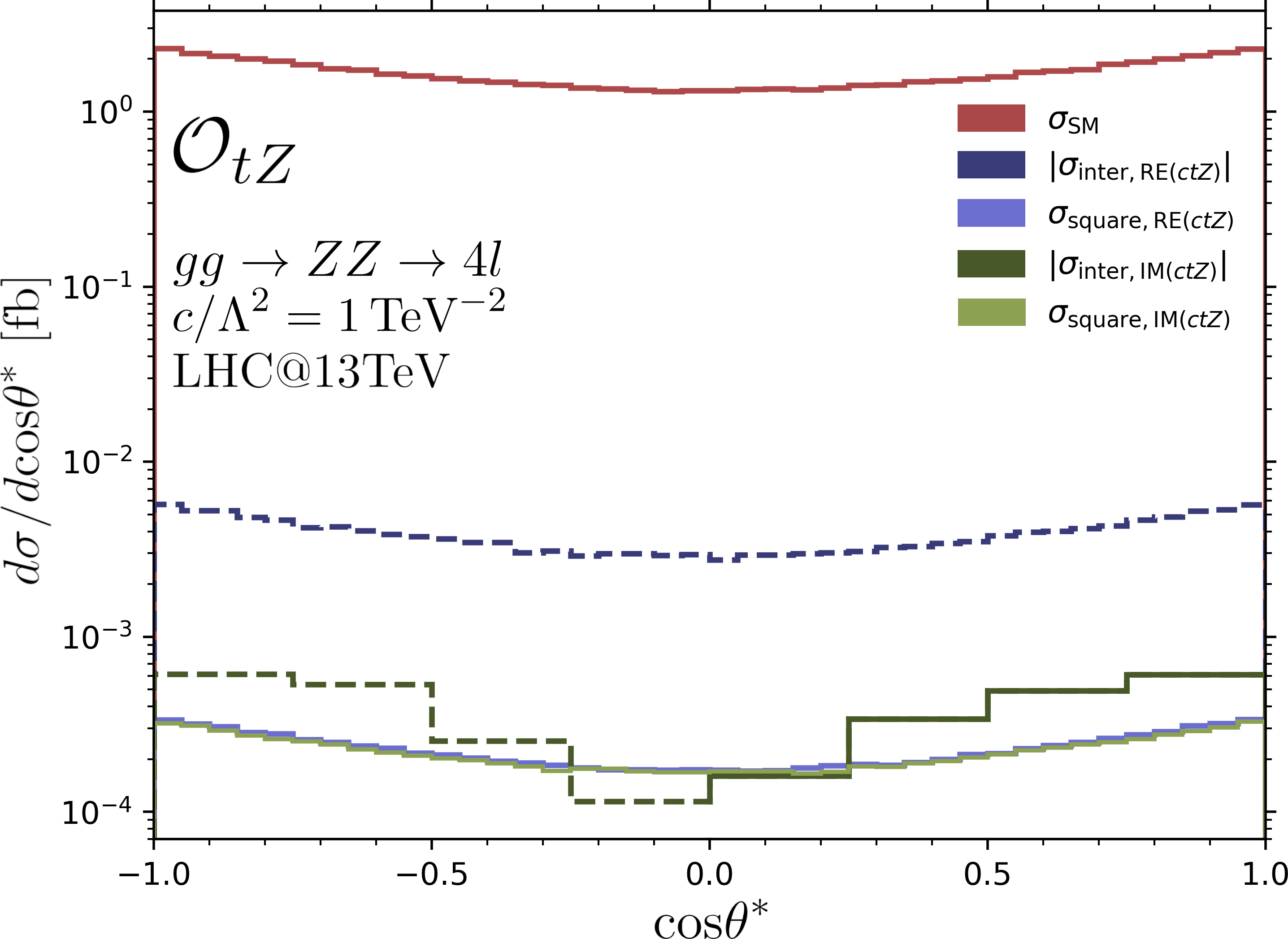}
\caption{}
\label{fig:time4}
\end{subfigure}

\caption{Angular distribution ($\text{cos}\theta^*$) of $gg \rightarrow ZZ \rightarrow e^+ e^- \mu^+ \mu^-$ in the presence of (a) $\OpW$ and $\OpWtil$, (b) $\mathcal{O}_{t \varphi}$, (c) $\mathcal{O}_{tG}$ and (d) $\mathcal{O}_{tZ}$. In (a), (c), and (d) the squared distributions overlap. For the interferences, a dashed line denotes a negative contribution. In addition, to avoid statistical fluctuations we only consider $8$ bins for the CP-odd interferences.}
\label{ctheta_smeftZZ}
\end{figure}

\begin{figure}
\centering

\begin{subfigure}{0.49\columnwidth}
\centering
\includegraphics[width=\textwidth]{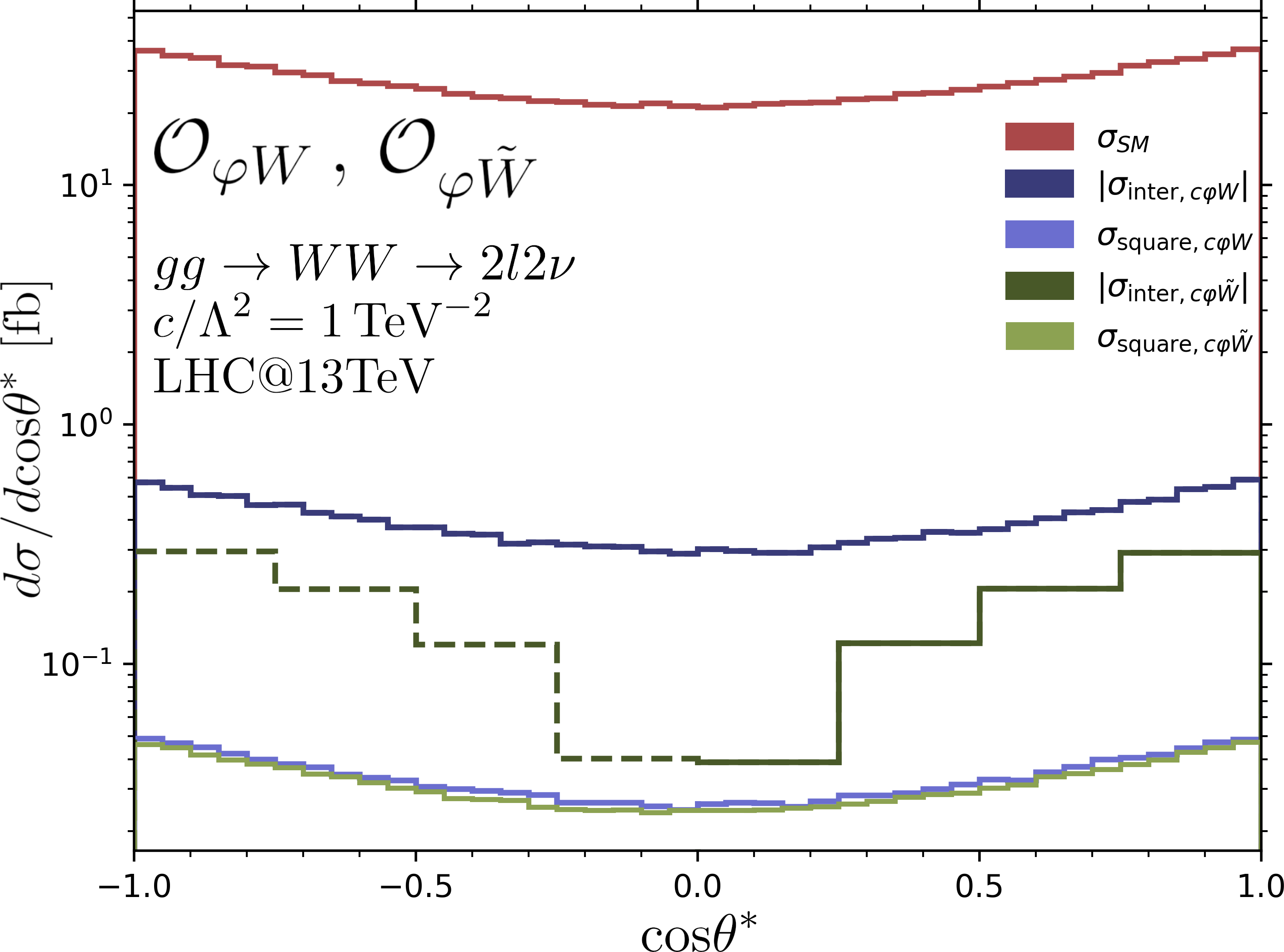}
\caption{}
\label{fig:ctheta2l2v_opw}
\end{subfigure}\hfill
\begin{subfigure}{0.49\columnwidth}
\centering
\includegraphics[width=\textwidth]{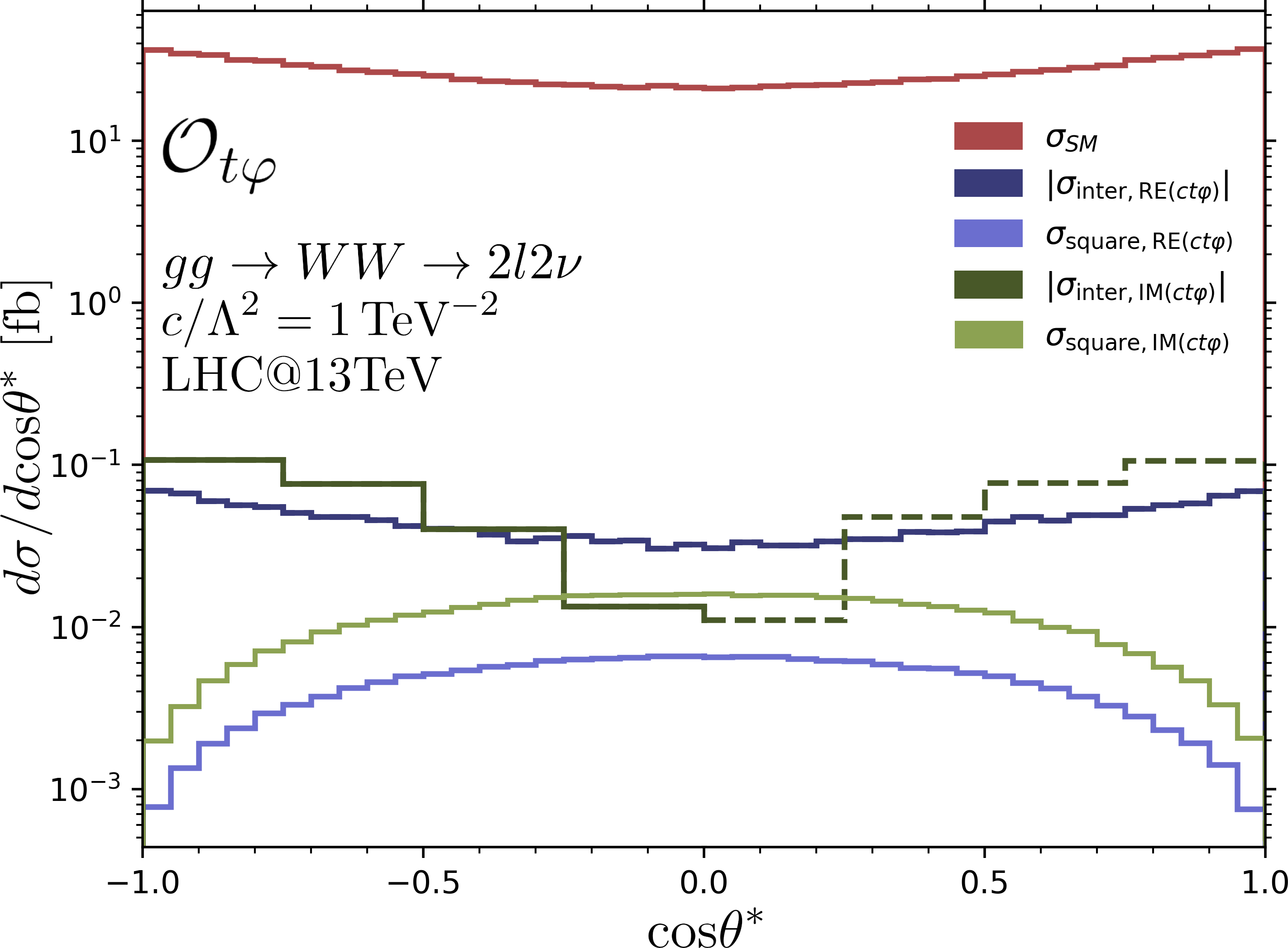}
\caption{}
\label{fig:ctheta2l2v_otp}
\end{subfigure}

\medskip

\begin{subfigure}{0.49\columnwidth}
\centering
\includegraphics[width=\textwidth]{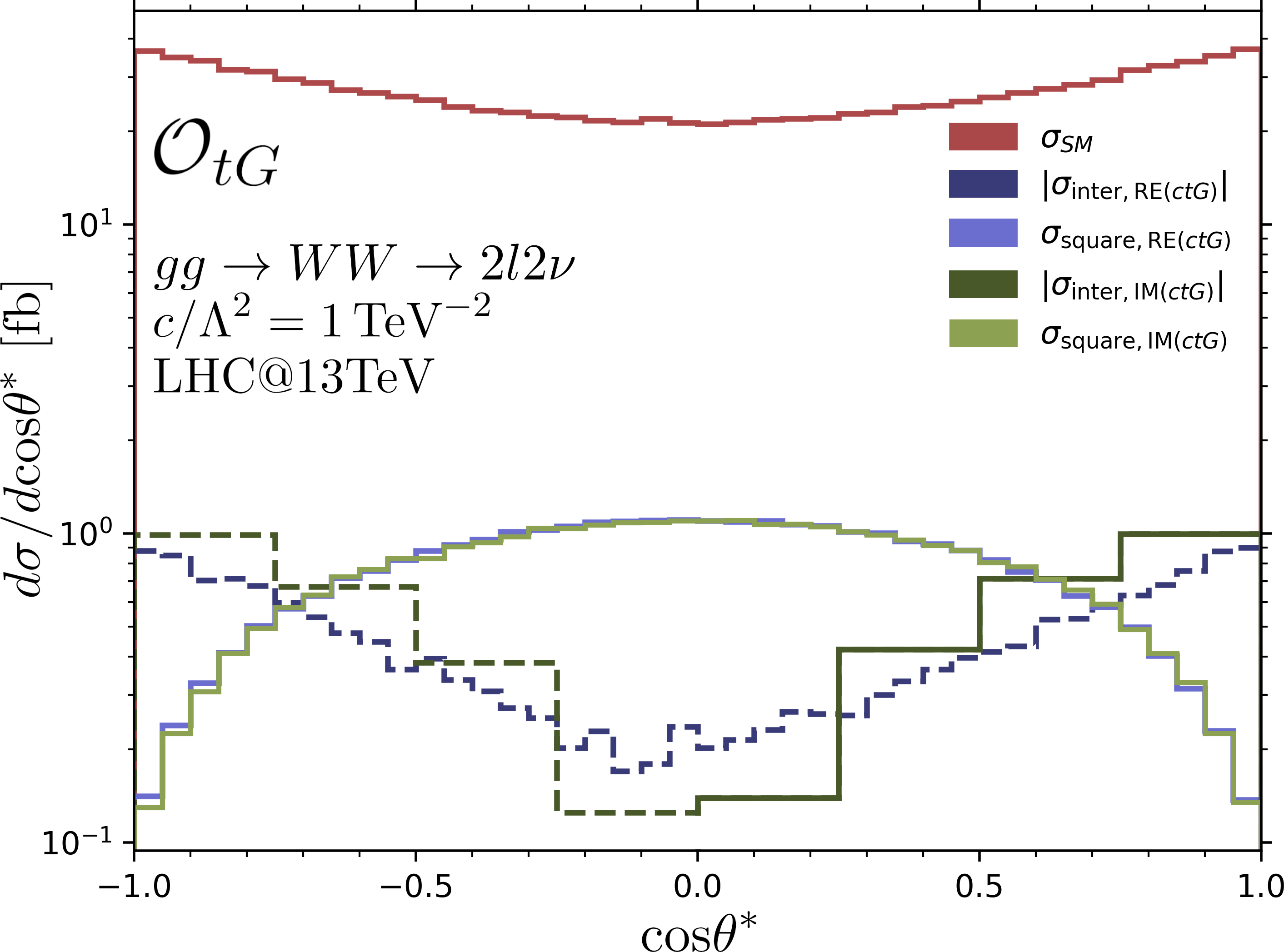}
\caption{}
\label{fig:ctheta2l2v_otg}
\end{subfigure}\hfill
\begin{subfigure}{0.49\columnwidth}
\centering
\includegraphics[width=\textwidth]{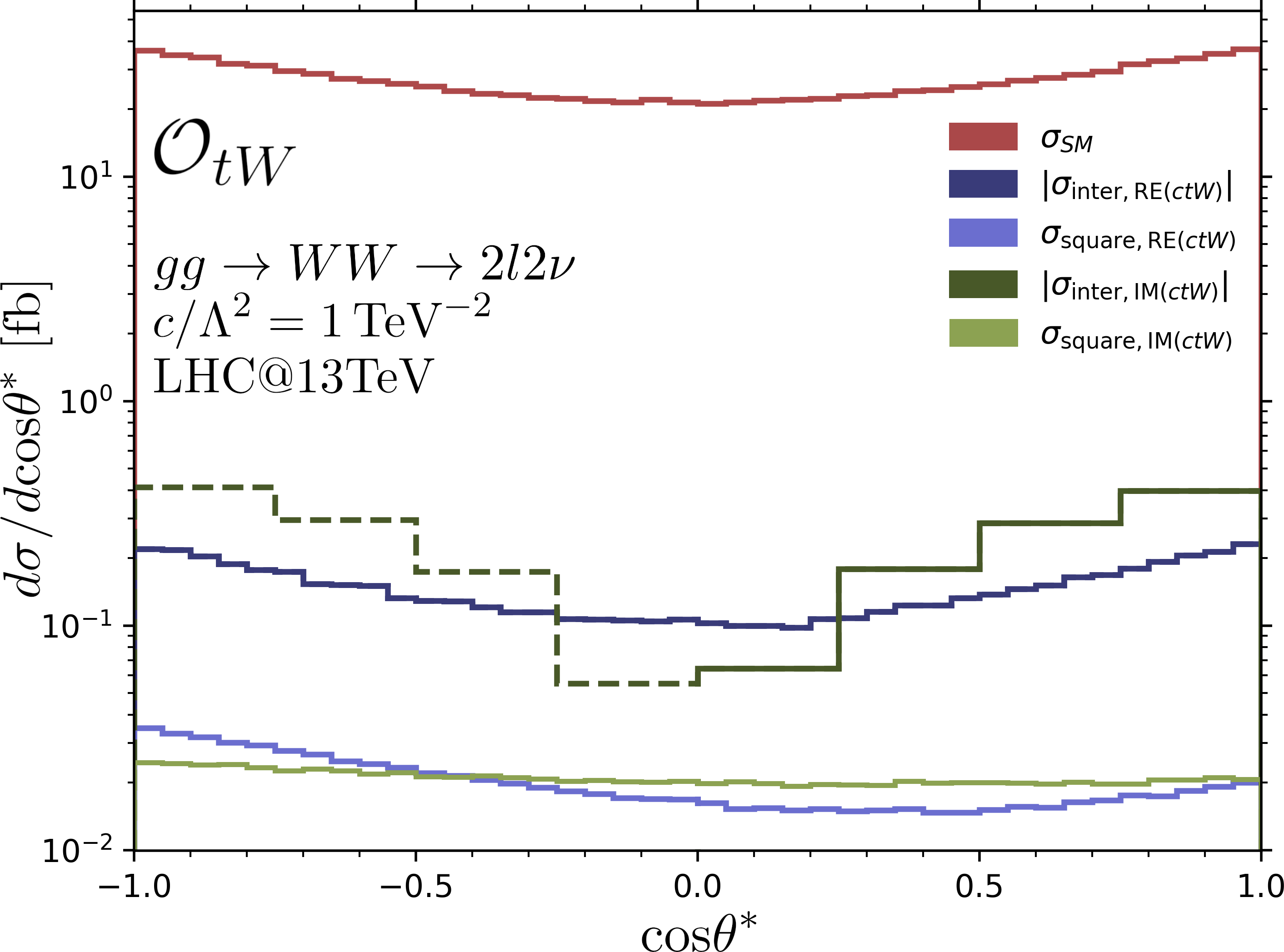}
\caption{}
\label{fig:ctheta2l2v_otw}
\end{subfigure}

\caption{Angular distribution ($\text{cos}\theta^*$) of $gg \rightarrow W^+W^- \rightarrow \mu^+ \nu_\mu \, e^- \bar{\nu}_e$ in the presence of (a) $\OpW$ and $\OpWtil$, (b) $\mathcal{O}_{t \varphi}$, (c) $\mathcal{O}_{tG}$ and (d) $\mathcal{O}_{tW}$. In (a) and (c) the squared distributions overlap. For the interferences, a dashed line denotes a negative contribution. In addition, to avoid statistical fluctuations we only consider $8$ bins for the CP-odd interferences.}
\label{ctheta_smeft2l2v}
\end{figure}

We start by studying the impact of CP-conserving and CP-violating SMEFT operators on the angular ($\text{cos}\theta^*$) distributions. The results are shown in Fig.~\ref{ctheta_smeftZZ} for $gg \rightarrow e^+ e^- \mu^+ \mu^-$ and in Fig.~\ref{ctheta_smeft2l2v} for $gg \rightarrow \mu^+ \nu_\mu e^- \bar{\nu}_e$ and were obtained setting  $c/\Lambda^2 = 1 \text{TeV}^{-2}$.

Remarkably, we see that the interferences of the CP-odd coefficients with the SM are now non-zero even though they vanished in the case of $gg \rightarrow ZZ, WW$. 
As was discussed earlier, in the $2 \rightarrow 2$ processes this is due to the CP-odd properties of the amplitudes, which lead to the exact cancellation between the different non-zero helicity amplitudes. In $gg \rightarrow 4l, 2l2\nu$, these cancellations do not happen anymore and the interference is non-zero.
The resulting angular distributions for the interferences of the CP-violating coefficients are odd around $\text{cos}\theta^* =0$, such that the inclusive cross-section is still zero. We now comment in more detail on the two processes considered.

\paragraph{ZZ production}

The $gg \rightarrow ZZ \rightarrow 4l$ angular distributions of Fig.~\ref{ctheta_smeftZZ} for the SM and the CP-even SMEFT operators are all symmetric around $\text{cos}\theta^* =0$, such that $\langle\text{cos}\theta^*\rangle = 0$ and as a consequence $f_L = f_R$. This can be verified at the amplitude squared level in the underlying $gg \rightarrow Z Z$ process by summing all the helicity amplitudes for a left-handed $Z$ and comparing with the sum of the amplitudes involving a right-handed $Z$. The left–right symmetry of the $Z$ boson production in the center-of-mass frame is due to Bose symmetry and the CP properties of the amplitude, as was noted in \cite{2107.06579} for the Standard Model and generalised to the CP-even dimension-$6$ operators considered here. In the presence of CP-odd operators these arguments need to be treated more carefully. Indeed, the total cross-section in each $\text{cos}\theta^*$ bin of one operator is given by the sum of the SM distribution with the interference and the squared contributions. Taking as an example $\OpWtil$ (though the argument holds for all the CP-odd operators), in the $\text{cos}\theta^*<0$ region the interference is negative. 
 Therefore the overall cross-section in the $\text{cos}\theta^*<0$ region is smaller than the overall cross-section in the $\text{cos}\theta^*>0$ region and $\langle\text{cos}\theta^*\rangle \neq 0$. A physical consequence is that for the CP-violating operators the transverse polarisation fractions $f_L$ and $f_R$ are not equal, which can also be seen at the amplitude level. As mentioned in Sec.~\ref{sec:ggHH} the non-zero helicity amplitudes of the underlying $gg \rightarrow ZZ$ process are related under CP transformations: for example $I_{\text{CPV}}(++--) = - I_{\text{CPV}}(--++) $ and $I_{\text{CPV}}(+---) = - I_{\text{CPV}}(-+++) $. Summing over all helicity configurations in which the first $Z$ boson is left-handed gives an interference which has the same magnitude but opposite sign to that coming from the sum of helicity configurations in which the $Z$ is right-handed. Adding to each sum the $\mathcal{O}(\Lambda^0)$ and $\mathcal{O}(\Lambda^{-4})$ contributions, for which CP transformations give $|A(++--)|^2 =  |A(--++)|^2 $, leads to the polarisation fractions $f_L$ and $f_R$ being different. Hence, the CP properties of the amplitudes in the presence of CP-odd operators in double $Z$ production directly imply that $\langle\text{cos}\theta^*\rangle \neq 0$ and $f_L \neq f_R$. These observations have been made also in the context of polarisation and spin correlation observables in Higgs decays to $W$ and $Z$ bosons \cite{2209.14033,2403.13942}. The amplitudes for these processes share the same properties with the amplitudes we consider here and also there CP violation leads to a left-right asymmetry. 

 \paragraph{$WW$ production}

In the SM, the $gg \rightarrow WW$ process receives contributions from triangle diagrams with a Higgs propagator and from boxes  as shown in Fig.~\ref{diagWW}. The contribution of the top triangle diagram respects the $f_L = f_R$ symmetry. The box diagrams involve contributions from the vector-vector (VV), axial-vector-axial-vector (AA) and vector-axial-vector (VA) couplings of the $W$ bosons. The VV and AA terms are non-zero for all three quark generations and their contribution to the amplitude also respects the left-right polarisation symmetry. The VA terms are non-zero only for the third generation as they contribute only when the masses of the two quarks in the loop are different \cite{PhysRevD.43.1555}.  These VA terms are the ones that break the $f_L = f_R$ symmetry. In the SM the third generation contributes to the total cross-section at the few percent level and hence this breaking of the left-right polarisation symmetry is negligible at the inclusive level. The situation is modified by the insertion of SMEFT operators which only contribute to a subset of diagrams. $\OpW, \OpWtil$ and $\Oth$ modify the triangle diagrams and as such lead to symmetric quadratic distributions. Then, $\Otg$ modifies the top-gluon interaction and enters in the third-generation boxes as well as in the diagram with a  $t\bar{t}gh$ vertex. The boxes lead to an asymmetry between the left and right polarisation, but in the presence of $\Otg$ this asymmetry is very small, explaining the symmetric quadratic distributions of Fig.~\ref{fig:ctheta2l2v_otg}. The electroweak dipole operator $\OtW$ only enters in the third-generation boxes and the resulting asymmetry in the angular distribution can be seen in both the $\re\ctW$ and $\im\ctW$ squared distributions of Fig.~\ref{fig:ctheta2l2v_otw}. Finally, similarly to double $Z$ production, the interference patterns of the underlying $gg \rightarrow WW$ process in the presence of CP-odd operators are linked to the CP transformations of the amplitudes, although with modified relations between the different helicity configurations since the $W$ is not its own anti-particle: for example $I_{\text{CPV}}(+---) = - I_{\text{CPV}}(+-++) $. Again CP violation leads to differences between $f_L$ and $f_R$, enhanced compared to the effect of the third generation boxes.

\subsection{Polarisation fractions in the presence of SMEFT operators} \label{sec:4l_polfrac}

The polarisation fractions of Eqs.~\eqref{eq::polfrac1}-\eqref{eq::polfrac2} are modified by the CP-conserving and CP-violating SMEFT operators that enter in $gg \rightarrow 4l, 2l2\nu$. In particular $\langle\text{cos}\theta^*\rangle$ and $\langle\text{cos}^2\theta^*\rangle$ can be parametrised as a function of the Wilson coefficient $c_i$:
\begin{align}
    &\langle\text{cos}\theta^*\rangle_i = \frac{\lambda_{SM} + c_i \lambda_{\text{inter}} + c_i^2 \lambda_{\text{sq}}}{\sigma_{SM}+ c_i\sigma_{\text{inter}} + c_i^2 \sigma_{\text{sq}}}\label{eq::ctheta1}\\
&\langle\text{cos}^2\theta^*\rangle_i = \frac{\lambda_{SM}^\prime + c_i \lambda_{\text{inter}}^\prime + c_i^2 \lambda_{\text{sq}}^\prime}{\sigma_{SM}+ c_i\sigma_{\text{inter}} + c_i^2 \sigma_{\text{sq}}}
\label{eq::ctheta2}
\end{align}
Note that in these expressions the Wilson coefficients are dimensionful as the $1/\Lambda^2$ factor is absorbed in their definition. The Standard Model, $\mathcal{O}(\Lambda^{-2})$, and $\mathcal{O}(\Lambda^{-4})$ cross-sections are denoted by $\sigma_{SM}, \sigma_{\text{inter}}$, and $\sigma_{\text{sq}}$ respectively.
 
\paragraph{$ZZ$ production}
 
 The values of $\lambda_{\text{inter}}$, $\lambda_{\text{inter}}^\prime$, $\lambda_{\text{sq}}$, $\lambda_{\text{sq}}^\prime$, $\sigma_{\text{inter}}$ and $\sigma_{\text{sq}}$ for the different Wilson coefficients entering in $gg \rightarrow ZZ \rightarrow e^+ e^- \mu^+ \mu^-$ are given in Table~\ref{PolFracTable}. 
 For reference, the Standard Model values are found to be:
 \begin{equation} \label{lsmZZ}
     \lambda_{SM} = 0.0 \,\, \text{fb}\quad , \qquad  \lambda_{SM}^\prime = 1.283(1) \,\, \text{fb}\quad , \qquad \sigma_{SM} = 3.314(1)\,\,\text{fb}
 \end{equation} 
such that:
 \begin{equation}
      \langle\text{cos}\theta^*\rangle_{SM} = 0.0, \qquad \langle\text{cos}^2\theta^*\rangle_{SM} = 0.3871(3).
 \end{equation}
 In Eq.~\eqref{lsmZZ}, $\lambda_{SM} = 0.0$ is expected from the symmetric SM angular distribution of Fig.~\ref{ctheta_smeftZZ}. Similarly, the symmetric distributions for the squares of all the coefficients shown in Fig.~\ref{ctheta_smeftZZ} imply $\lambda_\text{sq}= 0.0$, and the symmetric interferences of the CP-even coefficients imply $\lambda_\text{inter}= 0.0$ for those. This is however not the case for the CP-odd interferences and their anti-symmetric angular distributions which lead to non-zero $\lambda_\text{inter}$, as alluded to in Sec.~\ref{sec:4l_angdist}. 
 
\begin{table}
     \centering
     \resizebox{\textwidth}{!}{
     \renewcommand{\arraystretch}{1.5}
     \begin{tabular}{c | c  c  c c c c}
    \toprule
    Coefficient & $\lambda_\text{inter} \, [\text{fb}]$& $\lambda_\text{sq}\, [\text{fb}]$ &$\lambda_\text{inter}^\prime\, [\text{fb}]$& $\lambda_\text{sq}^\prime\, [\text{fb}]$ &$\sigma_{\text{inter}}\, [\text{fb}]$&$\sigma_{\text{sq}}\, [\text{fb}]$\\
    \midrule
    \midrule
    %%%%%%%%%%%%%%%%%%%%%%%%%%%%%%%%%%%%%%%%%%%%%%%%%%%%%%%%%%%%%%%%%%%%%%%%%%%%%%%%%%%%%%%%%%%%%%%%%%%%%%%%%%%
    $\cpW$ 
    &$0.0$&$0.0$
    &$1.641(5)\times10^{-2}$ &$1.274(2)\times10^{-3}$ &$4.071(7)\times10^{-2}$&$3.215(3)\times10^{-3}$\\
%%%%%%%%%%%%%%%%%%%%%%%%%%%%%%%%%%%%%%%%%%%%%%%%%%%%%%%%%%%%%%%%%%%%%%%%%%%%%%%%%%%%%%%%%%%%%%%%%%%%%%%%%%%
    $\cpWtil$
    &$1.31(7)\times 10^{-3}$&$0.0$&$0.0$
    &$1.221(2)\times10^{-3}$ &$0.0$&$3.046(2)\times10^{-3}$
    \\
    $\re\ctg$
    &$0.0$&$0.0$ &$-1.37(2)\times10^{-2}$&$2.124(5)\times10^{-2}  $
    &$1.8(3)\times10^{-3}$&$1.0078(6)\times10^{-1}$
    \\
%%%%%%%%%%%%%%%%%%%%%%%%%%%%%%%%%%%%%%%%%%%%%%%%%%%%%%%%%%%%%%%%%%%%%%%%%%%%%%%%%%%%%%%%%%%%%%%%%%%%%%%%%%%
    $\im\ctg$
    &$6.2(4)\times 10^{-3} $&$0.0$&$0.0$&$2.127(5)\times10^{-2} $&$0.0$&$1.0031(3)\times10^{-1}$
    \\
%%%%%%%%%%%%%%%%%%%%%%%%%%%%%%%%%%%%%%%%%%%%%%%%%%%%%%%%%%%%%%%%%%%%%%%%%%%%%%%%%%%%%%%%%%%%%%%%%%%%%%%%%%%
  $\re\cth$

&$0.0$&$0.0$&$1.04(2)\times10^{-3} $&$1.543(4)\times10^{-4} $&$-1.25(3)\times10^{-3}$&$7.269(7)\times10^{-4}$
\\
%%%%%%%%%%%%%%%%%%%%%%%%%%%%%%%%%%%%%%%%%%%%%%%%%%%%%%%%%%%%%%%%%%%%%%%%%%%%%%%%%%%%%%%%%%%%%%%%%%%%%%%%%%%
 $\im\cth$

&$-8.6(6)\times 10^{-4} $&$0.0$&$0.0$&
$3.771(9)\times10^{-4} $&$0.0$&$1.760(1)\times10^{-3}$
 \\
%%%%%%%%%%%%%%%%%%%%%%%%%%%%%%%%%%%%%%%%%%%%%%%%%%%%%%%%%%%%%%%%%%%%%%%%%%%%%%%%%%%%%%%%%%%%%%%%%%%%%%%%%%%
 $\re\ctZ$

&$0.0$&$0.0$&
$-3.095(9)\times10^{-3} $&$1.840(3)\times10^{-4} $&$-7.726(4)\times10^{-3}$&$4.589(2)\times10^{-4}$
\\
%%%%%%%%%%%%%%%%%%%%%%%%%%%%%%%%%%%%%%%%%%%%%%%%%%%%%%%%%%%%%%%%%%%%%%%%%%%%%%%%%%%%%%%%%%%%%%%%%%%%%%%%%%%
$\im\ctZ$

&$5.0(2)\times10^{-4} $&$0.0$&$0.0$&$1.762(3)\times10^{-4} $&$0.0$&$4.416(4)\times10^{-4}$
\\
%%%%%%%%%%%%%%%%%%%%%%%%%%%%%%%%%%%%%%%%%%%%%%%%%%%%%%%%%%%%%%%%%%%%%%%%%%%%%%%%%%%%%%%%%%%%%%%%%%%%%%%%%%%
    \bottomrule
    \end{tabular}}
    \caption{Cross-sections and $\lambda$ parameters defined in Eq.~\eqref{eq::ctheta1} and \eqref{eq::ctheta2} for the CP-even and CP-odd operators entering in $gg \rightarrow ZZ \rightarrow e^+ e^- \mu^+ \mu^-$. The statistical uncertainties are given in the brackets.}
    \label{PolFracTable}
\end{table}

 In addition, another consequence of the CP-violating nature of some coefficients is that $\lambda_{\text{inter}}^\prime = 0$ for those, that is there is no linear contribution in  $\langle\text{cos}^2\theta^*\rangle$ for the CP-odd coefficients (these coefficients also have $\sigma_{\text{inter}}=0$) and the longitudinal polarisation fraction $f_0$, defined in Eq.~\eqref{eq::polfrac2} also receives no linear contributions from the CP-odd coefficients. 
 As discussed earlier, this is because at the interference level the different helicity amplitudes related by CP transformations cancel each other out. Combining this with the Bose symmetry of the $Z$ bosons gives a vanishing contribution at the interference level when adding all the helicity configurations in which the $Z$ boson considered is longitudinally polarised.

\begin{table}
     \centering
     \resizebox{\textwidth}{!}{
     \renewcommand{\arraystretch}{1.5}
     \begin{tabular}{c | c  c  c c c c}
    \toprule
    Coefficient & $\lambda_\text{inter} \, [\text{fb}]$& $\lambda_\text{sq}\, [\text{fb}]$ &$\lambda_\text{inter}^\prime\, [\text{fb}]$& $\lambda_\text{sq}^\prime\, [\text{fb}]$ &$\sigma_{\text{inter}}\, [\text{fb}]$&$\sigma_{\text{sq}}\, [\text{fb}]$\\
    \midrule
    \midrule
    %%%%%%%%%%%%%%%%%%%%%%%%%%%%%%%%%%%%%%%%%%%%%%%%%%%%%%%%%%%%%%%%%%%%%%%%%%%%%%%%%%%%%%%%%%%%%%%%%%%%%%%%%%%
    $\cpW$ 
    &$0.0$ &$0.0$&$3.152(9)\times10^{-1}$ &$2.692(5)\times10^{-2} $&$7.881(5)\times10^{-1} $&$6.768(4)\times10^{-2} $\\
%%%%%%%%%%%%%%%%%%%%%%%%%%%%%%%%%%%%%%%%%%%%%%%%%%%%%%%%%%%%%%%%%%%%%%%%%%%%%%%%%%%%%%%%%%%%%%%%%%%%%%%%%%%
    $\cpWtil$
    &$2.22(1)\times10^{-1} $ &$0.0$&$0.0 $&$2.550(5)\times10^{-2} $&$0.0$&$6.385(4)\times10^{-2} $
    \\

%%%%%%%%%%%%%%%%%%%%%%%%%%%%%%%%%%%%%%%%%%%%%%%%%%%%%%%%%%%%%%%%%%%%%%%%%%%%%%%%%%%%%%%%%%%%%%%%%%%%%%%%%%%
    $\re\ctg$
    &$0.0$&$0.0$&$-4.29(2) \times10^{-1} $&$3.289(8) \times10^{-1}  $&$-8.99(2)\times10^{-1}$&$1.5307(7) $
    \\
%%%%%%%%%%%%%%%%%%%%%%%%%%%%%%%%%%%%%%%%%%%%%%%%%%%%%%%%%%%%%%%%%%%%%%%%%%%%%%%%%%%%%%%%%%%%%%%%%%%%%%%%%%%
    $\im\ctg$
    &$7.48(7) \times10^{-1} $&$0.0$&$0.0 $&$3.277(8)\times10^{-1}$&$0.0$&$1.5225(8) $
    \\
%%%%%%%%%%%%%%%%%%%%%%%%%%%%%%%%%%%%%%%%%%%%%%%%%%%%%%%%%%%%%%%%%%%%%%%%%%%%%%%%%%%%%%%%%%%%%%%%%%%%%%%%%%%
  $\re\cth$

&$0.0$&$0.0$&$3.64(2)\times10^{-2} $&$1.931(5)\times10^{-3} $&$8.92(2)\times10^{-2} $&$9.06(1)\times10^{-3}$
\\
%%%%%%%%%%%%%%%%%%%%%%%%%%%%%%%%%%%%%%%%%%%%%%%%%%%%%%%%%%%%%%%%%%%%%%%%%%%%%%%%%%%%%%%%%%%%%%%%%%%%%%%%%%%
 $\im\cth$

&$-8.13(9)\times10^{-2}  $&$0.0$&$0.0 $&$4.72(1)\times10^{-3} $&$0.0$&$2.213(2)\times10^{-2}$
 \\
%%%%%%%%%%%%%%%%%%%%%%%%%%%%%%%%%%%%%%%%%%%%%%%%%%%%%%%%%%%%%%%%%%%%%%%%%%%%%%%%%%%%%%%%%%%%%%%%%%%%%%%%%%%
 $\re\ctW$

&$0.0$ &$-5.08(5)\times10^{-3} $&$1.201(4)\times10^{-1} $&$1.540(3)\times10^{-2} $&$2.901(2)\times10^{-1}$&$3.996(2)\times10^{-2} $
\\
%%%%%%%%%%%%%%%%%%%%%%%%%%%%%%%%%%%%%%%%%%%%%%%%%%%%%%%%%%%%%%%%%%%%%%%%%%%%%%%%%%%%%%%%%%%%%%%%%%%%%%%%%%%
$\im\ctW$

&$3.10(1)\times10^{-1}  $&$-1.32(5)\times10^{-3}$&$0.0$&$1.450(3)\times10^{-2}  $&$0.0$&$4.195(2)\times10^{-2} $
\\
%%%%%%%%%%%%%%%%%%%%%%%%%%%%%%%%%%%%%%%%%%%%%%%%%%%%%%%%%%%%%%%%%%%%%%%%%%%%%%%%%%%%%%%%%%%%%%%%%%%%%%%%%%%
    \bottomrule
    \end{tabular}}
    \caption{Cross-sections and $\lambda$ parameters defined in Eq.~\eqref{eq::ctheta1} and \eqref{eq::ctheta2} for the CP-even and CP-odd operators entering in $gg \rightarrow W^+W^- \rightarrow \mu^+ \nu_\mu \, e^- \bar{\nu}_e$. The statistical uncertainties are given in the brackets. Only the numerical values of the parameters that differ from zero by more than $2 \sigma$ are reported.}
    \label{PolFracTable2l2v}
\end{table}

 \paragraph{$WW$ production}

 The values of $\lambda_{\text{inter}}$, $\lambda_{\text{inter}}^\prime$, $\lambda_{\text{sq}}$, $\lambda_{\text{sq}}^\prime$, $\sigma_{\text{inter}}$ and $\sigma_{\text{sq}}$ for the different Wilson coefficients entering in $gg \rightarrow W^+W^- \rightarrow \mu^+ \nu_\mu \, e^- \bar{\nu}_e$ are given in Table~\ref{PolFracTable2l2v}, where we only give the numerical values of the parameters that differ from zero by more than $2 \sigma$. 
For reference, the Standard Model values are found to be:
 \begin{equation}
     \lambda_{SM} = 0.19(5) \,\, \text{fb}\quad , \qquad  \lambda_{SM}^\prime = 20.67(2) \,\, \text{fb}\quad , \qquad \sigma_{SM} = 53.48(1) \,\,\text{fb}
\label{eq::SMlambda}
 \end{equation} 
 such that:
 \begin{equation}
      \langle\text{cos}\theta^*\rangle_{SM} = 0.0036(9), \qquad \langle\text{cos}^2\theta^*\rangle_{SM} = 0.3865(4).
 \end{equation}
 
The CP-conserving coefficients $\cpW$ and $\re\cth$ have $\lambda_\text{inter}$ equal to zero, implying that $\langle\text{cos}\theta^*\rangle =0 $ for the $\mathcal{O}\big(\Lambda^{-2}\big)$ angular distributions even though the SM amplitude allows for $\langle\text{cos}\theta^*\rangle \neq 0 $ through the vector-axial-vector couplings of the $W$ bosons. Nevertheless in the interference between the SM and these operators, only the vector-vector and axial-vector-axial-vector couplings enter 
leading to a distribution symmetric around $\text{cos}\theta^* = 0$. Regarding the CP-even dipole coefficients, as they enter in the third-generation box diagrams, they could in principle have a non-symmetric interference with the SM. However their $\mathcal{O}\big(\Lambda^{-2}\big)$ angular distributions shown in Fig.~\ref{ctheta_smeft2l2v} are symmetric around $\text{cos}\theta^* =0 $ and as a consequence $\lambda_\text{inter}$ of $\re\ctg$ and $\re\ctW$ is found compatible with zero within $2\sigma$. 

Following the discussion of  Sec.~\ref{sec:4l_angdist}, $\OpW, \OpWtil$ and $\Oth$ have $\mathcal{O}\big(\Lambda^{-4}\big)$ angular distributions which are symmetric around $\text{cos}\theta^* = 0$ such that $\langle\text{cos}\theta^*\rangle =0 $, and this is reflected by the fact that their corresponding $\lambda_{\text{sq}}$ terms vanish. While $\Otg$ enters in the third generation box diagrams, its quadratic angular distributions still appear symmetric, as shown in Fig.~\ref{fig:ctheta2l2v_otg} and $\lambda_{\text{sq}}$ for both the CP-even and the CP-odd coefficient is found to be compatible with zero within $2 \sigma$. The only operator with significantly asymmetric $\mathcal{O}\big(\Lambda^{-4}\big)$ angular distributions is thus the electroweak dipole $\OtW$.

Finally, the CP-odd coefficients have $\lambda_\text{inter}^\prime = 0$  such that there is no interference piece in  $\langle\text{cos}^2\theta^*\rangle$. For $\cpWtil$ and $\im\cth$, this can be verified by considering the longitudinal polarisation fraction $f_0$ in the underlying $gg \rightarrow W^+ W^-$ process at the amplitude squared level, as outlined above in the case of $ZZ$ production. In $WW$ production, $\cpWtil$ and $\im\cth$ only enter in the triangle diagrams shown in Fig.~\ref{diagWW} and the only non-zero helicity configurations with a longitudinal $W^-$ are $(++0\,0)$ and $(--0\,0)$ which are related under CP transformations by $I_\text{CPV}(++0\,0) = -I_\text{CPV}(--0\,0)$. Therefore, even though the Bose symmetry of the final state does not apply to $gg \rightarrow WW$, the cancellation among helicity amplitudes still applies here to show the vanishing of some $\lambda_\text{inter}^\prime$. Regarding the dipole coefficients $\im\ctg$ and $\im\ctW$, the interference piece of $f_0$ does not vanish at the amplitude squared level, but at the inclusive cross-section level and $\lambda_\text{inter}^\prime$ is consistent with zero  for $\im\ctg$ and $\im\ctW$ as can be inferred from the anti-symmetric $\mathcal{O}\big(\Lambda^{-2}\big)$ angular distributions of Fig.~\ref{ctheta_smeft2l2v}.

\subsection{Inclusive polarisation fractions}\label{sec:inclusivepolfra}

\begin{figure}
\centering

\begin{subfigure}{0.49\columnwidth}
\centering
\includegraphics[width=\textwidth]{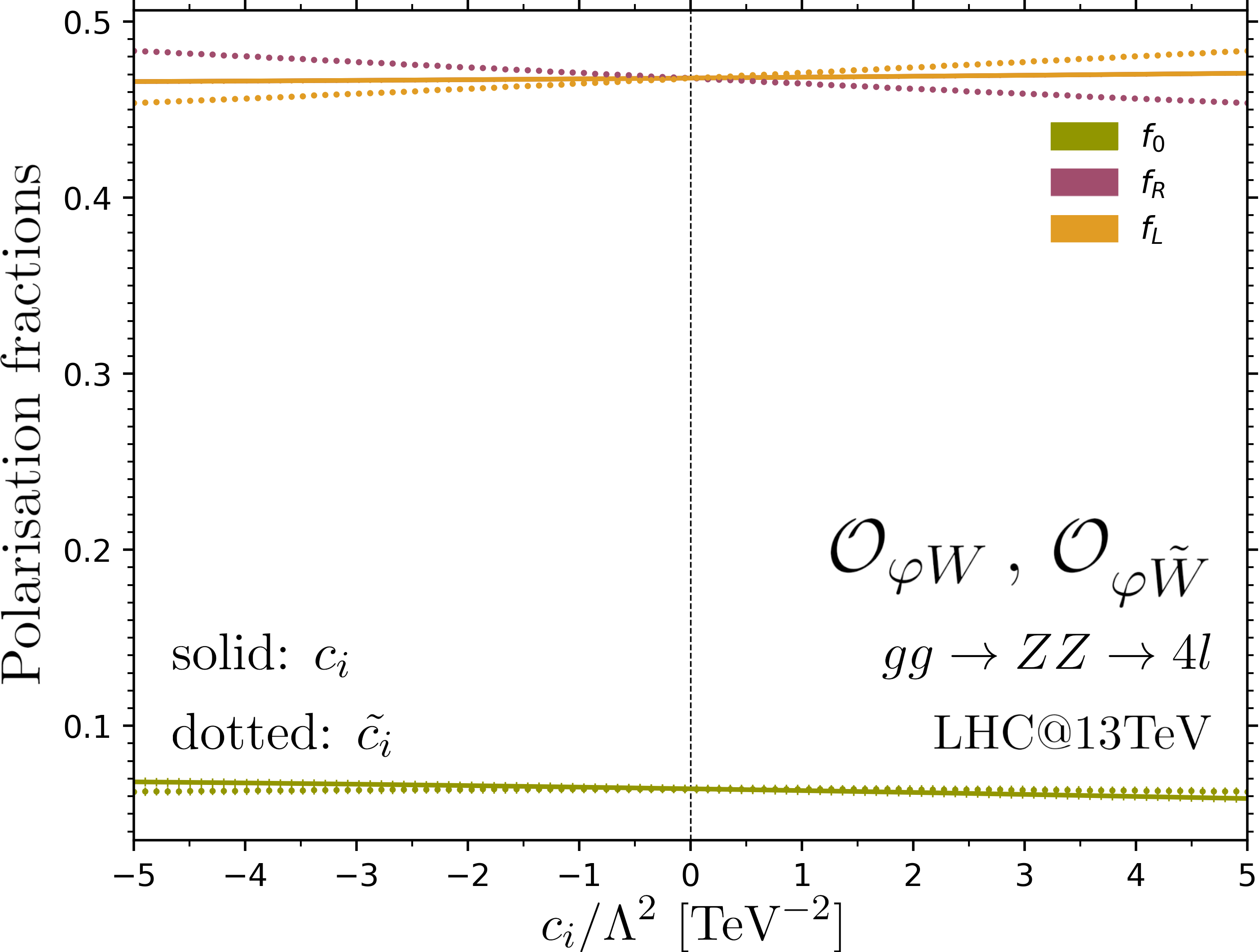}
\caption{}
\label{4l_inc_cpBB}
\end{subfigure}\hfill
\begin{subfigure}{0.49\columnwidth}
\centering
\includegraphics[width=\textwidth]{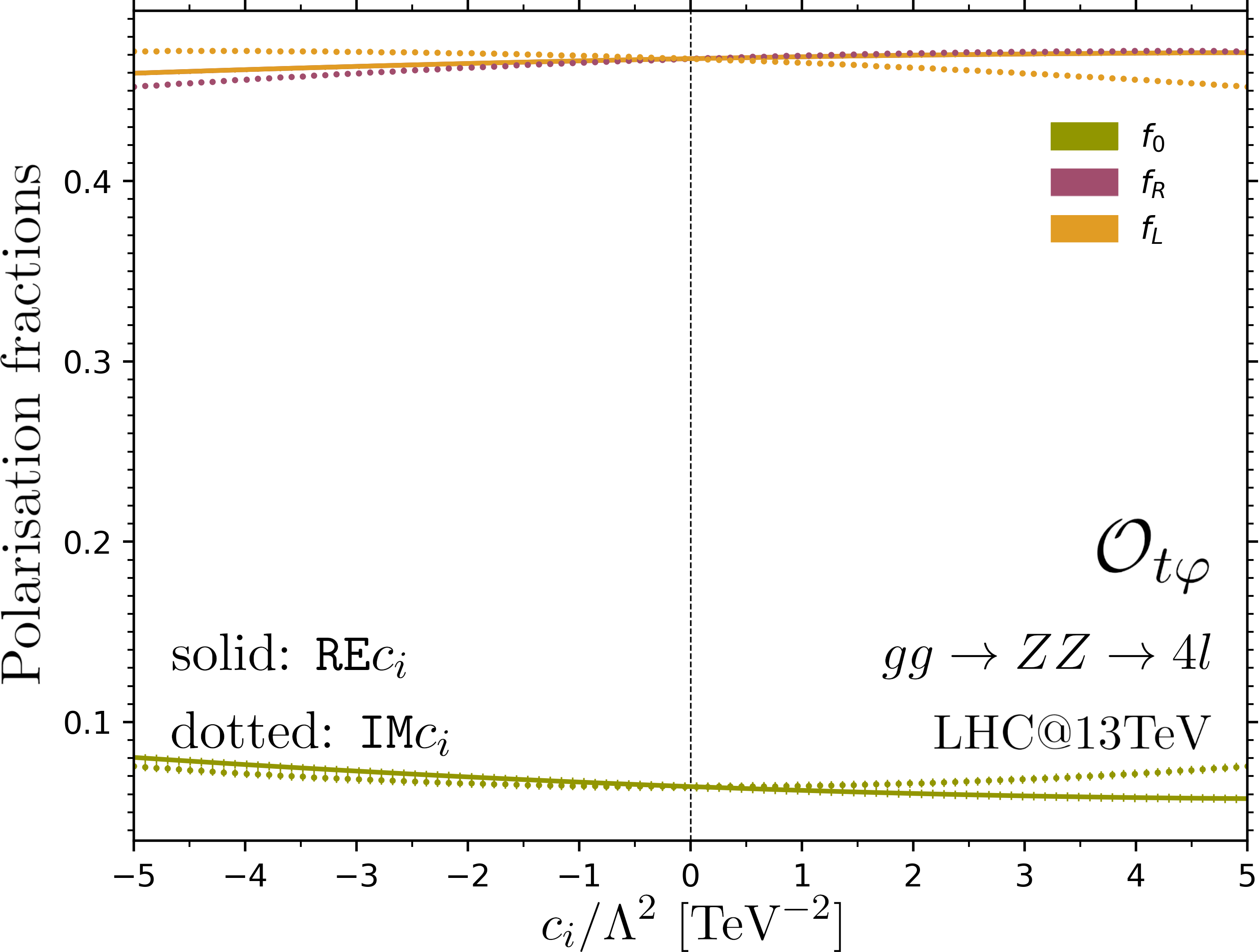}
\caption{}
\label{4l_inc_cpW}
\end{subfigure}

\medskip

\begin{subfigure}{0.49\columnwidth}
\centering
\includegraphics[width=\textwidth]{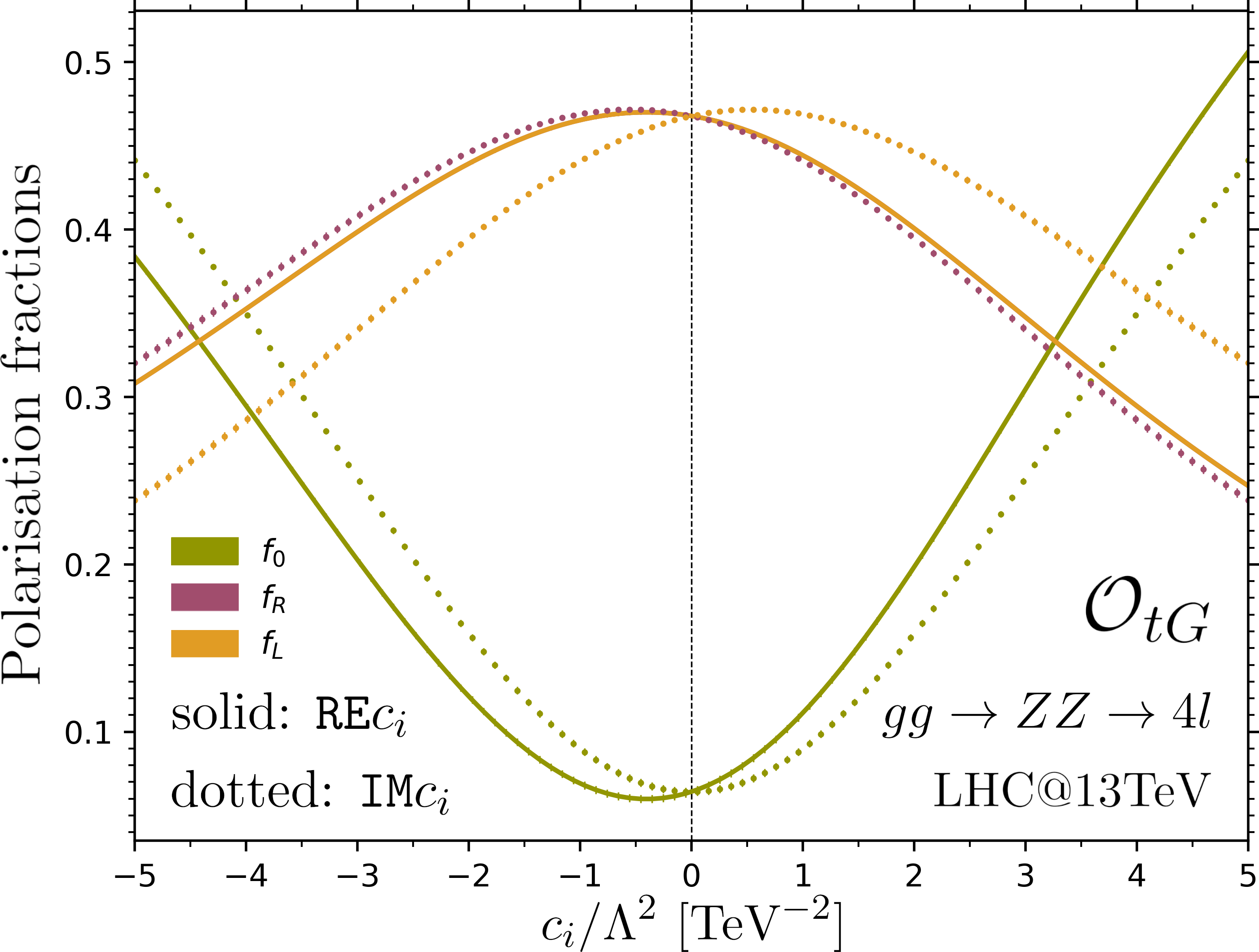}
\caption{}
\label{4l_inc_ctp}
\end{subfigure}\hfill
\begin{subfigure}{0.49\columnwidth}
\centering
\includegraphics[width=\textwidth]{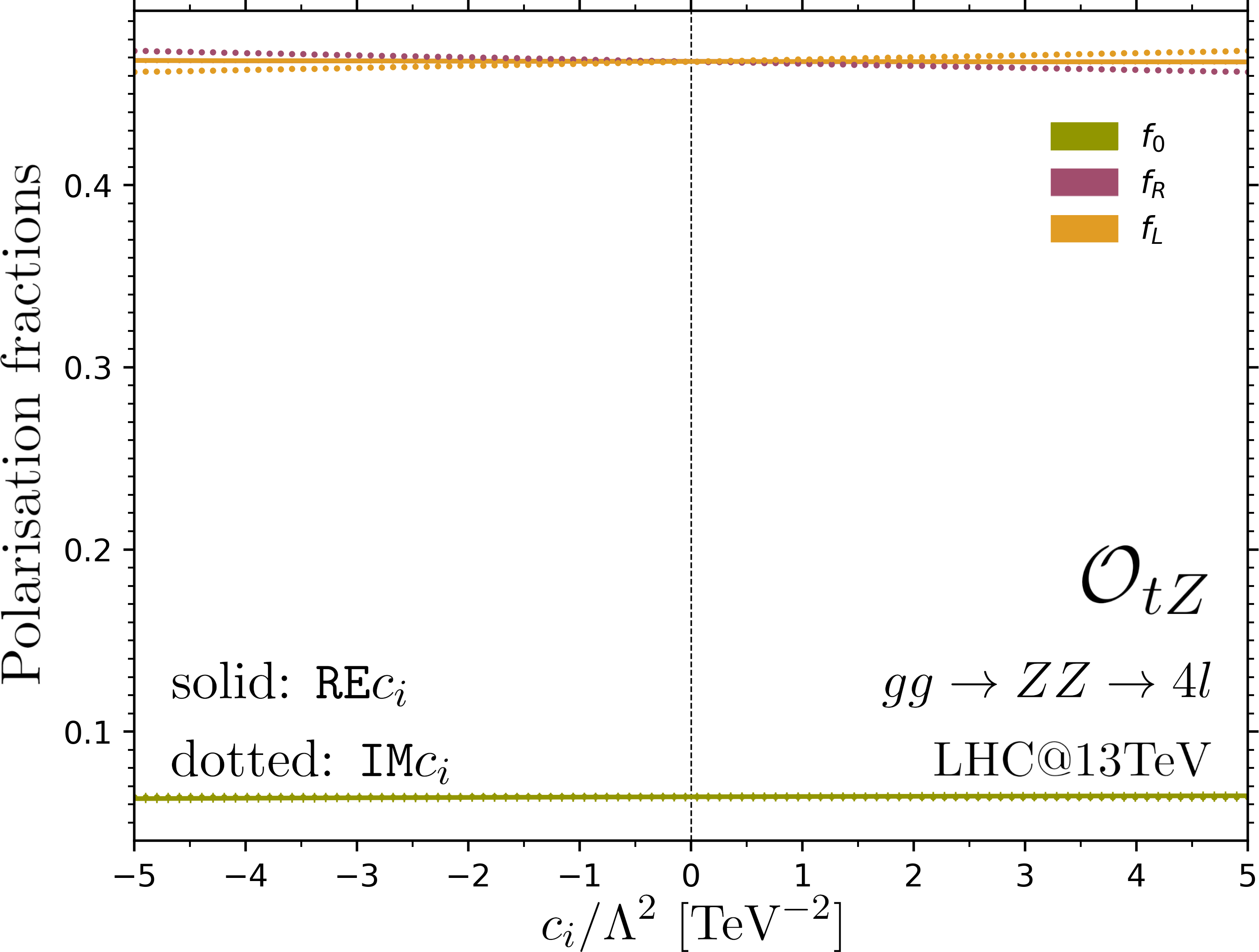}
\caption{}
\label{4l_inc_ctG}
\end{subfigure}

\caption{Polarisation fractions in $gg \rightarrow ZZ \rightarrow e^+ e^- \mu^+ \mu^-$ in the presence of (a) $\OpW$ and $\OpWtil$, (b) $\mathcal{O}_{t \varphi}$, (c) $\mathcal{O}_{tG}$ and (d) $\mathcal{O}_{tZ}$ for varying values of the Wilson coefficients. The solid (dotted) lines represent the results in the presence of CP-even (CP-odd) coefficients. The vertical line at $c_i = 0$ represents the Standard Model values.}
\label{inclusive4l}
\end{figure}

\begin{figure}
\centering

\begin{subfigure}{0.49\columnwidth}
\centering
\includegraphics[width=\textwidth]{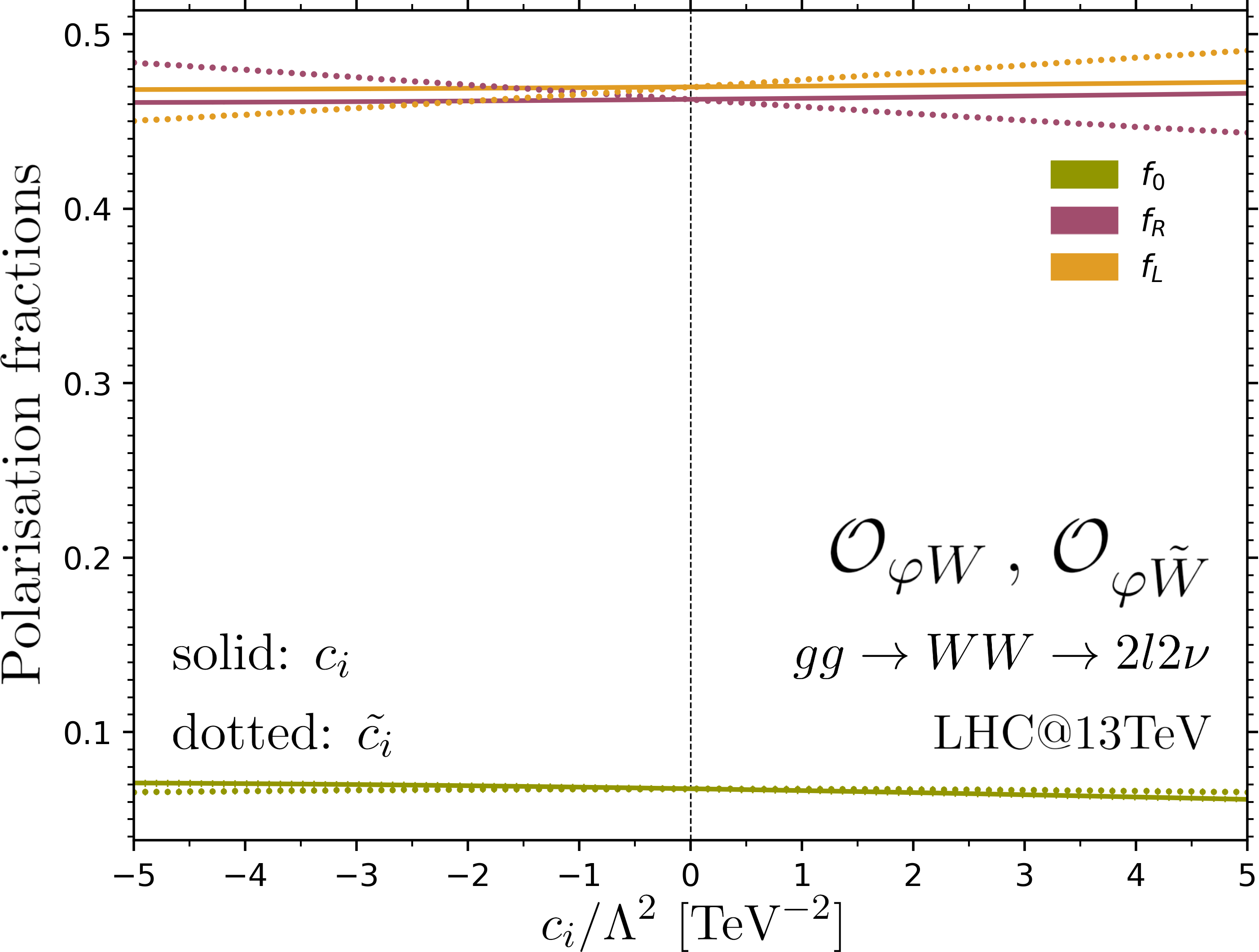}
\caption{}
\label{2l2v_inc_cpW}
\end{subfigure}\hfill
\begin{subfigure}{0.49\columnwidth}
\centering
\includegraphics[width=\textwidth]{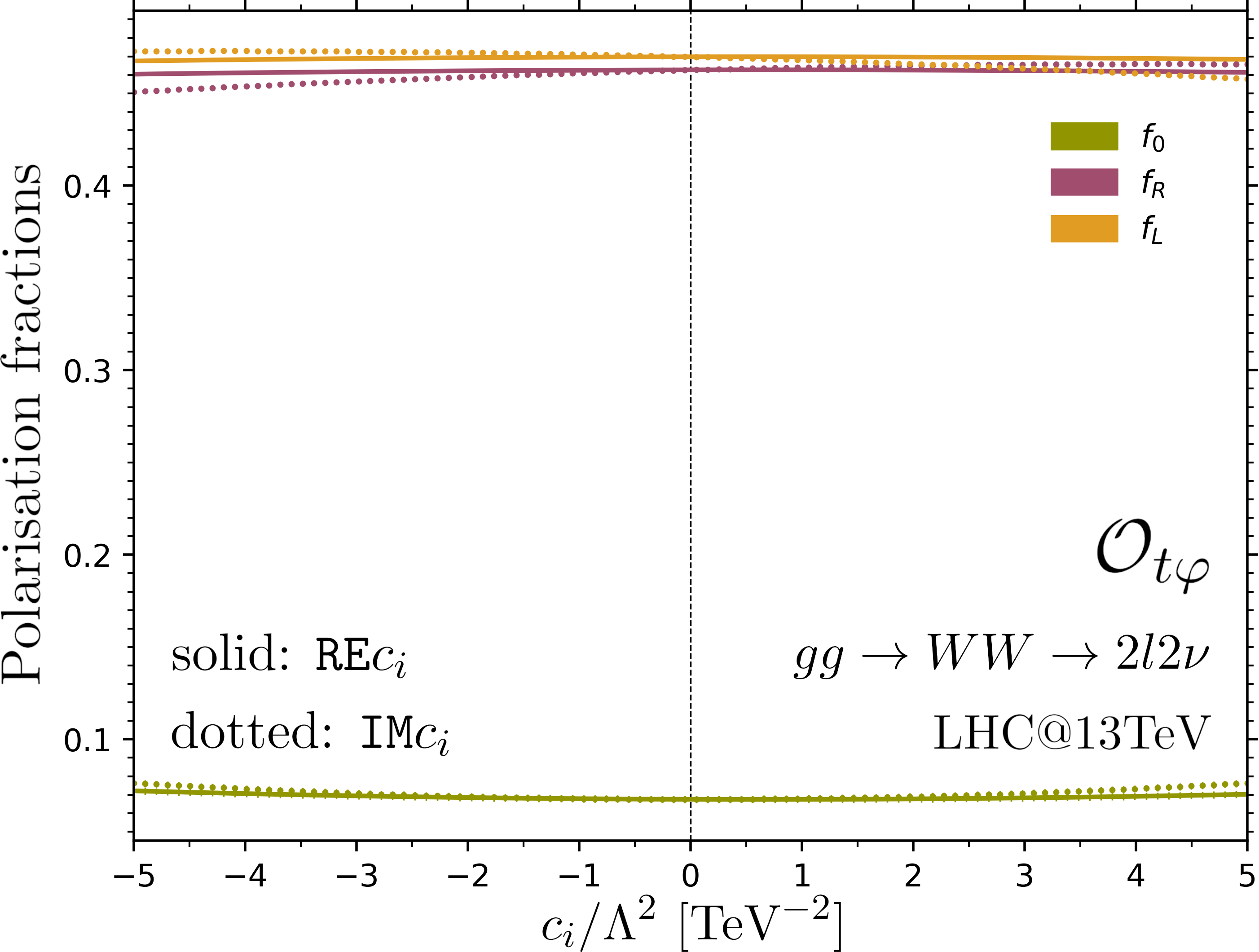}
\caption{}
\label{2l2v_inc_ctp}
\end{subfigure}

\medskip

\begin{subfigure}{0.49\columnwidth}
\centering
\includegraphics[width=\textwidth]{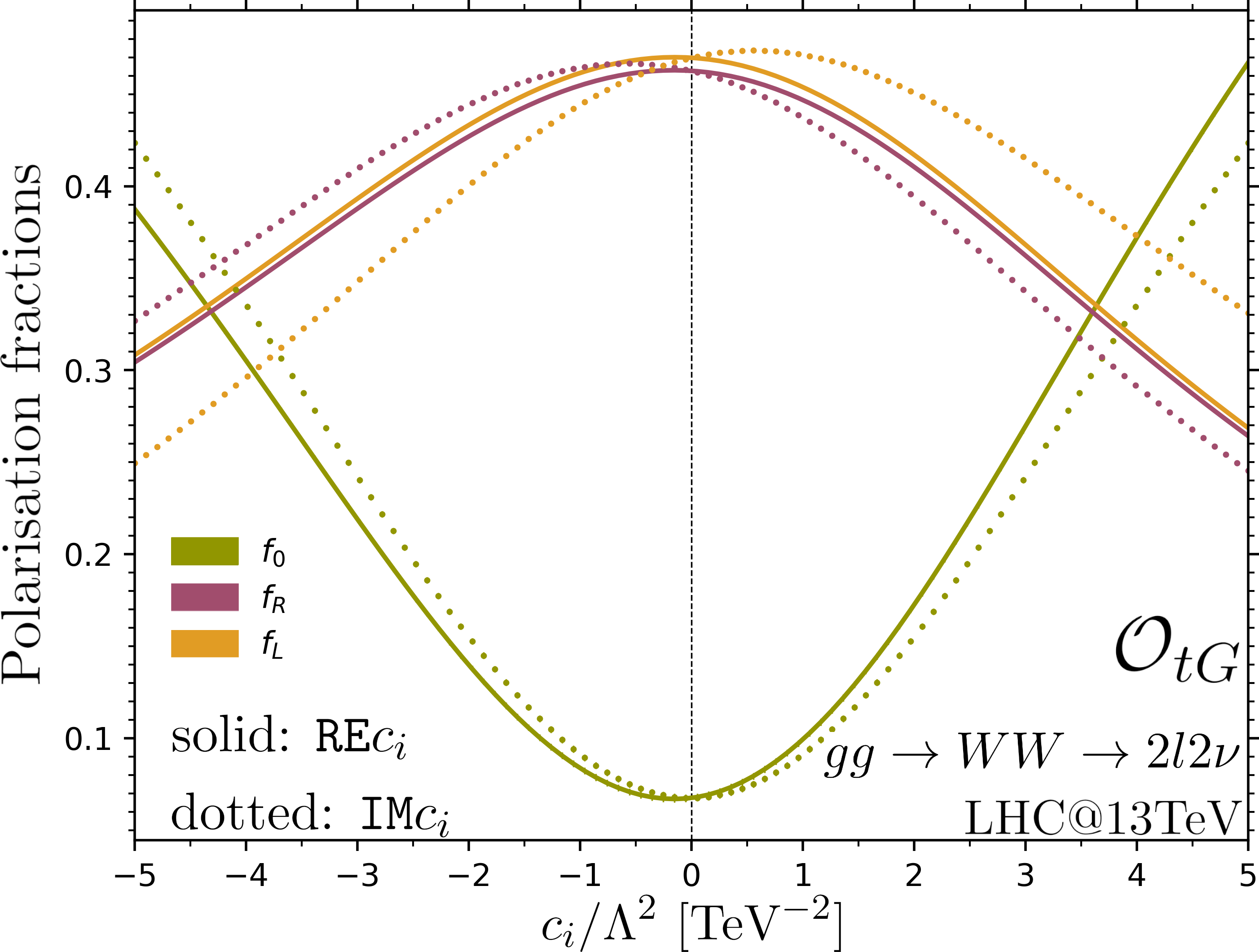}
\caption{}
\label{2l2v_inc_ctg}
\end{subfigure}\hfill
\begin{subfigure}{0.49\columnwidth}
\centering
\includegraphics[width=\textwidth]{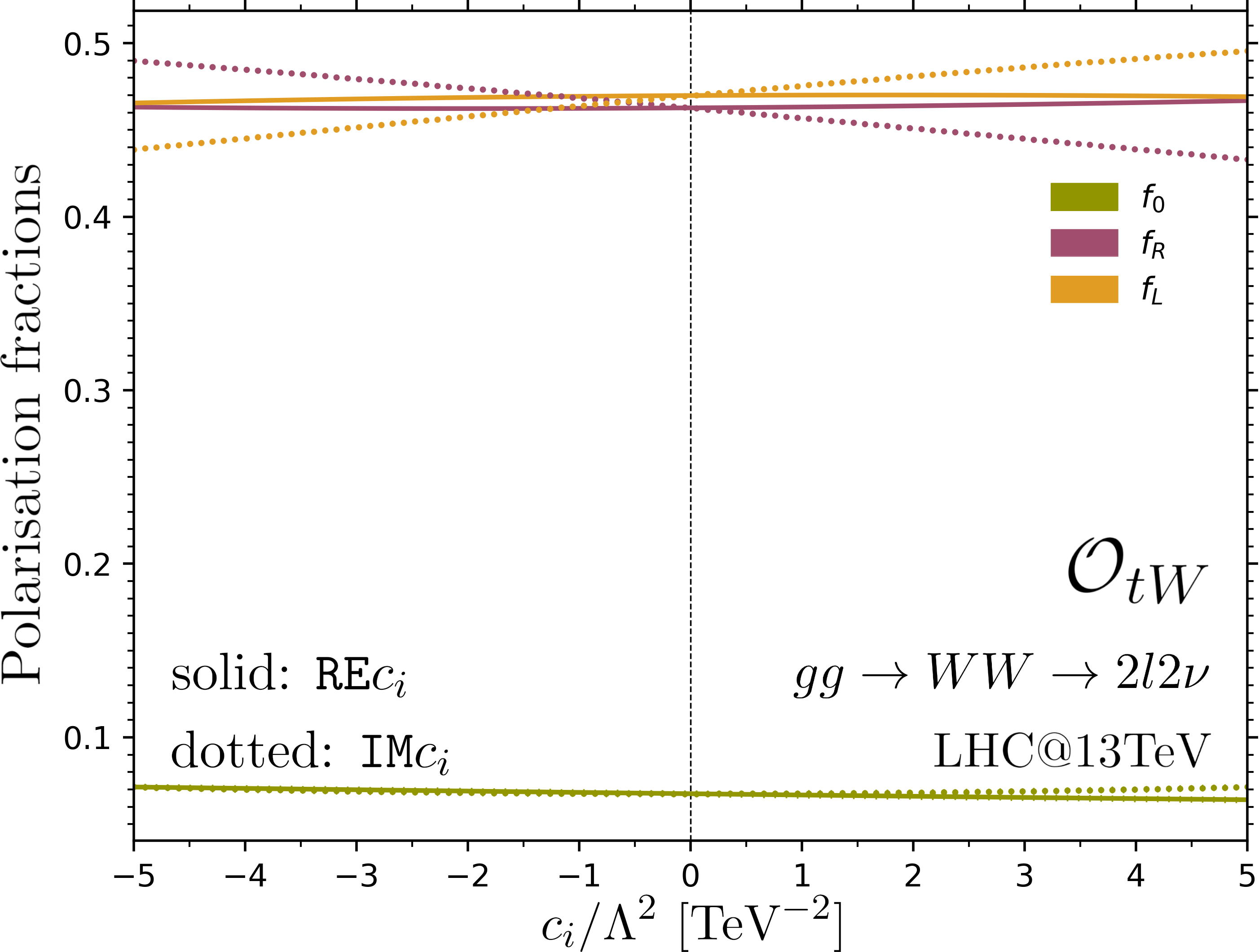}
\caption{}
\label{2l2v_inc_ctW}
\end{subfigure}

\caption{Polarisation fractions in $gg \rightarrow W^+W^- \rightarrow \mu^+ \nu_\mu \, e^- \bar{\nu}_e$ in the presence of (a) $\OpW$ and $\OpWtil$, (b) $\mathcal{O}_{t \varphi}$, (c) $\mathcal{O}_{tG}$ and (d) $\mathcal{O}_{tW}$ for varying values of the Wilson coefficients. The solid (dotted) lines represent the results in the presence of CP-even (CP-odd) coefficients. The vertical line at $c_i = 0$ represents the Standard Model values.}
\label{inclusive2l2v}
\end{figure}

The parametrisation of Eqs.~\eqref{eq::ctheta1}-\eqref{eq::ctheta2} can be combined with Eqs.~\eqref{eq::polfrac1}-\eqref{eq::polfrac2} and the numerical results of Tables~\ref{PolFracTable}-\ref{PolFracTable2l2v} to obtain the values of the polarisation fractions for any value of the Wilson coefficients. Following this approach, we show here the behaviour of the polarisation fractions for changing values of the CP-even and CP-odd coefficients. The results are presented in Fig.~\ref{inclusive4l} for $gg\rightarrow 4l$ and in Fig.~\ref{inclusive2l2v} for $gg \rightarrow 2l2\nu$.

\paragraph{$ZZ$ production}
The inclusive polarisation fractions are only moderately impacted by the presence of SMEFT coefficients, except for $\Otg$ which shows large deviations from the SM even at low values of the CP-even and CP-odd coefficients. As expected, the transverse polarisation fractions $f_L$ and $f_R$ of the CP-even operators are identical for all the values of the coefficients, while for the CP-odd operators $f_L$ and $f_R$ they become unequal for non-zero values of the coefficients. Only $\Oth$ and $\Otg$ impact the longitudinal polarisation fraction, for large values of the Wilson coefficients. Indeed these two coefficients lead to growing helicity amplitudes in the underlying $gg \rightarrow ZZ$ process when the $Z$ bosons are longitudinally polarised. These effects are either seen at high energies or, in the inclusive setup, for large values of the Wilson coefficients.

\paragraph{$WW$ production}
The longitudinal polarisation fraction $f_0$ is only moderately impacted by the presence of the Wilson coefficients except for the case of $\Otg$, which leads to growing amplitudes in the underlying $gg \rightarrow WW$ process when the $W$ bosons are longitudinally polarised. Whilst similarly to double $Z$ production, $\Oth$ also leads to a growing helicity amplitude when both final state $W$s are longitudinally polarised, the impact of the growth is not as significant as in $gg \rightarrow ZZ$. 
In all cases shown in Fig.~\ref{inclusive2l2v}, $f_L$ and $f_R$ are different. For the CP-even coefficients that do not enter in third-generation box diagrams, this effect is driven by the SM contribution. The transverse polarisation fractions in the presence of the CP-even electroweak dipole $\OtW$ seem to converge for large coefficient values, but we verified that they simply cross such that for larger values of the Wilson coefficients not shown on this plot $f_L$ and $f_R$ become different again. Furthermore, the CP-odd contribution of  $\OtW$ has a much larger impact on the transverse polarisation fractions than $\OtZ$ does in $gg \rightarrow 4l$. The fast differentiation of $f_L$ and $f_R$ in $gg \rightarrow 2l2v$ is indeed due to the combination of the anti-symmetric interference (also present for the CP-violating $\OtZ$) and the fact that the $gg \rightarrow WW$ boxes inherently lead to $f_L \neq f_R$ when the quarks running in the loop do not have the same mass. 

 \subsection{Energy behaviour of the polarisation fractions}

\begin{figure}
\centering

\begin{subfigure}{0.49\columnwidth}
\centering
\includegraphics[width=\textwidth]{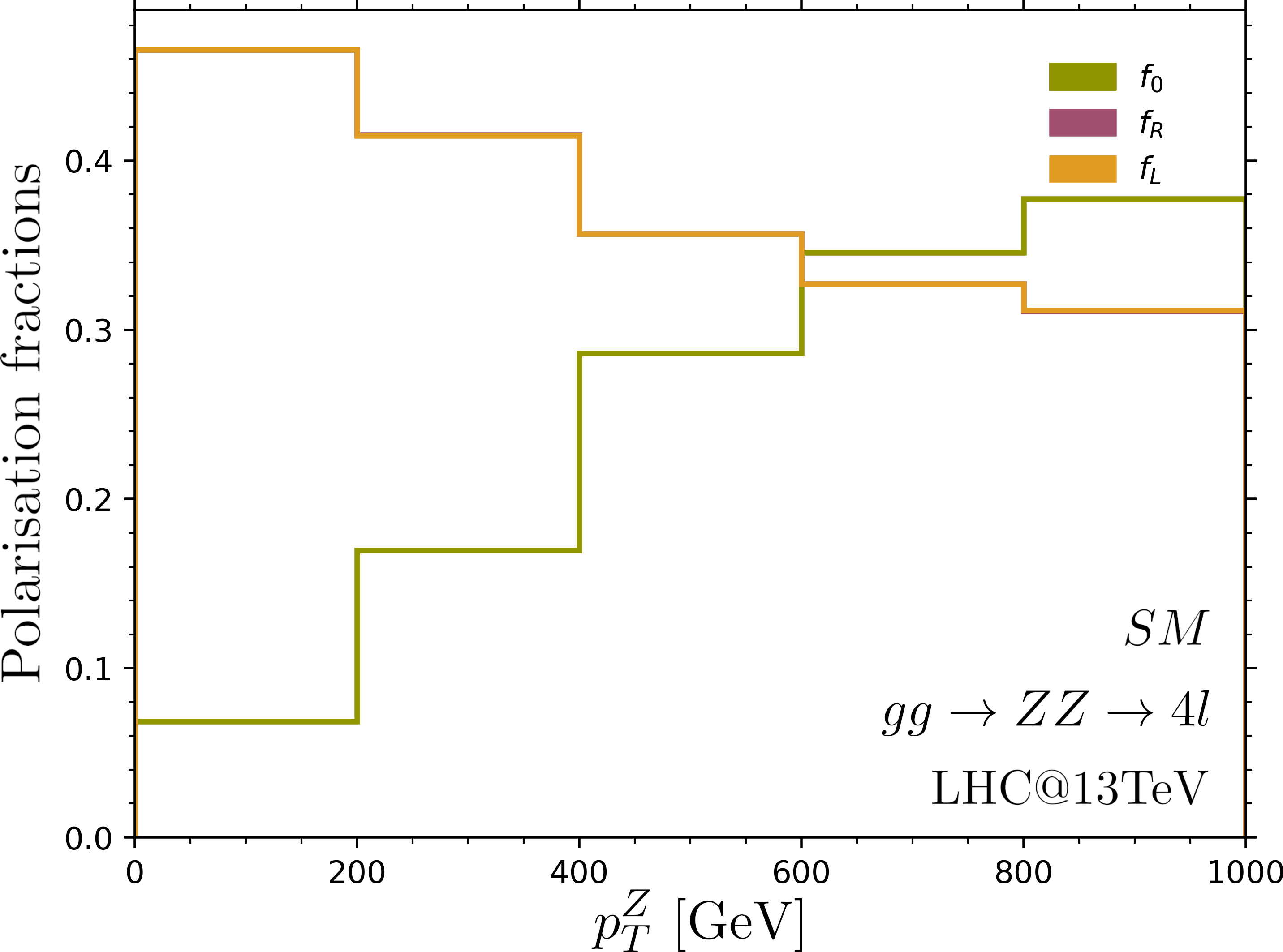}
\caption{}
\label{fig:pTZZ_SM}
\end{subfigure}
\begin{subfigure}{0.49\columnwidth}
\centering
\includegraphics[width=\textwidth]{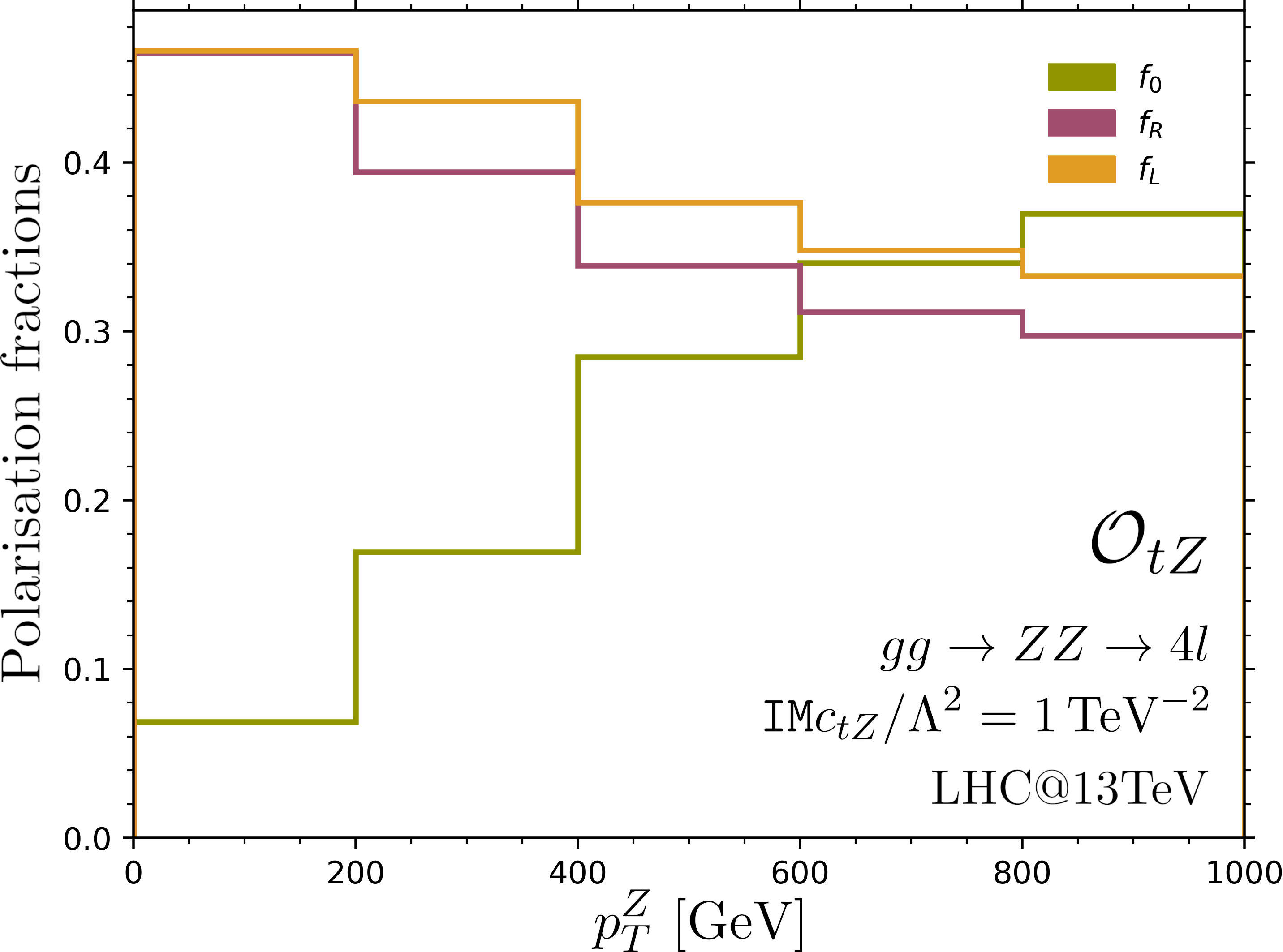}
\caption{}
\label{fig:pTZZ_IMctZ}
\end{subfigure}\hfill

\medskip

 \begin{subfigure}{0.49\columnwidth}
 \centering
 \includegraphics[width=\textwidth]{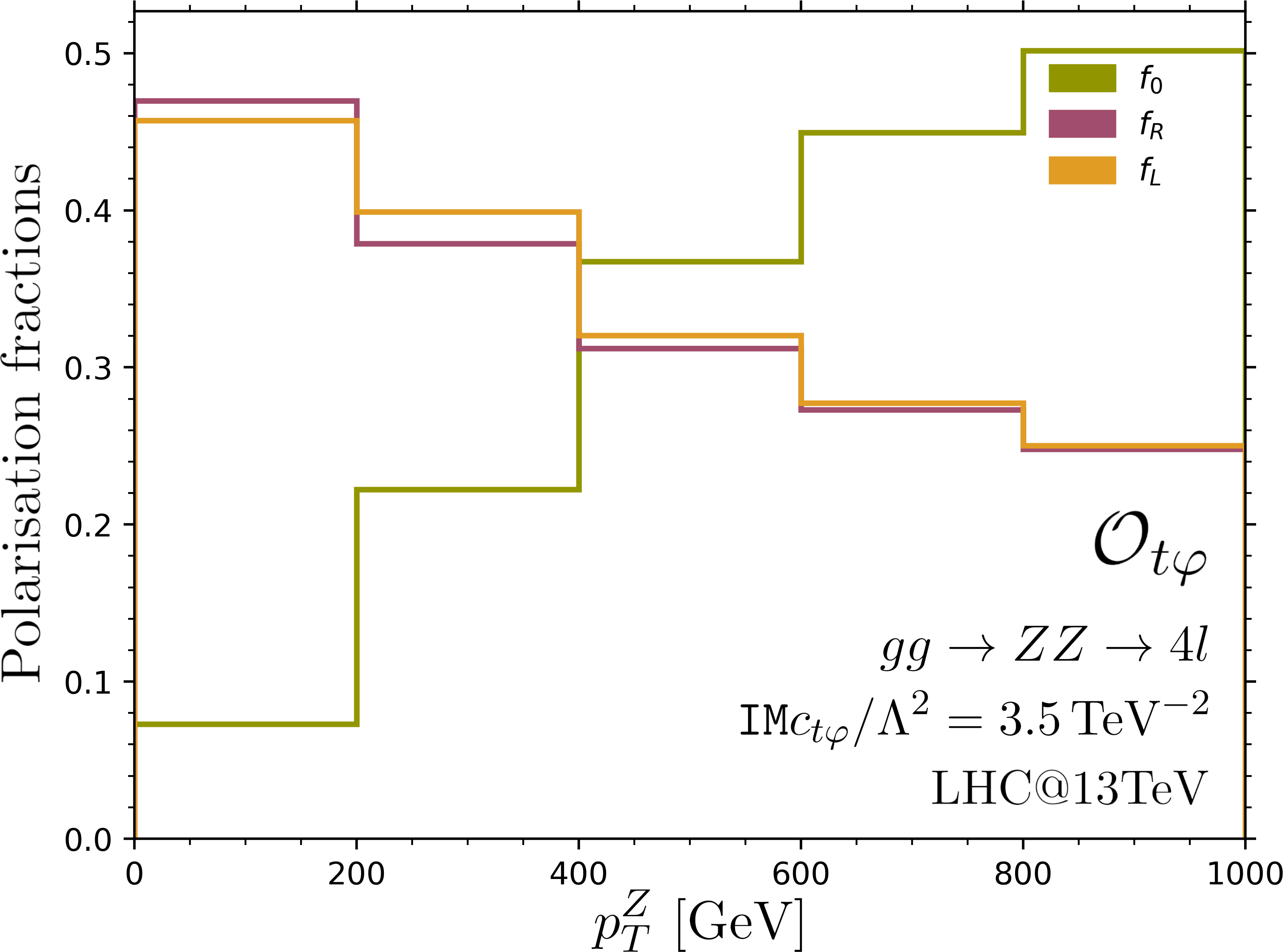}
 \caption{}
 \label{fig:pTZZ_IMctp}
 \end{subfigure}\hfill
 \begin{subfigure}{0.49\columnwidth}
\centering
\includegraphics[width=\textwidth]{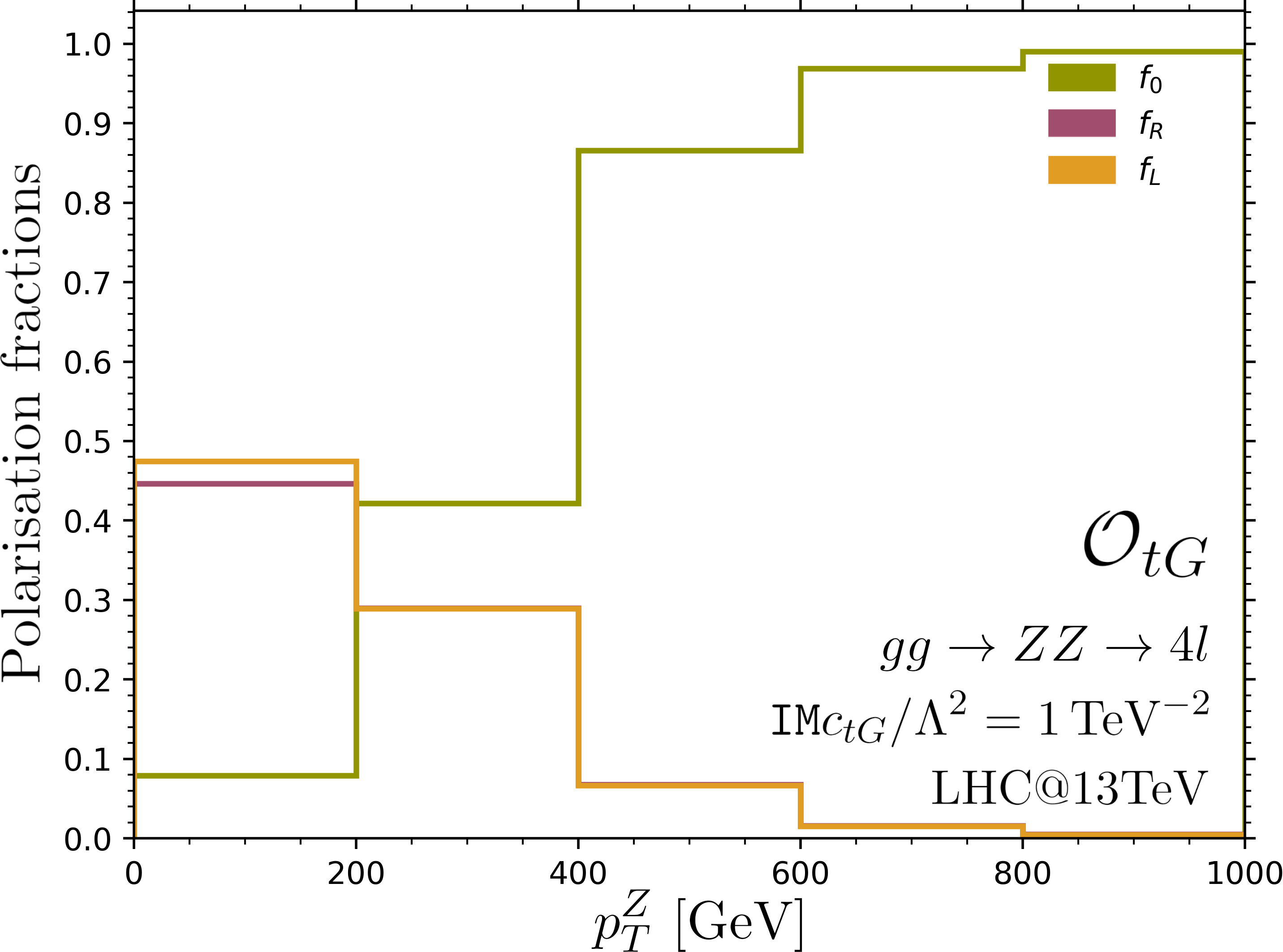}
\caption{}
\label{fig:pTZZ_IMctG}
\end{subfigure}

\caption{Polarisation fractions as a function of $p_T^{Z}$ for $gg\rightarrow 4l$ (a) in the SM and for (b) $\im\ctZ = 1 \, \text{TeV}^{-2}$, (c) $\im\cth = 3.5 \, \text{TeV}^{-2}$ and (d) $\im\ctg = 1 \, \text{TeV}^{-2}$. The Monte Carlo uncertainties, represented by shaded regions, are smaller than the width of the histogram lines and cannot be seen. Due to Bose symmetry and the CP properties of the SM, the $f_L$ and $f_R$ distributions overlap. See the text for more details.}
\label{fig:pTZZ}
\end{figure}

\begin{figure}[htp]
\centering

\begin{subfigure}{0.49\columnwidth}
\centering
\includegraphics[width=\textwidth]{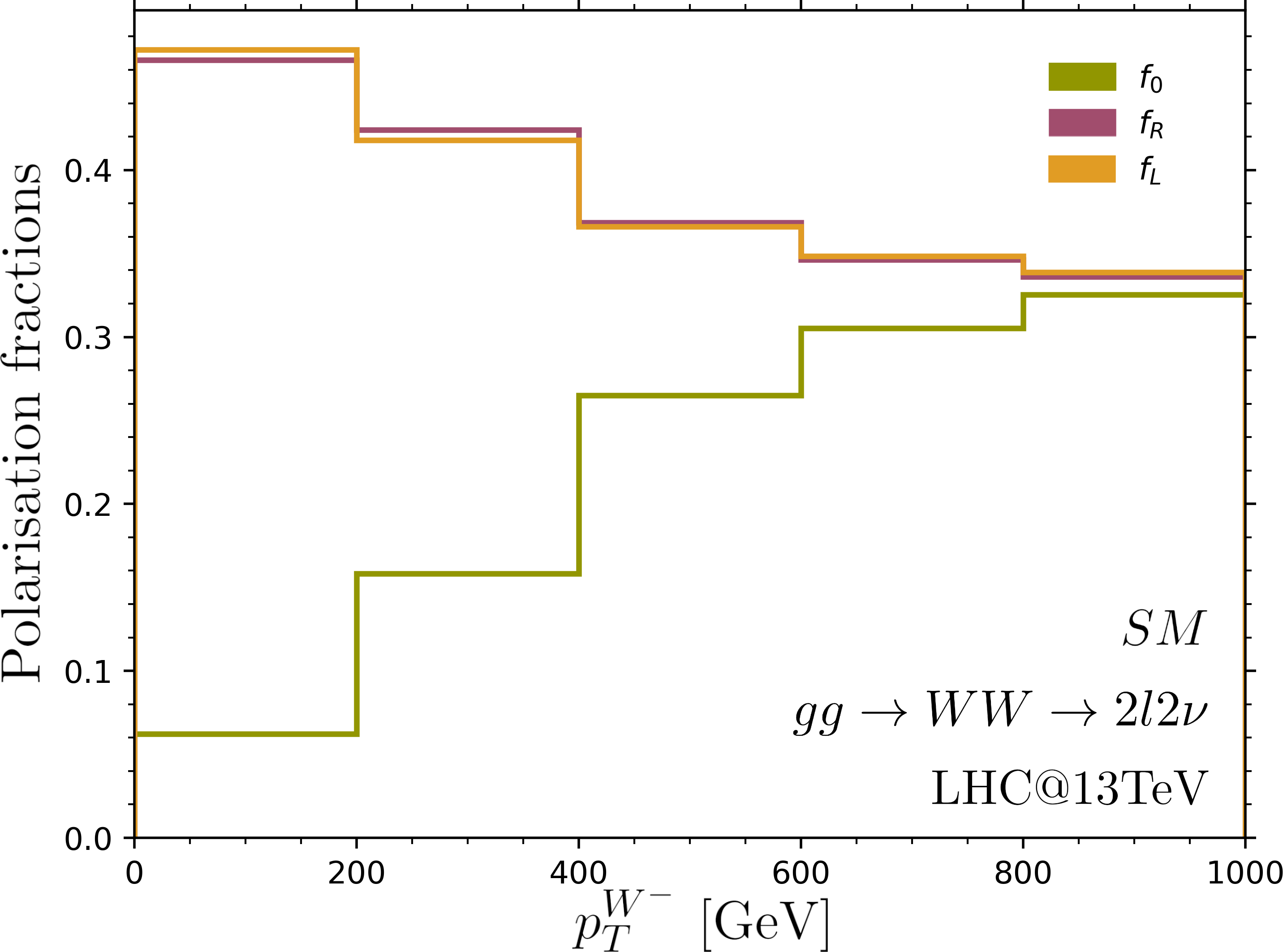}
\caption{}
\label{fig:pTWW_SM}
\end{subfigure}
\begin{subfigure}{0.49\columnwidth}
\centering
\includegraphics[width=\textwidth]{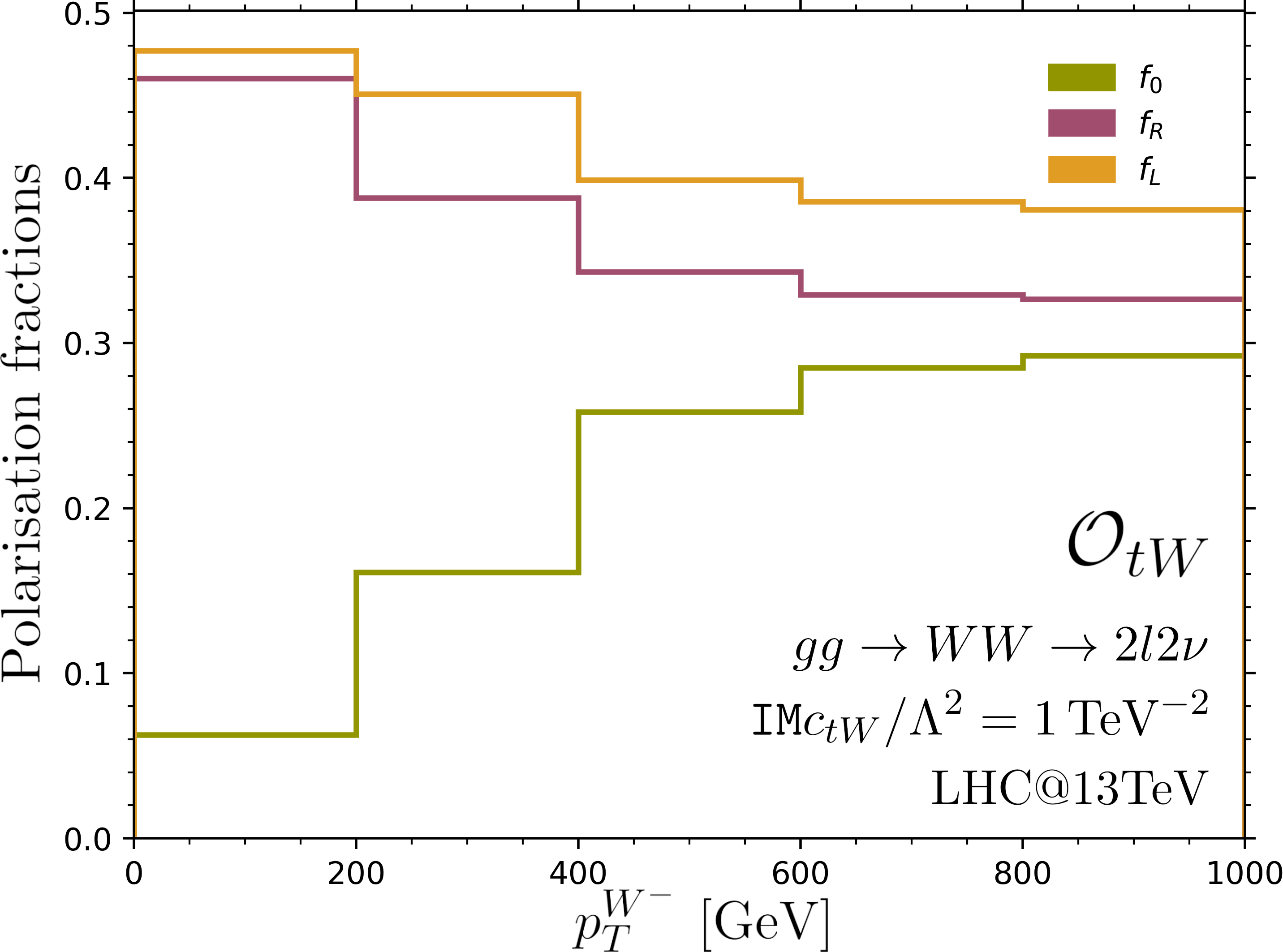}
\caption{}
\label{fig:pTWW_IMctW}
\end{subfigure}\hfill

\medskip

 \begin{subfigure}{0.49\columnwidth}
 \centering
 \includegraphics[width=\textwidth]{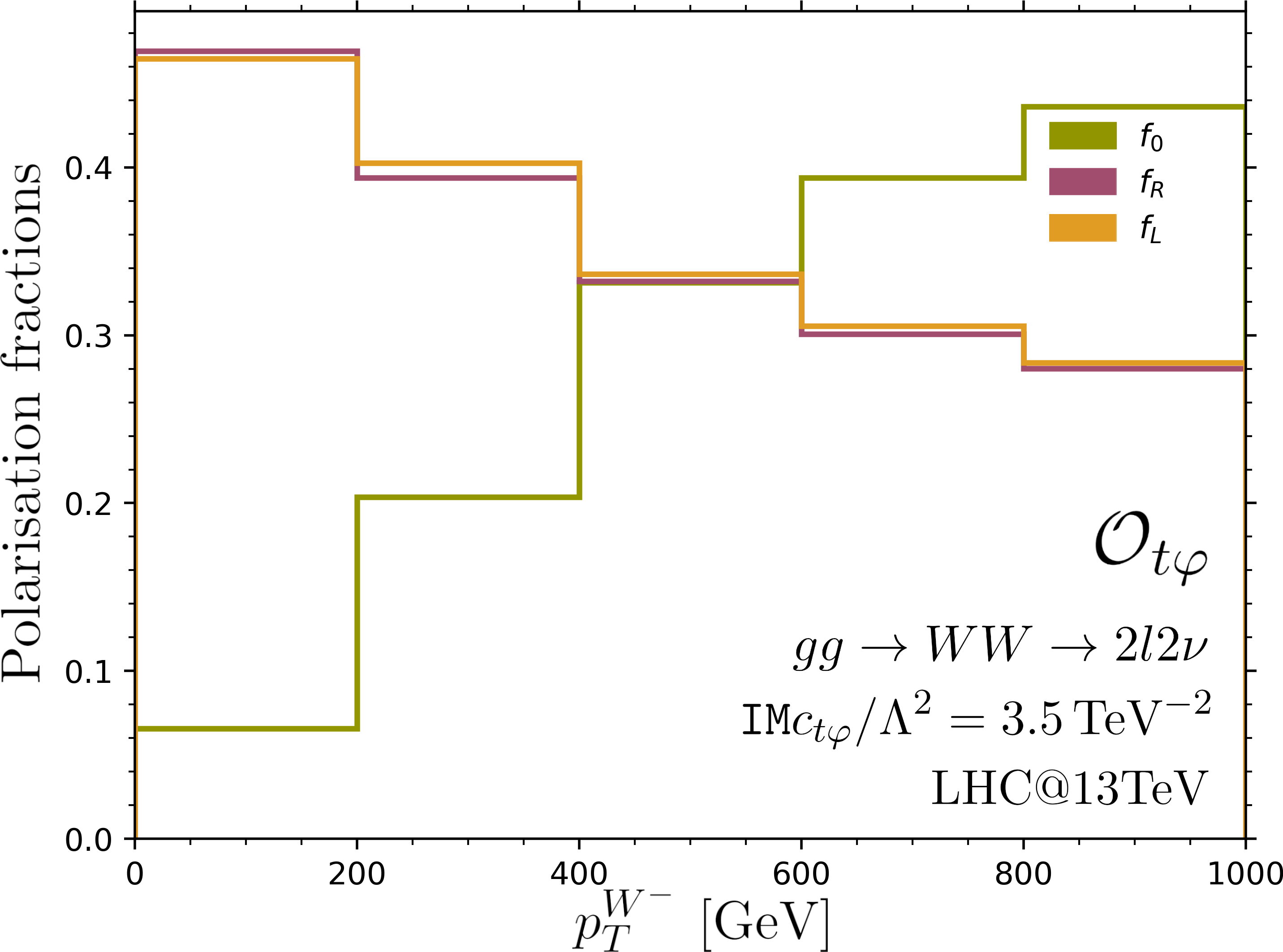}
 \caption{}
 \label{fig:pTWW_IMctp}
 \end{subfigure}\hfill
 \begin{subfigure}{0.49\columnwidth}
\centering
\includegraphics[width=\textwidth]{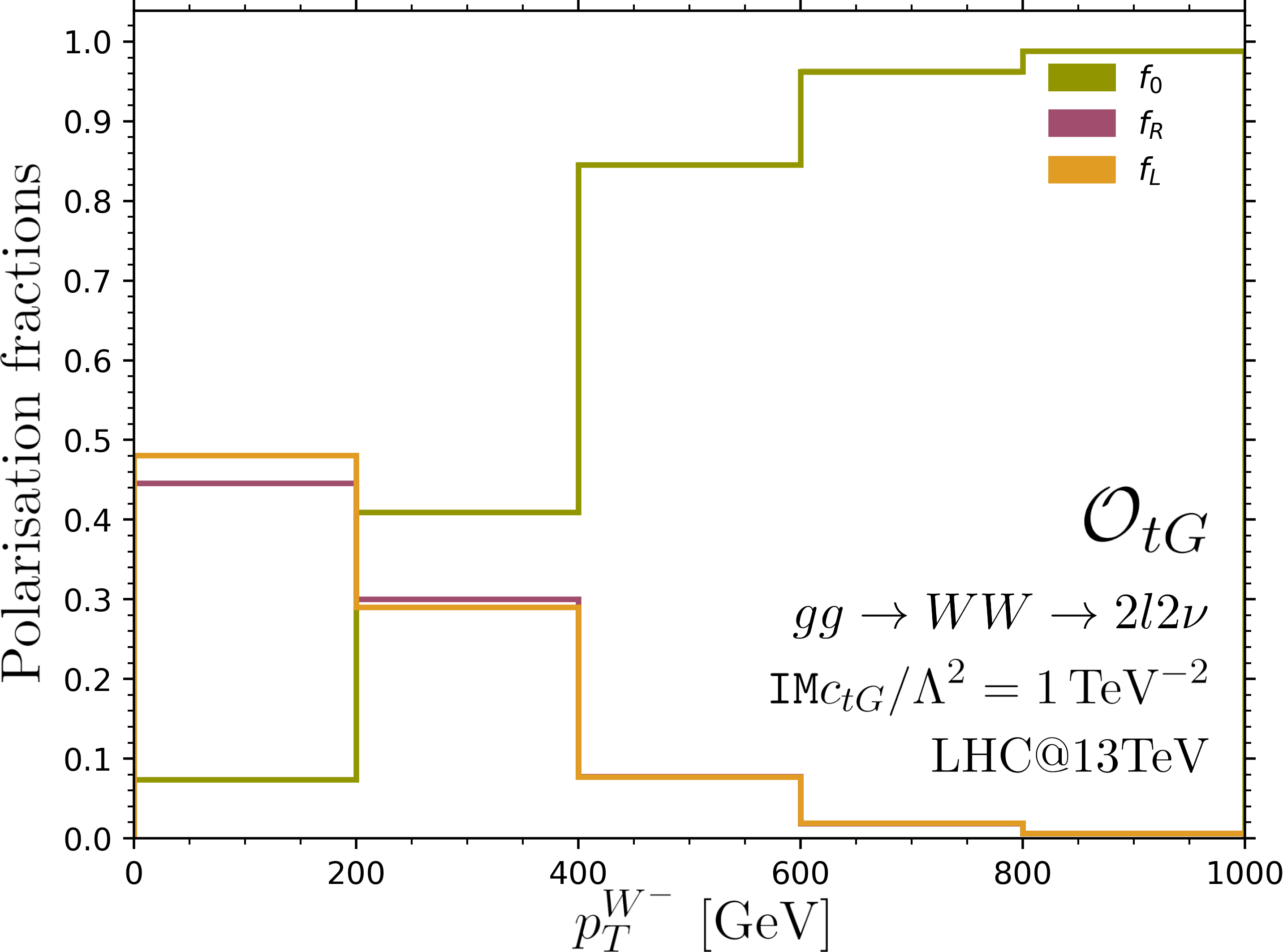}
\caption{}
\label{fig:pTWW_IMctG}
\end{subfigure}

\caption{Polarisation fractions as a function of $p_T^{W^-}$ for $gg\rightarrow 2l2\nu$ (a) in the SM and for (b) $\im\ctW = 2 \, \text{TeV}^{-2}$, (c) $\im\cth = 3.5 \, \text{TeV}^{-2}$, and (d) $\im\ctg = 1 \, \text{TeV}^{-2}$. The Monte Carlo uncertainties are represented by shaded regions but are smaller than the width of the histogram lines.}
\label{fig:pTWW}
\end{figure}

Whilst the effect of the operators on the polarisation fractions is already visible at the inclusive level, we expect more pronounced effects at higher energies where the EFT contributions are expected to be enhanced. To explore these effects we study the behaviour of the polarisation fractions as a function of the transverse momentum of the $Z$ and $W$ bosons and present the results for the SM and a representative sample of the dimension-$6$ operators in Fig.~\ref{fig:pTZZ} for $gg \rightarrow 4l$ and in Fig.~\ref{fig:pTWW} for $gg \rightarrow 2l2\nu$.
We selected the CP-odd coefficients which enter directly in the loop part of the process. We set all the Wilson coefficients, except $\im\cth$, to the value of $1 \,\text{TeV}^{-2}$. The CP-even top Yukawa coefficient is loosely constrained by global fits of collider data, with values as large as $\re\cth= 3.5 \,\text{TeV}^{-2}$ being allowed \cite{2404.12809} and experimental measurements from the LHC set even looser constraints on the CP-odd coefficient \cite{2303.05974}. We thus pick $\im\cth = 3.5 \,\text{TeV}^{-2}$ in the following.  The large values of the CP-odd coefficients chosen here only aim at having a first view at the effects these coefficients can have on the polarisation fractions. For these results we have also verified our polarised cross-sections by considering only $gg \rightarrow ZZ$ and $gg \rightarrow WW$  with the modified truncated-propagator method implemented in \code{Madgraph5\_aMC@NLO} \cite{1912.01725}. 

\paragraph{$ZZ$ production}

We comment first on the Standard Model, shown in Fig.~\ref{fig:pTZZ_SM}. As discussed earlier, Bose symmetry and the CP properties of the SM lead to the transverse polarisation fractions $f_L$ and $f_R$ being equal at all energies. At low transverse momentum the longitudinal mode is highly suppressed, but its contribution grows with energy and even surpasses the transverse polarisation fractions for $p_T > 600$ GeV, which reproduces the results of \cite{2004.02031} \footnote{Ref. \cite{2004.02031} shows the polarisation fractions for $200 < m_{ZZ} < 1000$ GeV, a region in which the longitudinal mode, while growing with energy, is always much smaller than the transverse ones. This energy region corresponds approximately to the events present in our first two $p_T$ bins so $0 < p_T^{Z_1} < 400$ GeV. The events present in our last $p_T$ bin, for which the longitudinal mode is dominant, have $1600<m_{ZZ}<5000$ GeV.}.

Next we consider polarisation fractions in the presence of the CP-odd coefficients. For those, it has been shown in Sec.~\ref{sec:4l_angdist} that $f_L$ and $f_R$ can be different and this is observed in Fig.~\ref{fig:pTZZ}. Starting with the electroweak dipole operator $\OtZ$ shown in Fig.~\ref{fig:pTZZ_IMctZ}, the longitudinal polarisation fraction reproduces the SM behaviour while the transverse ones differ for $p_T > 200$ GeV where they stop overlapping. At higher transverse momenta, $f_L$ and $f_R$ are still suppressed but their values are different, unlike the SM case. For all coefficients the values of the polarisation fractions at low $p_T$ can be compared with the inclusive values found in Sec.~\ref{sec:inclusivepolfra} as the low $p_T$ region dominates the total cross-section. For the $p_T < 200$ GeV region, EFT effects for $c_i = 1 \, \text{TeV}^{-2}$ are often very moderate, motivating differential studies that can focus on high-energy regions where the EFT effects are enhanced.

The CP-odd top Yukawa operator is shown in Fig.~\ref{fig:pTZZ_IMctp}, where at low $p_T$ the longitudinal polarisation fraction is suppressed as in the SM scenario. However for transverse momenta larger that $200$ GeV, $f_0$ grows much faster than in the SM case, and in the last $p_T$ bin this polarisation mode reaches $50 \%$. We commented earlier on the energy behaviour of the helicity amplitudes of  $gg \rightarrow ZZ$ in the presence of $\im \cth$, and in particular we found that only the $(+ + 0 \,0)$ configuration grows with energy, with the exact expression of the leading term given in Eq.~\eqref{eq:ggZZ_IMctp}, in agreement with the $f_0$ polarisation fraction enhancement at high energies.

Finally the polarisation fractions in the presence of the CP-violating top chromoelectric operator are presented in Fig.~\ref{fig:pTZZ_IMctG}. Once again, while at low $p_T$ $f_0$ reproduce the SM behaviour, for $p_T > 200$ GeV the longitudinal polarisation fraction grows faster. However here the suppression of the transverse modes is much faster than in the presence of $\im\cth$, as for $p_T > 200$ GeV the longitudinal polarisation already surpasses the transverse ones. This is because even though the dominant helicity configuration of the $gg \rightarrow ZZ$ process is still $(+ + 0 \,0)$, its amplitude grows quadratically with energy when modified by $\im\ctg$ and hence $f_0$ becomes rapidly dominant.

\paragraph{ $WW$ production}

The Standard Model results for $W$ pair production are shown in Fig.~\ref{fig:pTWW_SM}. As explained in Sec.~\ref{sec:4l_angdist}, the breaking of the left-right polarisation symmetry is small in the SM and hence the $f_L$ and $f_R$ distributions differ slightly at low $p_T$ before overlapping when $p_T > 400$ GeV. At low $p_T$ the transverse polarisation fractions dominate but these contributions decrease with energy as the longitudinal contribution grows. In the highest energy bin $800 < p_T^{W^-} < 1000$ GeV, all three polarisation fractions are similar.

The polarisation fractions in the presence of the CP-violating electroweak dipole $\OtW$ are given in Fig.~\ref{fig:pTWW_IMctW}. This operator only enters in the third-generation boxes such that as expected, the left and right polarisation fractions are not equal at all energies. We observe here that $f_L > f_R$, implying that $\langle\text{cos}\theta^*\rangle > 0$. By comparing to the  distribution in Fig.~\ref{fig:ctheta2l2v_otw}, we conclude that the linear term has a larger impact on the polarisation fractions than the quadratic term for the chosen value of the coefficient. Furthermore, at low energies the longitudinal polarisation fraction is  subdominant but grows with energy while the transverse polarisation fractions decrease. This behaviour changes slightly for $p_T>600$ GeV where the $f_0$ growth slows down. Calculating the helicity amplitudes of the underlying $gg \rightarrow W^+W^-$ process, we find that similarly to the CP-even case \cite{2306.09963}, the only growing amplitudes in the presence of $\im\ctW$ are $A(+ + - +), A(+ + - -)$ and $A(+ - - -)$, that is helicity amplitudes where the $W^-$ is transversely polarised. Hence at high energies transverse helicity configurations dominate.

Next, we present the polarisation fraction distributions in the presence of $\im\cth$ in Fig.~\ref{fig:pTWW_IMctp}. The growth of the longitudinal polarisation is much faster than in the SM case, and $f_0$ is even equal to $f_L$ and $f_R$ in the third $p_T$ bin where $400 < p_T^{W^-} < 600$ GeV. In the presence of $\im\cth$, only the $(++00)$ $gg \rightarrow W^+W^-$  helicity amplitude grows logarithmically %\cite{2306.09963} 
and thus at high energies the longitudinal polarisation dominates. We note however, that similarly to what was observed in the case of the inclusive polarisation fractions in Figs.~\ref{4l_inc_ctp} and \ref{2l2v_inc_ctp}, the impact of the growing $(++00)$ helicity amplitude is less important in the case of $2l2\nu$ production compared to $4l$ production: at high energies $f_0$ modified by $\im\cth$ is larger in $4l$ production than in $2l2\nu$ production even though in both processes $f_0$ is enhanced compared to the SM scenario.

Finally the polarisation fractions in the presence of the CP-odd top chromoelectric operator are shown in Fig.~\ref{fig:pTWW_IMctG}. Similarly to $\im\cth$, in the presence of $\im\ctg$ the longitudinal polarisation fraction grows with energy and surpasses the transverse modes. However, even though $\im\ctg/\Lambda^2$ is set to a smaller value than $\im\cth/\Lambda^2$, the former coefficient leads to a much faster growth of the longitudinal polarisation which dominates the $W^-$ production for $p_T>400$ GeV. This is because in the presence of $\im\ctg$, the total amplitude for $gg \rightarrow W^+W^-$ is completely dominated by the $(++00)$ helicity configuration which grows quadratically with energy, leading to a strong suppression of the transverse polarisations.\\

We conclude this section by emphasising that whilst we have shown that interesting effects arise in the polarisation of the gauge bosons produced in the gluon fusion channel, the size of these effects has to be assessed also in the presence of the quark-initiated channel which dominates the inclusive cross-section for diboson production. The polarisation patterns of the gauge bosons produced by quark-anti-quark annihilation are well known and extensively studied in the literature both within the SM \cite{2006.14867, 2102.13583,2107.06579,2311.05220,2401.17365,2409.06396} but also in the presence of new physics effects \cite{2303.10493,2405.19083,2409.00168}. A more detailed phenomenological study including both channels and estimating the corresponding experimental sensitivity is thus highly desirable.

\section{Conclusions}
\label{sec:Conclusions}

In this paper we have studied the impact of CP-violating dimension-$6$ SMEFT operators entering in diboson production from gluon fusion. We considered CP-odd operators modifying the coupling of the Higgs to the $Z$ and $W$ bosons ($\OpBtil$, $\OpWtil$, $\OpWBtil$), and operators modifying top interactions ($\Otg, \Oth, \OtW$ and $\OtZ$). More specifically we have presented the impact of the operators in double Higgs, double $Z$ and double $W$ production, probing different observables. 

The relevant CP-violating operators have been added to the \code{SMEFTatNLO} UFO along with the necessary rational terms and the UV counterterms. Out of all the processes and operators considered, only the top chromoelectric dipole operator gives non vanishing rational terms, for $ggH$ and  $ggHH$. 
In addition, $\im\ctg$ introduces UV divergent diagrams, which can be renormalised with $\cpgtil$. We verified our implementation by calculating analytically in Mathematica the full amplitude of the $gg \rightarrow HH, WW, ZZ, \gamma Z, \gamma \gamma$ processes in the presence of one operator at a time, and comparing the numerical values obtained from our analytical predictions with the numerical predictions given by the modified version of \code{SMEFTatNLO} in \code{Madgraph5\_aMC@NLO}. 

We then analysed the effects of CP-violating operators in $2 \rightarrow 2$ processes, focusing on $HH, ZZ$ and $WW$ production. The diboson invariant mass distributions of these processes in the presence of the dimension-$6$ operators reveal that the interference of the CP-odd operators with the SM always vanishes. Regarding the quadratic distributions, the CP-even operators and their CP-odd counterparts behave similarly in all cases, often displaying overlapping distributions. This behaviour can be understood from the underlying helicity amplitudes: for each pair of CP-even operator and its CP-odd counterpart, the helicity configurations which dominate the total amplitudes either coincide up to a phase, or converge in the high energy limit. 

Having observed that kinematic observables in $2 \rightarrow 2$ processes do not allow the distinction between CP-even and CP-odd contributions, we extended the analysis to a more realistic setup by exploring the potential of angular and polarisation observables in leptonic decays of the $Z$ and $W$ bosons.  We found that even though the inclusive interference cross-section of the CP-odd coefficients is still zero,  $\theta^*$, the opening angle between the three-momentum of the charged lepton in the $V$ rest frame and the $V$ three-momentum in the center-of-mass frame of the $VV$ pair proves to be a sensitive observable. Indeed the $\mathcal{O}(\Lambda^{-2})$ distributions of the CP-odd coefficients are anti-symmetric around $\text{cos}\theta^* = 0$ as expected from the CP properties of the underlying amplitudes. 
All the other distributions, that is the SM, the interference of the CP-even coefficients, and the squared distributions of all coefficients, are symmetric around $\text{cos}\theta^* =0$.

The polarisation fractions of the gauge bosons are directly related to these angular distributions, and can thus also be modified in the presence of CP-even and CP-odd SMEFT operators. We have provided numerical results to parametrise the modification of the polarisation fractions as a function of the Wilson coefficients. As a result of the asymmetric contributions of the CP-odd interference to the angular distribution we have found that CP-violating interactions lead to a left-right asymmetry in both the $ZZ$ and $WW$ processes. We have also observed that in $ZZ$ production the left and right polarisation fractions are identical for the SM and the CP-even operators, a consequence of the Bose symmetry of the final state combined with the CP properties of the amplitudes. This is not the case in $WW$ production due to the presence of box diagrams with propagators of different masses in the loop. In the SM this leads to a small asymmetry between left and right which becomes more enhanced in the presence of the dipole operator coefficients $\re\ctW$ and $\im\ctW$.

Finally we showed how the polarisation fractions change as a function of the gauge boson transverse momentum. We found that the polarisation fractions can be significantly modified in the presence of the SMEFT coefficients, in particular in the high-energy region.  For both processes the values of the longitudinal and transverse polarisation fractions in the presence of SMEFT operators differ from the SM values, and this can be directly related to the high energy behaviour of the helicity amplitudes of the underlying $gg \rightarrow ZZ$ and $gg \rightarrow WW$ processes.

Our study constitutes an important step towards a more comprehensive approach to CP-violation in loop-induced processes. Several research directions remain open and deserve further exploration. The inclusion of the gluon fusion processes presented here in a full phenomenological analysis considering also the quark-initiated channel is an obvious next step. A more detailed phenomenological analysis employing LHC Run 3 and HL-LHC projections could then establish the prospects of employing this class of processes and the corresponding differential observables to extract information on CP-violating interactions. In particular the prospects of experimentally extracting the polarisation fractions using LHC data need to be established and their potential sensitivity to new physics effects can be compared to those extracted from other observables. Additionally, the study of angular observables beyond the one considered here can be envisioned, also in the light of recent efforts to explore the full spin density in diboson production.

\section*{Acknowledgments}
M.T. thanks H. El Faham, V. Miralles, A. Rossia, C. Severi and G. Ventura for the insightful conversations. M. T. and E.V. acknowledge useful discussions with H.S. Shao and Giovanni Pelliccioli. The work of  M.T. and E.V. is supported by a Royal Society University Research Fellowship through grant URF/R1/201553. E. V. is supported by the European Research Council (ERC) under the European Union’s Horizon 2020 research and innovation programme (Grant agreement No. 949451).

\bibliographystyle{JHEP.bst}
\bibliography{main.bib}

\providecommand{\href}[2]{#2}\begingroup\raggedright\begin{thebibliography}{10}

\bibitem{1706.08945}
I.~Brivio and M.~Trott, \emph{{The Standard Model as an Effective Field
  Theory}}, \href{https://doi.org/10.1016/j.physrep.2018.11.002}{\emph{Phys.
  Rept.} {\bfseries 793} (2019) 1}
  [\href{https://arxiv.org/abs/1706.08945}{{\ttfamily 1706.08945}}].

\bibitem{1608.00977}
A.~Azatov, C.~Grojean, A.~Paul and E.~Salvioni, \emph{{Resolving gluon fusion
  loops at current and future hadron colliders}},
  \href{https://doi.org/10.1007/JHEP09(2016)123}{\emph{JHEP} {\bfseries 09}
  (2016) 123} [\href{https://arxiv.org/abs/1608.00977}{{\ttfamily
  1608.00977}}].

\bibitem{2004.02031}
Q.-H.~Cao, B.~Yan, C.P.~Yuan and Y.~Zhang, \emph{{Probing $Zt\bar{t}$ couplings
  using $Z$ boson polarization in $ZZ$ production at hadron colliders}},
  \href{https://doi.org/10.1103/PhysRevD.102.055010}{\emph{Phys. Rev. D}
  {\bfseries 102} (2020) 055010}
  [\href{https://arxiv.org/abs/2004.02031}{{\ttfamily 2004.02031}}].

\bibitem{2306.09963}
A.~Rossia, M.~Thomas and E.~Vryonidou, \emph{{Diboson production in the SMEFT
  from gluon fusion}},
  \href{https://doi.org/10.1007/JHEP11(2023)132}{\emph{JHEP} {\bfseries 11}
  (2023) 132} [\href{https://arxiv.org/abs/2306.09963}{{\ttfamily
  2306.09963}}].

\bibitem{1612.01808}
F.~Ferreira, B.~Fuks, V.~Sanz and D.~Sengupta, \emph{{Probing ${CP}$-violating
  Higgs and gauge-boson couplings in the Standard Model effective field
  theory}}, \href{https://doi.org/10.1140/epjc/s10052-017-5226-6}{\emph{Eur.
  Phys. J. C} {\bfseries 77} (2017) 675}
  [\href{https://arxiv.org/abs/1612.01808}{{\ttfamily 1612.01808}}].

\bibitem{1804.01477}
M.~Chiesa, A.~Denner and J.-N.~Lang, \emph{{Anomalous triple-gauge-boson
  interactions in vector-boson pair production with RECOLA2}},
  \href{https://doi.org/10.1140/epjc/s10052-018-5949-z}{\emph{Eur. Phys. J. C}
  {\bfseries 78} (2018) 467}
  [\href{https://arxiv.org/abs/1804.01477}{{\ttfamily 1804.01477}}].

\bibitem{1810.11657}
R.~Rahaman and R.K.~Singh, \emph{{Anomalous triple gauge boson couplings in
  $ZZ$ production at the LHC and the role of $Z$ boson polarizations}},
  \href{https://doi.org/10.1016/j.nuclphysb.2019.114754}{\emph{Nucl. Phys. B}
  {\bfseries 948} (2019) 114754}
  [\href{https://arxiv.org/abs/1810.11657}{{\ttfamily 1810.11657}}].

\bibitem{2009.13394}
S.~Das~Bakshi, J.~Chakrabortty, C.~Englert, M.~Spannowsky and P.~Stylianou,
  \emph{{$CP$ violation at ATLAS in effective field theory}},
  \href{https://doi.org/10.1103/PhysRevD.103.055008}{\emph{Phys. Rev. D}
  {\bfseries 103} (2021) 055008}
  [\href{https://arxiv.org/abs/2009.13394}{{\ttfamily 2009.13394}}].

\bibitem{2102.01115}
A.~Biek\"otter, P.~Gregg, F.~Krauss and M.~Sch\"onherr, \emph{{Constraining CP
  violating operators in charged and neutral triple gauge couplings}},
  \href{https://doi.org/10.1016/j.physletb.2021.136311}{\emph{Phys. Lett. B}
  {\bfseries 817} (2021) 136311}
  [\href{https://arxiv.org/abs/2102.01115}{{\ttfamily 2102.01115}}].

\bibitem{1503.01114}
C.-Y.~Chen, S.~Dawson and Y.~Zhang, \emph{{Complementarity of LHC and EDMs for
  Exploring Higgs CP Violation}},
  \href{https://doi.org/10.1007/JHEP06(2015)056}{\emph{JHEP} {\bfseries 06}
  (2015) 056} [\href{https://arxiv.org/abs/1503.01114}{{\ttfamily
  1503.01114}}].

\bibitem{1510.00725}
Y.T.~Chien, V.~Cirigliano, W.~Dekens, J.~de~Vries and E.~Mereghetti,
  \emph{{Direct and indirect constraints on CP-violating Higgs-quark and
  Higgs-gluon interactions}},
  \href{https://doi.org/10.1007/JHEP02(2016)011}{\emph{JHEP} {\bfseries 02}
  (2016) 011} [\href{https://arxiv.org/abs/1510.00725}{{\ttfamily
  1510.00725}}].

\bibitem{1603.03049}
V.~Cirigliano, W.~Dekens, J.~de~Vries and E.~Mereghetti, \emph{{Is there room
  for CP violation in the top-Higgs sector?}},
  \href{https://doi.org/10.1103/PhysRevD.94.016002}{\emph{Phys. Rev. D}
  {\bfseries 94} (2016) 016002}
  [\href{https://arxiv.org/abs/1603.03049}{{\ttfamily 1603.03049}}].

\bibitem{1903.03625}
V.~Cirigliano, A.~Crivellin, W.~Dekens, J.~de~Vries, M.~Hoferichter and
  E.~Mereghetti, \emph{{CP Violation in Higgs-Gauge Interactions: From Tabletop
  Experiments to the LHC}},
  \href{https://doi.org/10.1103/PhysRevLett.123.051801}{\emph{Phys. Rev. Lett.}
  {\bfseries 123} (2019) 051801}
  [\href{https://arxiv.org/abs/1903.03625}{{\ttfamily 1903.03625}}].

\bibitem{2109.15085}
J.~Kley, T.~Theil, E.~Venturini and A.~Weiler, \emph{{Electric dipole moments
  at one-loop in the dimension-6 SMEFT}},
  \href{https://doi.org/10.1140/epjc/s10052-022-10861-5}{\emph{Eur. Phys. J. C}
  {\bfseries 82} (2022) 926}
  [\href{https://arxiv.org/abs/2109.15085}{{\ttfamily 2109.15085}}].

\bibitem{2008.11743}
C.~Degrande, G.~Durieux, F.~Maltoni, K.~Mimasu, E.~Vryonidou and C.~Zhang,
  \emph{{Automated one-loop computations in the standard model effective field
  theory}}, \href{https://doi.org/10.1103/PhysRevD.103.096024}{\emph{Phys. Rev.
  D} {\bfseries 103} (2021) 096024}
  [\href{https://arxiv.org/abs/2008.11743}{{\ttfamily 2008.11743}}].

\bibitem{2012.11343}
I.~Brivio, \emph{{SMEFTsim 3.0 \textemdash{} a practical guide}},
  \href{https://doi.org/10.1007/JHEP04(2021)073}{\emph{JHEP} {\bfseries 04}
  (2021) 073} [\href{https://arxiv.org/abs/2012.11343}{{\ttfamily
  2012.11343}}].

\bibitem{1901.04821}
A.~Azatov, D.~Barducci and E.~Venturini, \emph{{Precision diboson measurements
  at hadron colliders}},
  \href{https://doi.org/10.1007/JHEP04(2019)075}{\emph{JHEP} {\bfseries 04}
  (2019) 075} [\href{https://arxiv.org/abs/1901.04821}{{\ttfamily
  1901.04821}}].

\bibitem{2405.19083}
H.~El~Faham, G.~Pelliccioli and E.~Vryonidou, \emph{{Triple-gauge couplings in
  LHC diboson production: a SMEFT view from every angle}},
  \href{https://doi.org/10.1007/JHEP08(2024)087}{\emph{JHEP} {\bfseries 08}
  (2024) 087} [\href{https://arxiv.org/abs/2405.19083}{{\ttfamily
  2405.19083}}].

\bibitem{1306.6464}
P.~Artoisenet et~al., \emph{{A framework for Higgs characterisation}},
  \href{https://doi.org/10.1007/JHEP11(2013)043}{\emph{JHEP} {\bfseries 11}
  (2013) 043} [\href{https://arxiv.org/abs/1306.6464}{{\ttfamily 1306.6464}}].

\bibitem{1407.5089}
F.~Demartin, F.~Maltoni, K.~Mawatari, B.~Page and M.~Zaro, \emph{{Higgs
  characterisation at NLO in QCD: CP properties of the top-quark Yukawa
  interaction}},
  \href{https://doi.org/10.1140/epjc/s10052-014-3065-2}{\emph{Eur. Phys. J. C}
  {\bfseries 74} (2014) 3065}
  [\href{https://arxiv.org/abs/1407.5089}{{\ttfamily 1407.5089}}].

\bibitem{1705.05314}
R.~Grober, M.~Muhlleitner and M.~Spira, \emph{{Higgs Pair Production at NLO QCD
  for CP-violating Higgs Sectors}},
  \href{https://doi.org/10.1016/j.nuclphysb.2017.10.002}{\emph{Nucl. Phys. B}
  {\bfseries 925} (2017) 1} [\href{https://arxiv.org/abs/1705.05314}{{\ttfamily
  1705.05314}}].

\bibitem{Degrande:2020evl}
C.~Degrande, G.~Durieux, F.~Maltoni, K.~Mimasu, E.~Vryonidou and C.~Zhang,
  \emph{{Automated one-loop computations in the standard model effective field
  theory}}, \href{https://doi.org/10.1103/PhysRevD.103.096024}{\emph{Phys. Rev.
  D} {\bfseries 103} (2021) 096024}
  [\href{https://arxiv.org/abs/2008.11743}{{\ttfamily 2008.11743}}].

\bibitem{0609007}
G.~Ossola, C.G.~Papadopoulos and R.~Pittau, \emph{{Reducing full one-loop
  amplitudes to scalar integrals at the integrand level}},
  \href{https://doi.org/10.1016/j.nuclphysb.2006.11.012}{\emph{Nucl. Phys. B}
  {\bfseries 763} (2007) 147}
  [\href{https://arxiv.org/abs/hep-ph/0609007}{{\ttfamily hep-ph/0609007}}].

\bibitem{1103.0621}
V.~Hirschi, R.~Frederix, S.~Frixione, M.V.~Garzelli, F.~Maltoni and R.~Pittau,
  \emph{{Automation of one-loop QCD corrections}},
  \href{https://doi.org/10.1007/JHEP05(2011)044}{\emph{JHEP} {\bfseries 05}
  (2011) 044} [\href{https://arxiv.org/abs/1103.0621}{{\ttfamily 1103.0621}}].

\bibitem{0802.1876}
G.~Ossola, C.G.~Papadopoulos and R.~Pittau, \emph{{On the Rational Terms of the
  one-loop amplitudes}},
  \href{https://doi.org/10.1088/1126-6708/2008/05/004}{\emph{JHEP} {\bfseries
  05} (2008) 004} [\href{https://arxiv.org/abs/0802.1876}{{\ttfamily
  0802.1876}}].

\bibitem{0903.0356}
P.~Draggiotis, M.V.~Garzelli, C.G.~Papadopoulos and R.~Pittau, \emph{{Feynman
  Rules for the Rational Part of the QCD 1-loop amplitudes}},
  \href{https://doi.org/10.1088/1126-6708/2009/04/072}{\emph{JHEP} {\bfseries
  04} (2009) 072} [\href{https://arxiv.org/abs/0903.0356}{{\ttfamily
  0903.0356}}].

\bibitem{0910.3130}
M.V.~Garzelli, I.~Malamos and R.~Pittau, \emph{{Feynman rules for the rational
  part of the Electroweak 1-loop amplitudes}},
  \href{https://doi.org/10.1007/JHEP10(2010)097}{\emph{JHEP} {\bfseries 01}
  (2010) 040} [\href{https://arxiv.org/abs/0910.3130}{{\ttfamily 0910.3130}}].

\bibitem{1106.5483}
H.-S.~Shao, Y.-J.~Zhang and K.-T.~Chao, \emph{{Dijet Invariant Mass
  Distribution in Top Quark Hadronic Decay with QCD Corrections}},
  \href{https://doi.org/10.1103/PhysRevD.84.094021}{\emph{Phys. Rev. D}
  {\bfseries 84} (2011) 094021}
  [\href{https://arxiv.org/abs/1106.5483}{{\ttfamily 1106.5483}}].

\bibitem{Kreimer:1989ke}
D.~Kreimer, \emph{{The $\gamma$(5) Problem and Anomalies: A Clifford Algebra
  Approach}}, \href{https://doi.org/10.1016/0370-2693(90)90461-E}{\emph{Phys.
  Lett. B} {\bfseries 237} (1990) 59}.

\bibitem{Korner:1991sx}
J.G.~Korner, D.~Kreimer and K.~Schilcher, \emph{{A Practicable gamma(5) scheme
  in dimensional regularization}},
  \href{https://doi.org/10.1007/BF01559471}{\emph{Z. Phys. C} {\bfseries 54}
  (1992) 503}.

\bibitem{9401354}
D.~Kreimer, \emph{{The Role of gamma(5) in dimensional regularization}},
  \href{https://arxiv.org/abs/hep-ph/9401354}{{\ttfamily hep-ph/9401354}}.

\bibitem{tHooft:1972tcz}
G.~'t~Hooft and M.J.G.~Veltman, \emph{{Regularization and Renormalization of
  Gauge Fields}},
  \href{https://doi.org/10.1016/0550-3213(72)90279-9}{\emph{Nucl. Phys. B}
  {\bfseries 44} (1972) 189}.

\bibitem{Breitenlohner:1976te}
P.~Breitenlohner and D.~Maison, \emph{{Dimensionally Renormalized Green's
  Functions for Theories with Massless Particles. 2.}},
  \href{https://doi.org/10.1007/BF01609071}{\emph{Commun. Math. Phys.}
  {\bfseries 52} (1977) 55}.

\bibitem{Breitenlohner:1977hr}
P.~Breitenlohner and D.~Maison, \emph{{Dimensional Renormalization and the
  Action Principle}}, \href{https://doi.org/10.1007/BF01609069}{\emph{Commun.
  Math. Phys.} {\bfseries 52} (1977) 11}.

\bibitem{2111.11449}
E.~Mereghetti, C.J.~Monahan, M.D.~Rizik, A.~Shindler and P.~Stoffer,
  \emph{{One-loop matching for quark dipole operators in a gradient-flow
  scheme}}, \href{https://doi.org/10.1007/JHEP04(2022)050}{\emph{JHEP}
  {\bfseries 04} (2022) 050}
  [\href{https://arxiv.org/abs/2111.11449}{{\ttfamily 2111.11449}}].

\bibitem{2211.01379}
J.~Aebischer, M.~Pesut and Z.~Polonsky, \emph{{Dipole operators in Fierz
  identities}},
  \href{https://doi.org/10.1016/j.physletb.2023.137968}{\emph{Phys. Lett. B}
  {\bfseries 842} (2023) 137968}
  [\href{https://arxiv.org/abs/2211.01379}{{\ttfamily 2211.01379}}].

\bibitem{2304.00985}
J.~B\"uhler and P.~Stoffer, \emph{{One-loop matching of CP-odd four-quark
  operators to the gradient-flow scheme}},
  \href{https://doi.org/10.1007/JHEP08(2023)194}{\emph{JHEP} {\bfseries 08}
  (2023) 194} [\href{https://arxiv.org/abs/2304.00985}{{\ttfamily
  2304.00985}}].

\bibitem{2012.08506}
J.~Fuentes-Martin, M.~K\"onig, J.~Pag\`es, A.E.~Thomsen and F.~Wilsch,
  \emph{{SuperTracer: A Calculator of Functional Supertraces for One-Loop EFT
  Matching}}, \href{https://doi.org/10.1007/JHEP04(2021)281}{\emph{JHEP}
  {\bfseries 04} (2021) 281}
  [\href{https://arxiv.org/abs/2012.08506}{{\ttfamily 2012.08506}}].

\bibitem{2211.09144}
J.~Fuentes-Mart\'\i{}n, M.~K\"onig, J.~Pag\`es, A.E.~Thomsen and F.~Wilsch,
  \emph{{Evanescent operators in one-loop matching computations}},
  \href{https://doi.org/10.1007/JHEP02(2023)031}{\emph{JHEP} {\bfseries 02}
  (2023) 031} [\href{https://arxiv.org/abs/2211.09144}{{\ttfamily
  2211.09144}}].

\bibitem{1108.2040}
C.~Degrande, C.~Duhr, B.~Fuks, D.~Grellscheid, O.~Mattelaer and T.~Reiter,
  \emph{{UFO - The Universal FeynRules Output}},
  \href{https://doi.org/10.1016/j.cpc.2012.01.022}{\emph{Comput. Phys. Commun.}
  {\bfseries 183} (2012) 1201}
  [\href{https://arxiv.org/abs/1108.2040}{{\ttfamily 1108.2040}}].

\bibitem{Mertig:1990an}
R.~Mertig, M.~Bohm and A.~Denner, \emph{{FEYN CALC: Computer algebraic
  calculation of Feynman amplitudes}},
  \href{https://doi.org/10.1016/0010-4655(91)90130-D}{\emph{Comput. Phys.
  Commun.} {\bfseries 64} (1991) 345}.

\bibitem{1601.01167}
V.~Shtabovenko, R.~Mertig and F.~Orellana, \emph{{New Developments in FeynCalc
  9.0}}, \href{https://doi.org/10.1016/j.cpc.2016.06.008}{\emph{Comput. Phys.
  Commun.} {\bfseries 207} (2016) 432}
  [\href{https://arxiv.org/abs/1601.01167}{{\ttfamily 1601.01167}}].

\bibitem{2001.04407}
V.~Shtabovenko, R.~Mertig and F.~Orellana, \emph{{FeynCalc 9.3: New features
  and improvements}},
  \href{https://doi.org/10.1016/j.cpc.2020.107478}{\emph{Comput. Phys. Commun.}
  {\bfseries 256} (2020) 107478}
  [\href{https://arxiv.org/abs/2001.04407}{{\ttfamily 2001.04407}}].

\bibitem{1611.06793}
V.~Shtabovenko, \emph{{FeynHelpers: Connecting FeynCalc to FIRE and
  Package-X}}, \href{https://doi.org/10.1016/j.cpc.2017.04.014}{\emph{Comput.
  Phys. Commun.} {\bfseries 218} (2017) 48}
  [\href{https://arxiv.org/abs/1611.06793}{{\ttfamily 1611.06793}}].

\bibitem{1503.01469}
H.H.~Patel, \emph{{Package-X: A Mathematica package for the analytic
  calculation of one-loop integrals}},
  \href{https://doi.org/10.1016/j.cpc.2015.08.017}{\emph{Comput. Phys. Commun.}
  {\bfseries 197} (2015) 276}
  [\href{https://arxiv.org/abs/1503.01469}{{\ttfamily 1503.01469}}].

\bibitem{10.1007/BF01621031}
G.J.~van Oldenborgh and J.A.M.~Vermaseren, \emph{{New Algorithms for One Loop
  Integrals}}, \href{https://doi.org/10.1007/BF01621031}{\emph{Z. Phys. C}
  {\bfseries 46} (1990) 425}.

\bibitem{9807565}
T.~Hahn and M.~Perez-Victoria, \emph{{Automatized one loop calculations in
  four-dimensions and D-dimensions}},
  \href{https://doi.org/10.1016/S0010-4655(98)00173-8}{\emph{Comput. Phys.
  Commun.} {\bfseries 118} (1999) 153}
  [\href{https://arxiv.org/abs/hep-ph/9807565}{{\ttfamily hep-ph/9807565}}].

\bibitem{hep-ph/0012260}
T.~Hahn, \emph{{Generating Feynman diagrams and amplitudes with FeynArts 3}},
  \href{https://doi.org/10.1016/S0010-4655(01)00290-9}{\emph{Comput. Phys.
  Commun.} {\bfseries 140} (2001) 418}
  [\href{https://arxiv.org/abs/hep-ph/0012260}{{\ttfamily hep-ph/0012260}}].

\bibitem{1308.2627}
E.E.~Jenkins, A.V.~Manohar and M.~Trott, \emph{{Renormalization Group Evolution
  of the Standard Model Dimension Six Operators I: Formalism and lambda
  Dependence}}, \href{https://doi.org/10.1007/JHEP10(2013)087}{\emph{JHEP}
  {\bfseries 10} (2013) 087} [\href{https://arxiv.org/abs/1308.2627}{{\ttfamily
  1308.2627}}].

\bibitem{1310.4838}
E.E.~Jenkins, A.V.~Manohar and M.~Trott, \emph{{Renormalization Group Evolution
  of the Standard Model Dimension Six Operators II: Yukawa Dependence}},
  \href{https://doi.org/10.1007/JHEP01(2014)035}{\emph{JHEP} {\bfseries 01}
  (2014) 035} [\href{https://arxiv.org/abs/1310.4838}{{\ttfamily 1310.4838}}].

\bibitem{1312.2014}
R.~Alonso, E.E.~Jenkins, A.V.~Manohar and M.~Trott, \emph{{Renormalization
  Group Evolution of the Standard Model Dimension Six Operators III: Gauge
  Coupling Dependence and Phenomenology}},
  \href{https://doi.org/10.1007/JHEP04(2014)159}{\emph{JHEP} {\bfseries 04}
  (2014) 159} [\href{https://arxiv.org/abs/1312.2014}{{\ttfamily 1312.2014}}].

\bibitem{1410.8849}
{\scshape NNPDF} collaboration, \emph{{Parton distributions for the LHC Run
  II}}, \href{https://doi.org/10.1007/JHEP04(2015)040}{\emph{JHEP} {\bfseries
  04} (2015) 040} [\href{https://arxiv.org/abs/1410.8849}{{\ttfamily
  1410.8849}}].

\bibitem{2406.06670}
F.~Maltoni, G.~Ventura and E.~Vryonidou, \emph{{Impact of SMEFT renormalisation
  group running on Higgs production at the LHC}},
  \href{https://arxiv.org/abs/2406.06670}{{\ttfamily 2406.06670}}.

\bibitem{2409.00168}
A.N.~Rossia and E.~Vryonidou, \emph{{CP-odd effects at NLO in SMEFT $WH$ and
  $ZH$ production}},  \href{https://arxiv.org/abs/2409.00168}{{\ttfamily
  2409.00168}}.

\bibitem{2203.02418}
A.~Azatov et~al., \emph{{Off-shell Higgs Interpretations Task Force: Models and
  Effective Field Theories Subgroup Report}},
  \href{https://arxiv.org/abs/2203.02418}{{\ttfamily 2203.02418}}.

\bibitem{1204.6427}
W.J.~Stirling and E.~Vryonidou, \emph{{Electroweak gauge boson polarisation at
  the LHC}}, \href{https://doi.org/10.1007/JHEP07(2012)124}{\emph{JHEP}
  {\bfseries 07} (2012) 124} [\href{https://arxiv.org/abs/1204.6427}{{\ttfamily
  1204.6427}}].

\bibitem{2311.09715}
{\scshape ATLAS} collaboration, \emph{{Measurement of ZZ production
  cross-sections in the four-lepton final state in pp collisions at s=13.6TeV
  with the ATLAS experiment}},
  \href{https://doi.org/10.1016/j.physletb.2024.138764}{\emph{Phys. Lett. B}
  {\bfseries 855} (2024) 138764}
  [\href{https://arxiv.org/abs/2311.09715}{{\ttfamily 2311.09715}}].

\bibitem{2409.16731}
M.~Grossi, G.~Pelliccioli and A.~Vicini, \emph{{From angular coefficients to
  quantum observables: a phenomenological appraisal in di-boson systems}},
  \href{https://arxiv.org/abs/2409.16731}{{\ttfamily 2409.16731}}.

\bibitem{1107.5569}
J.M.~Campbell, R.K.~Ellis and C.~Williams, \emph{{Gluon-Gluon Contributions to
  $W^+ W^-$ Production and Higgs Interference Effects}},
  \href{https://doi.org/10.1007/JHEP10(2011)005}{\emph{JHEP} {\bfseries 10}
  (2011) 005} [\href{https://arxiv.org/abs/1107.5569}{{\ttfamily 1107.5569}}].

\bibitem{2107.06579}
A.~Denner and G.~Pelliccioli, \emph{{NLO EW and QCD corrections to polarized ZZ
  production in the four-charged-lepton channel at the LHC}},
  \href{https://doi.org/10.1007/JHEP10(2021)097}{\emph{JHEP} {\bfseries 10}
  (2021) 097} [\href{https://arxiv.org/abs/2107.06579}{{\ttfamily
  2107.06579}}].

\bibitem{2209.14033}
J.A.~Aguilar-Saavedra, \emph{{Laboratory-frame tests of quantum entanglement in
  H\textrightarrow{}WW}},
  \href{https://doi.org/10.1103/PhysRevD.107.076016}{\emph{Phys. Rev. D}
  {\bfseries 107} (2023) 076016}
  [\href{https://arxiv.org/abs/2209.14033}{{\ttfamily 2209.14033}}].

\bibitem{2403.13942}
J.A.~Aguilar-Saavedra, \emph{{Tripartite entanglement in
  H\textrightarrow{}ZZ,WW decays}},
  \href{https://doi.org/10.1103/PhysRevD.109.113004}{\emph{Phys. Rev. D}
  {\bfseries 109} (2024) 113004}
  [\href{https://arxiv.org/abs/2403.13942}{{\ttfamily 2403.13942}}].

\bibitem{PhysRevD.43.1555}
C.~Kao and D.A.~Dicus, \emph{{Production of W+ W- from gluon fusion}},
  \href{https://doi.org/10.1103/PhysRevD.43.1555}{\emph{Phys. Rev. D}
  {\bfseries 43} (1991) 1555}.

\bibitem{2404.12809}
E.~Celada, T.~Giani, J.~ter Hoeve, L.~Mantani, J.~Rojo, A.N.~Rossia et~al.,
  \emph{{Mapping the SMEFT at high-energy colliders: from LEP and the (HL-)LHC
  to the FCC-ee}}, \href{https://doi.org/10.1007/JHEP09(2024)091}{\emph{JHEP}
  {\bfseries 09} (2024) 091}
  [\href{https://arxiv.org/abs/2404.12809}{{\ttfamily 2404.12809}}].

\bibitem{2303.05974}
{\scshape ATLAS} collaboration, \emph{{Probing the CP nature of the
  top\textendash{}Higgs Yukawa coupling in tt\textasciimacron{}H and tH events
  with H\textrightarrow{}bb\textasciimacron{} decays using the ATLAS detector
  at the LHC}},
  \href{https://doi.org/10.1016/j.physletb.2024.138469}{\emph{Phys. Lett. B}
  {\bfseries 849} (2024) 138469}
  [\href{https://arxiv.org/abs/2303.05974}{{\ttfamily 2303.05974}}].

\bibitem{1912.01725}
D.~Buarque~Franzosi, O.~Mattelaer, R.~Ruiz and S.~Shil, \emph{{Automated
  predictions from polarized matrix elements}},
  \href{https://doi.org/10.1007/JHEP04(2020)082}{\emph{JHEP} {\bfseries 04}
  (2020) 082} [\href{https://arxiv.org/abs/1912.01725}{{\ttfamily
  1912.01725}}].

\bibitem{2006.14867}
A.~Denner and G.~Pelliccioli, \emph{{Polarized electroweak bosons in ${\bf
  \text{W}^+\text{W}^-}$ production at the LHC including NLO QCD effects}},
  \href{https://doi.org/10.1007/JHEP09(2020)164}{\emph{JHEP} {\bfseries 09}
  (2020) 164} [\href{https://arxiv.org/abs/2006.14867}{{\ttfamily
  2006.14867}}].

\bibitem{2102.13583}
R.~Poncelet and A.~Popescu, \emph{{NNLO QCD study of polarised ${\bf
  \text{W}^+\text{W}^-}$ production at the LHC}},
  \href{https://doi.org/10.1007/JHEP07(2021)023}{\emph{JHEP} {\bfseries 07}
  (2021) 023} [\href{https://arxiv.org/abs/2102.13583}{{\ttfamily
  2102.13583}}].

\bibitem{2311.05220}
G.~Pelliccioli and G.~Zanderighi, \emph{{Polarised-boson pairs at the LHC with
  NLOPS accuracy}},
  \href{https://doi.org/10.1140/epjc/s10052-023-12347-4}{\emph{Eur. Phys. J. C}
  {\bfseries 84} (2024) 16} [\href{https://arxiv.org/abs/2311.05220}{{\ttfamily
  2311.05220}}].

\bibitem{2401.17365}
M.~Javurkova, R.~Ruiz, R.C.L.~de~S\'a and J.~Sandesara, \emph{{Polarized ZZ
  pairs in gluon fusion and vector boson fusion at the LHC}},
  \href{https://doi.org/10.1016/j.physletb.2024.138787}{\emph{Phys. Lett. B}
  {\bfseries 855} (2024) 138787}
  [\href{https://arxiv.org/abs/2401.17365}{{\ttfamily 2401.17365}}].

\bibitem{2409.06396}
T.N.~Dao and D.N.~Le, \emph{{Polarized $W^+W^-$ pairs at the LHC: Effects from
  bottom-quark induced processes at NLO QCD+EW}},
  \href{https://arxiv.org/abs/2409.06396}{{\ttfamily 2409.06396}}.

\bibitem{2303.10493}
C.~Degrande and H.-L.~Li, \emph{{Impact of dimension-8 SMEFT operators on
  diboson productions}},
  \href{https://doi.org/10.1007/JHEP06(2023)149}{\emph{JHEP} {\bfseries 06}
  (2023) 149} [\href{https://arxiv.org/abs/2303.10493}{{\ttfamily
  2303.10493}}].

\end{thebibliography}\endgroup

\end{document}